\documentclass[a4paper,11pt]{article}
\usepackage[english]{babel}
\usepackage{jheppub}
\usepackage{subfigure}
\usepackage{xspace}

\usepackage{amsmath,amssymb,graphicx,textcomp,enumerate,alltt,xspace,multirow}

\newcommand{\fb}{\mathrm{fb}}

\newcommand{\MEPSatLO}{{\rmfamily\scshape MEPS@LO}\xspace}
\newcommand{\OneLOop}{{\rmfamily\scshape OneLOop}\xspace}
\newcommand{\CutTools}{{\rmfamily\scshape CutTools}\xspace}
\newcommand{\FastJet}{{\rmfamily\scshape FastJet}\xspace}
\newcommand{\Powhel}{{\rmfamily\scshape Powhel}\xspace}
\newcommand{\Powheg}{{\rmfamily\scshape Powheg}\xspace}
\newcommand{\PowhegBox}{{\rmfamily\scshape Powheg-Box}\xspace}
\newcommand{\PowhegBoxRes}{{\rmfamily\scshape Powheg-Box-Res}\xspace}
\newcommand{\PowhegHooks}{{\rmfamily\scshape PowhegHooks}\xspace}
\newcommand{\PowhegOpenLoops}{{\rmfamily\scshape Powheg+OpenLoops}\xspace}
\newcommand{\PowhegPythia}{{\rmfamily\scshape Powheg+Pythia}\xspace}
\newcommand{\PowhegHerwig}{{\rmfamily\scshape Powheg+Herwig}\xspace}

\newcommand{\Collier}{{\rmfamily\scshape Collier}\xspace}
\newcommand{\Sherpa}{{\rmfamily\scshape Sherpa}\xspace}
\newcommand{\SherpaOpenLoops}{{\rmfamily\scshape Sherpa+OpenLoops}\xspace}
\newcommand{\OpenLoops}{{\rmfamily\scshape OpenLoops}\xspace}
\newcommand{\MadgraphaMC}{{\rmfamily\scshape Madgraph5aMC@NLO}\xspace}
\newcommand{\Pythia}{{\rmfamily\scshape Pythia}\xspace}
\newcommand{\Herwig}{{\rmfamily\scshape Herwig}\xspace}
\newcommand{\Rivet}{{\rmfamily\scshape Rivet}\xspace}

\newcommand{\mathd}{\mathrm{d}}

\newcommand{\tmop}[1]{\ensuremath{\operatorname{#1}}}

\newcommand{\GeV}{\text{GeV}\xspace}

\newcommand{\rB}{\mathrm{B}}
\newcommand{\rad}{\mathrm{rad}}
\newcommand{\rT}{\mathrm{T}}
\newcommand{\rF}{\mathrm{F}}
\newcommand{\pt}{p_{\rT}}

\newcommand{\qcut}{Q_{\mathrm{cut}}}
\newcommand{\Nmax}{N_{\mathrm{max}}}
\newcommand{\IS}{\mathrm{IS}}
\newcommand{\FS}{\mathrm{FS}}

\newcommand{\xis}{x_{\mathrm{IS}}}
\newcommand{\xfs}{x_{\mathrm{FS}}}
\newcommand{\yis}{y_{\mathrm{IS}}}

\newcommand{\ttbar}{\ensuremath{t \bar t}\xspace}
\newcommand{\bbbar}{\ensuremath{b \bar b}\xspace}
\newcommand\ttbb{\ensuremath{t \bar t b\bar b}\xspace}
\newcommand{\Nb}{N_b}

\newcommand{\muF}{\ensuremath{\mu_{\mathrm{F}}}}
\newcommand{\muR}{\ensuremath{\mu_{\mathrm{R}}}}

\def\beq{\begin{equation}}
\def\beqn{\begin{eqnarray}}
\def\eeq{\end{equation}}
\def\eeqn{\end{eqnarray}}

\def\({\left(} 
\def\){\right)} 
 
\newcommand     \MSB            {\ifmmode {\overline{\rm MS}} \else
                                 $\overline{\rm MS}$\fi}

\newcommand\as{\alpha_{\rm S}}
\newcommand\asfour{\alpha^{(4\rF)}_{\rm S}}
\newcommand\asfive{\alpha^{(5\rF)}_{\rm S}}

\newcommand\muf{\mu_{\rm F}}
\newcommand\mur{\mu_{\rm R}}
\newcommand\xir{\xi_{\rm R}}
\newcommand\xif{\xi_{\rm F}}

\newcommand\pT{\ensuremath{p_{\rm T}}}

\newcount\minutes 
\newcount\scratch 
\def\timestamp{%
\scratch=\time 
\divide\scratch by 60 
\edef\hours{\the\scratch} 
\multiply\scratch by 60 
\minutes=\time 
\advance\minutes by -\scratch 
---$\,$\hours:\null 
\ifnum\minutes< 10 0\fi 
\the\minutes}

\def\refeq#1{\mbox{(\ref{#1})}}
\def\refeqs#1#2{\mbox{(\ref{#1})--(\ref{#2})}}
\def\reffi#1{\mbox{Fig.~\ref{#1}}}
\def\reffis#1#2{\mbox{Figures~\ref{#1}--\ref{#2}}}
\def\refta#1{\mbox{Table~\ref{#1}}}

\def\refse#1{\mbox{Section~\ref{#1}}}
\def\refses#1#2{\mbox{Sections~\ref{#1}--\ref{#2}}}

\def\citere#1{\mbox{Ref.~\cite{#1}}}
\def\citeres#1{\mbox{Refs.~\cite{#1}}}

\def\rd{\mathrm{d}}

\newcommand{\sing}{\mathrm{s}}
\newcommand{\fin}{\mathrm{f}}
\newcommand{\damp}{\mathrm{damp}}
\newcommand{\hdamp}{\ensuremath{h_{\mathrm{damp}}}\xspace}
\newcommand{\bzd}{\mathrm{bzd}}
\newcommand{\hbzd}{\ensuremath{h_{\mathrm{bzd}}}\xspace}
\newcommand{\calR}{\mathcal{R}}
\newcommand{\calK}{\mathcal{K}}

\newcommand{\ie}{i.e.~}

\newcommand{\pp}[2]{(p_{#1}\cdot p_{#2})}

\preprint{
\begin{flushright}
IPPP/18/7\\
ZU-TH 06/18   
\end{flushright}
}

\title{New NLOPS predictions for $\boldsymbol{\ttbar+b}$-jet production at the LHC}

\author[a]{Tom\'a\v{s} Je\v{z}o,} 
\author[b]{Jonas M. Lindert,}
\author[a]{Niccolo Moretti,}
\author[a]{and Stefano Pozzorini}
\emailAdd{tomas.jezo@physik.uzh.ch}
\emailAdd{jonas.m.lindert@durham.ac.uk}
\emailAdd{moretti@physik.uzh.ch}
\emailAdd{pozzorin@physik.uzh.ch}

\affiliation[a]{
  Physics Institute, Universit\"at Z\"urich, Z\"urich, Switzerland
}
\affiliation[b]{
  Institute for Particle Physics Phenomenology, Durham University, South Rd, Durham DH1 3LE, UK
}

\abstract{Measurements of $\ttbar H$ production in the $H\to \bbbar$ channel
depend in a critical way on the theoretical uncertainty 
associated with the irreducible $\ttbar+b$-jet
background.
In this paper, analysing the various topologies that 
account for $b$-jet production in association with a
$\ttbar$ pair, we demonstrate 
that the process at hand is largely driven by 
final-state $g\to \bbbar$ splittings.
We also show that in 
five-flavour simulations based on $\ttbar+$multi-jet merging,
$b$-jet production is mostly driven by the parton shower,
while matrix elements play only a marginal role in the description of $g\to \bbbar$
splittings.
Based on these observations we advocate the use of NLOPS simulations of
$pp\to \ttbb$ in the four-flavour scheme, and we present a new \Powheg
generator of this kind.
Predictions and uncertainties for $\ttbar+b$-jet observables at the
13\,TeV LHC are presented both for the case of stable top quarks and with
spin-correlated top decays.
Besides QCD scale variations we consider also 
theoretical uncertainties related to the \Powheg matching method
and to the parton shower modelling,
with emphasis on $g\to \bbbar$ splittings.
In general, matching and shower uncertainties turn out to be 
remarkably small. This is confirmed also by 
a consistent comparison against \SherpaOpenLoops.
}


\begin{document}
\maketitle

\flushbottom

\section{Introduction}

At the Large Hadron Collider, searches for $\ttbar H$ production 
in the $H\to \bbbar$ channel 
are plagued by a large QCD background, which is dominated by 
$\ttbb$ production,
and  the availability of precise theoretical predictions 
for this multi-particle background process is of crucial
importance for the 
sensitivity of $\ttbar H(\bbbar)$ analyses.
The process $pp\to \ttbb$ is also very interesting on its own, as it
provides a unique laboratory to explore the QCD dynamics of heavy-quark
production and to test state-of-the-art Monte Carlo predictions in a
nontrivial multi-scale environment.

As a result of its $\as^4$ dependence, the leading-order (LO) $\ttbb$
cross section is highly sensitive to variations of the renormalisation
scale.  The uncertainty corresponding to standard factor-two scale
variations amounts to 70-80\% at LO, and the inclusion of next-to-leading
order (NLO) QCD corrections~\cite{Bredenstein:2009aj,Bevilacqua:2009zn,Bredenstein:2010rs} 
is mandatory.  At NLO, the scale dependence goes down 
to 20--30\%, and 
in order to avoid excessively  large $K$-factors and potentially large 
corrections beyond NLO, the 
renormalisation scale should be chosen in a way that 
accounts for the fact that the typical energies of the 
$b$-jet system are far below the hardness of the 
underlying  $pp\to\ttbar$ process~\cite{Bredenstein:2010rs}.

The first NLOPS simulation of $pp\to \ttbb$
was carried out in \Powhel~\cite{TROCSANYI:2014lha,Garzelli:2014aba}
by combining NLO matrix elements in the five-flavour (5F) scheme
with parton showers by means of the \Powheg method~\cite{Nason:2004rx, Frixione:2007vw}.
Shortly after, an NLOPS generator based on four-flavour (4F)
$pp\to \ttbb$ matrix elements became available in the
\SherpaOpenLoops framework~\cite{Cascioli:2013era}, which
implements an improved version~\cite{Hoeche:2011fd}
of the MC@NLO matching method~\cite{Frixione:2002ik}.
Thanks to the inclusion of $b$-mass effects, 
$\ttbb$ matrix elements in the 4F scheme are applicable 
to the full $b$-quark phase space, including regions 
where one $b$-quark remains unresolved.
Thus the 4F scheme guarantees a consistent NLOPS description of 
inclusive  $\ttbar+b$-jet production 
with one or more $b$-jets.
On the contrary, NLOPS $\ttbb$ generators based on 5F matrix 
elements with massless $b$-quarks suffer from collinear $g\to \bbbar$
singularities that require ad-hoc restrictions of the physical phase space 
through generation cuts.

In~\citere{Cascioli:2013era} it was pointed out that 
matching and shower effects play an
unexpectedly important role in $\ttbar+b$-jet production.  
This is due to the fact that two 
hard $b$-jets can arise from two hard jets involving 
each a collinear $g\to \bbbar$ splitting. 
In NLOPS simulations of $pp\to \ttbb$, 
such configurations result from the combination of 
a $g\to \bbbar$ splitting that is described at NLO accuracy 
through $\ttbb$ matrix elements together with 
a second $g\to \bbbar$ splitting generated by the parton shower.
The impact of this so-called double-splitting mechanism can 
have similar magnitude to the $\ttbar H(\bbbar)$ signal,
and the thorough understanding of the related 
matching and shower uncertainties is very important for 
$\ttbar H$ analyses.

A first assessment of NLOPS uncertainties 
was presented in~\citere{deFlorian:2016spz} 
through a tuned comparison of NLOPS $\ttbb$ simulations in 
\Powhel~\cite{TROCSANYI:2014lha,Garzelli:2014aba},
\SherpaOpenLoops~\cite{Cascioli:2013era} and \MadgraphaMC~\cite{Alwall:2014hca}.
On the one hand, this study has revealed significant differences
between the two generators based on the MC@NLO matching method\footnote{Here one should 
keep in mind that the \MadgraphaMC and
\Sherpa implementations are not identical. 
For instance, only the latter guarantees 
an exact $\mathcal{O}(\alpha_S)$ description of soft radiation, which
is achieved by upgrading the first shower emission and 
the related Catani-Seymour counterterm to full-colour accuracy.
}, 
i.e. \Sherpa and \MadgraphaMC. Such differences were found to be related to a
pronounced dependence on the shower starting scale in 
\MadgraphaMC.
On the other hand, in spite of the fact that 
\SherpaOpenLoops and \Powhel implement different matching methods 
and different parton showers,
the predictions of these two generators turned out to be quite consistent.
However, due to the limitations related to the use of the 5F scheme in
\Powhel---which have been overcome only very recently with the 4F upgrade 
of \Powhel~\cite{Bevilacqua:2017cru}---the agreement between \Powhel
and \SherpaOpenLoops did not allow to draw any firm conclusion in the
study of~\citere{deFlorian:2016spz}.

To date the assessment of theoretical uncertainties in 
$\ttbb$ production remains an important open problem.
In this context one should address the question of which 
of the various NLOPS methods and tools
on the market are more or less appropriate to describe 
the process at hand.
Moreover, in order to 
address such issues in a systematic way,
it is desirable to develop a better picture of the QCD dynamics that 
drive $\ttbar+b$-jet production.
In this spirit, this paper starts with a discussion of the various possible frameworks for
theoretical simulations of $\ttbar+b$-jet production at NLOPS accuracy.  In
particular, we present detailed studies on the role of $g\to \bbbar$
splittings and discuss the advantages and disadvantages of the 4F and 5F
schemes.
To this end we quantify the relative importance of $g\to \bbbar$ splittings
of initial-state (IS) and final-state (FS) type by using approximations based on collinear
QCD factorisation, as well as by decomposing $pp\to \ttbb$ matrix
elements into diagrams involving IS and FS $g\to \bbbar$ splittings.
These studies demonstrate that $\ttbar+b$-jet production is widely dominated
by $pp \to \ttbar g$ followed by FS $g\to \bbbar$ splittings.  This holds
also for observables where initial-state splittings are expected to be enhanced, such as
in regions with a single resolved $b$-jet.
These findings support the use of NLOPS generators based on $pp\to \ttbb$
matrix elements in the 4F scheme, where $b$-mass
effects guarantee a consistent treatment of FS $g\to \bbbar$ splittings.  
We also consider more inclusive simulations of $\ttbar+$jets production based
on multi-jet merging~\cite{Catani:2001cc,Mangano:2001xp,Hoeche:2009rj,Hoeche:2012yf,Lonnblad:2012ix,Frederix:2012ps}.  
In this case we find that $\ttbar+b$-jet
observables suffer from an unexpectedly strong dependence on the
parton-shower modeling of $g\to \bbbar$ splittings.

Motivated by the above findings we present a new \Powheg generator for
$pp\to \ttbb$ in the 4F scheme.\footnote{Preliminary results of this project have been 
presented in 2017 at the 25th International Workshop on Deep Inelastic Scattering and Related Topics~\cite{talkDIS17}
and QCD@LHC 2017~\cite{talkQCDatLHC17} conferences.}
At variance with the \Powhel generator of
\citere{Bevilacqua:2017cru}, this new \Powheg generator is implemented in the
\PowhegBoxRes framework~\cite{Jezo:2015aia} using \OpenLoops, which guarantees a
very fast evaluation of the required $2\to 4$ and $2\to 5$ matrix elements.
The new generator supports also top-quark decays including 
spin-correlation effects.
Moreover,  in order to guarantee
a more consistent resummation of QCD radiation,
the separation of the so-called singular and finite parts 
in the \PowhegBox 
is not restricted to initial-state radiation as in~\citere{Bevilacqua:2017cru}
but is applied also to final-state radiation (see \refse{se:technical}).

For what concerns the \Powheg methodology we pay particular attention to
issues related to the multi-scale nature of the process at hand.  In
particular we point out that the treatment of the recoil associated with NLO
radiation can induce sizeable distortions of the underlying $\ttbb$ cross
section. This technical inconvenience
restricts the domain of applicability of QCD factorisation 
in a way that can jeopardise the efficiency of event generation 
and can also lead to unphysical resummation effects.
Fortunately, such issues can be avoided by means of a \PowhegBox mechanism
that restricts the resummation of real radiation to kinematic regions
where QCD factorisation is fulfilled within reasonably good accuracy.

Predictions for  $pp\to \ttbar+b$-jets at the 13\,TeV LHC are presented for
various cross sections and distributions with emphasis on the discussion of
theoretical uncertainties.  Besides QCD scale variations, also uncertainties
related to the matching method and intrinsic shower uncertainties are
analysed in detail.  In particular, we consider different approximations for the
modelling of $g\to \bbbar$ splitting as well as $\as$ scale uncertainties in
\Pythia.  Moreover we compare \Pythia to \Herwig.
Finally, to gain further insights into the size and the nature of matching
and shower uncertainties we present a consistent comparison of \PowhegPythia
generators of $\ttbb$ and inclusive $\ttbar$ production, against 
corresponding generators based on \SherpaOpenLoops.

The new  $\ttbb$ generator will be soon publicly available 
on the \PowhegBox web site~\cite{PWGBpage}.

The paper is organised as follows.  In \refse{se:gbbstudies} we study the 
role of $g\to \bbbar$ splittings in $\ttbar+b$-jet production, and we
point out the advantages of Monte Carlo generators based on $pp\to \ttbb$ matrix
elements in the 4F scheme as compared to more inclusive generators 
of $\ttbar$ production in the 5F scheme.
Technical aspects of the new \Powheg generator and the setup for numerical
simulations are discussed in \refse{se:technical}.  In particular, in
\refse{se:methods} we review aspects of the \Powheg method that can play a
critical role for multi-scale process like $pp\to \ttbb$.
Detailed predictions and uncertainty estimates for cross sections and
distributions for $pp\to \ttbb$ with stable and unstable top quarks can be
found in Sections \ref{se:stabletops} and \ref{se:decayedtops},
respectively.  
Our main findings are summarised in \refse{se:conclusions}.

\section{Anatomy of $\boldsymbol{\ttbar+b}$-jet production and
$\boldsymbol{g\to \bbbar}$ splittings}
\label{se:gbbstudies}

Events with $\ttbar+b$-jets final states arise from an underlying
$pp\to \ttbar$ process that takes place at scales of the order of 500\,GeV
and is accompanied by the production of $b$-jets at 
typical transverse momenta of a few tens of GeV.
The production of $b$-jets is governed by IS or FS $g\to \bbbar$
splittings and is enhanced in kinematic regions where
the $p_\rT$ of individual $b$-quarks becomes small
or the $\bbbar$ pair becomes collinear and, possibly, also soft.

The understanding of the QCD dynamics that governs $b$-jet production is a crucial prerequisite 
for a reliable theoretical description of $\ttbar+b$-jet production and related 
uncertainties.  In this spirit, this section compares various
theoretical frameworks for the description of $\ttbar+b$-jet production at
NLO QCD accuracy with a special focus on the role of $g\to \bbbar$ splittings.
Specifically, we compare inclusive or merged simulations of
$\ttbar+$multi-jet production in the 5F scheme against a
description based on $\ttbb$ matrix elements in the 4F
scheme, pointing out the advantages of the latter.

Numerical studies presented in this section  
are based on the setup specified in Sections~\ref{se:input} and~\ref{se:cuts}
and have been performed with \SherpaOpenLoops.

\subsection[NLOPS $\ttbar$ simulations in the five-flavour scheme]{NLOPS $\boldsymbol{\ttbar}$ simulations in the five-flavour scheme}
\label{se:ttnlops}

\begin{figure}[t!]
\centering
\includegraphics[width=0.21\textwidth]{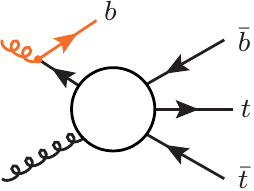}
\hspace{.15\textwidth}
\includegraphics[width=0.21\textwidth]{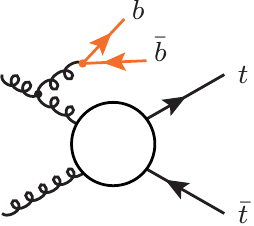}
\caption{Sample $\ttbb$ diagrams with IS (left) and FS (right)
$g\to \bbbar$ splittings. In NLOPS simulations of inclusive $\ttbar$ production
in the 5F scheme the black subtopologies are described in terms of tree matrix elements,
while the orange lines correspond to parton shower emissions.}
\label{fig:ttinctopologies}
\end{figure}

Inclusive NLOPS generators of $\ttbar$
production~\cite{Frixione:2003ei,Frixione:2007nw} are based on $pp\to
\ttbar$ NLO matrix elements matched to partons showers in the 5F scheme.  In
this framework, as illustrated in \reffi{fig:ttinctopologies},
$\ttbar+b$-jet events are generated starting from $2\to 3$ tree matrix
elements of type $gb\to \ttbar b$ or $gg/q\bar q\to \ttbar g$.
In the latter case, $\ttbb$ events arise 
via FS $g\to \bbbar$ shower splittings.
Instead, in the case of $gb\to \ttbar b$ 
the final-state $b$-quark emerges from the matrix element,
while $g\to \bbbar$ splittings generate 
the initial-state $b$-quark through the evolution of the 5F PDFs. 
The unresolved spectator $b$-quark associated with such IS $g\to \bbbar$ splittings
is emitted by the parton shower via backward evolution.
The main advantage of the 5F scheme lies in the resummation of potentially
large $\as\ln(m_t/m_b)$ terms associated with the evolution of the $b$-quark
density.  However, such logarithmic effects are typically rather mild at the
LHC~\cite{Maltoni:2012pa}.  Moreover, as we will show in \refse{se:ttbbISvsFS},
$\ttbar+b$-jet production is largely dominated by topologies with FS
$g\to \bbbar$ splittings.
For this reason, $\ttbar+b$-jet predictions 
based on NLOPS $\ttbar$ generators suffer from the twofold disadvantage
given by the direct dependence on the parton-shower modelling of
FS $g\to \bbbar$ splittings plus
the LO nature of the underlying $\ttbar g$ matrix element.

\subsection[$\ttbar+$multi-jet merging in the five-flavour scheme]{$\boldsymbol{\ttbar+}$multi-jet merging in the five-flavour scheme}
\label{se:mepslo}

\def\Neff{N_{\mathrm{eff}}}
\def\Nres{N}

As a possible strategy to reduce the sensitivity to the parton shower and
increase the accuracy of theoretical predictions
we consider $\ttbar+$multi-jet merging in the 5F scheme.
In this approach, a tower of NLOPS simulations
for $\ttbar+0,1,\dots,N$\,jet production is merged into a single inclusive 
sample~\cite{Hoeche:2012yf,Lonnblad:2012ix,Frederix:2012ps}.
This is achieved by clustering QCD partons into jets with a certain
$k_\rT$-resolution,  $\qcut$, which is known as merging scale.  
At LO, the phase-space regions with $\Nres=0,1,\dots, \Nmax$
resolved jets ($k_T> \qcut$) are described in terms of
\mbox{$\ttbar+\Nres$}-jet LOPS simulations.
The LOPS simulation  with $\Nmax$ jets fills also the 
phase space with $\Nres>\Nmax$ resolved jets by means 
of the parton shower.
At NLO, the resolution criterion used to separate regions of different 
jet multiplicity exactly the same as for LO, while 
the
basic difference with respect to LO merging lies in the fact that 
\mbox{$\ttbar+N$-}jet LOPS simulations are replaced by corresponding NLOPS
simulations. 
Thus in NLO (LO) merging the effective number of resolved jets that 
is described at NLOPS (LOPS) accuracy is $\Neff=\mathrm{min}\{\Nres,N_\mathrm{\max}\}$,
while the $(\Neff+1)^{\mathrm{th}}$ resolved or unresolved jet is
described at LOPS (pure PS) accuracy, and all remaining resolved or unresolved jets 
are described at pure PS accuracy.

Multi-jet merging for $\ttbar+$jet at NLO can be performed in a fully
automated way within the 
\Sherpa~\cite{Gleisberg:2008ta} and 
\MadgraphaMC~\cite{Alwall:2014hca} frameworks.
However, the fact that $\ttbar+b$-jet events constitute only a small 
fraction of a $\ttbar+$jets sample poses very high requirements 
in terms of Monte Carlo statistics. 
Moreover, in order to minimise the dependence on parton-shower modelling,
such simulations should be performed using a small merging scale
and including a sufficiently high number of NLO jets, $\Nmax$.
The required CPU resources grow very fast at large $\Nmax$ and small
$\qcut$, and state-of-the-art merged simulations can 
handle up to two jets at NLO~\cite{Hoeche:2014qda} at present.

\begin{figure}[t!]
\centering
\subfigure[]{
\includegraphics[width=0.2\textwidth]{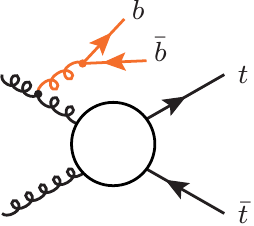}
}
\subfigure[]{
\includegraphics[width=0.2\textwidth]{graphs/ggttbbMEPSb}
}\hspace{.05\textwidth}
\subfigure[]{
\includegraphics[width=0.2\textwidth]{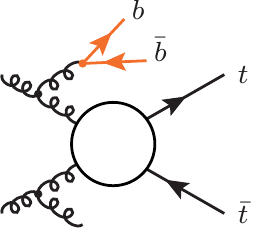}
}\hspace{.05\textwidth}
\subfigure[]{
\includegraphics[width=0.2\textwidth]{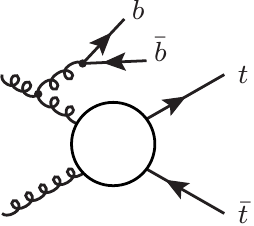}
}\hspace{.05\textwidth}
\caption{Sample diagrams describing the interplay between 
matrix elements (black) and parton shower (orange)
for $\ttbar+b$-jet events 
in merged LOPS simulations of $\ttbar+0,1,2$\,jet production in the 5F scheme.
In regions where the $\bbbar$ system and its parent gluon are produced below the 
merging scale, the event is described through hard $\ttbar$ matrix elements
plus parton-shower branchings (a). When the parent gluon becomes harder,
the parton shower is used only to model 
collinear $g\to\bbbar$ splittings (b). Above the merging scale,
if $g\to \bbbar$ splittings do not belong to the two 
hardest branchings they are still left to the parton shower, 
while $2\to 4$ matrix elements are used to account for other 
light jets (c). Otherwise, the event is described through $\ttbb$ matrix elements 
(d). 
}
\label{fig:MEPStopologies}
\end{figure}

The multi-jet merging description of $\ttbb$ events with FS
$g\to
\bbbar$ splittings is sketched in \reffi{fig:MEPStopologies}
for the case of $\ttbar+0,1,2$-jet merging at LO.
In regions where the $\bbbar$ pair and/or the parent gluon are emitted at
small scales, $g\to \bbbar$ splittings are expected to be generated by the
parton shower, while hard $b$-jet pairs are expected to arise from $\ttbb$
matrix elements.
However, as we will see, typical $\ttbar+b$-jet events involve additional
light jets that are emitted at harder scales with respect to the $g\to
\bbbar$ branching.  In that case, $2\to 4$ matrix elements account only for
light jets, and $g\to \bbbar$ splittings are left to the parton shower.
In general, the relative importance of matrix elements and parton shower
depends on the resolution scale $\qcut$, and using a finite resolution is
mandatory in the 5F scheme, since collinear $g\to \bbbar$ splittings are
divergent, \ie $\ttbb$ matrix elements cannot be used in the full phase
space.

\begin{figure}[t!]
\centering
\subfigure[]{
\includegraphics[width=0.4\textwidth]{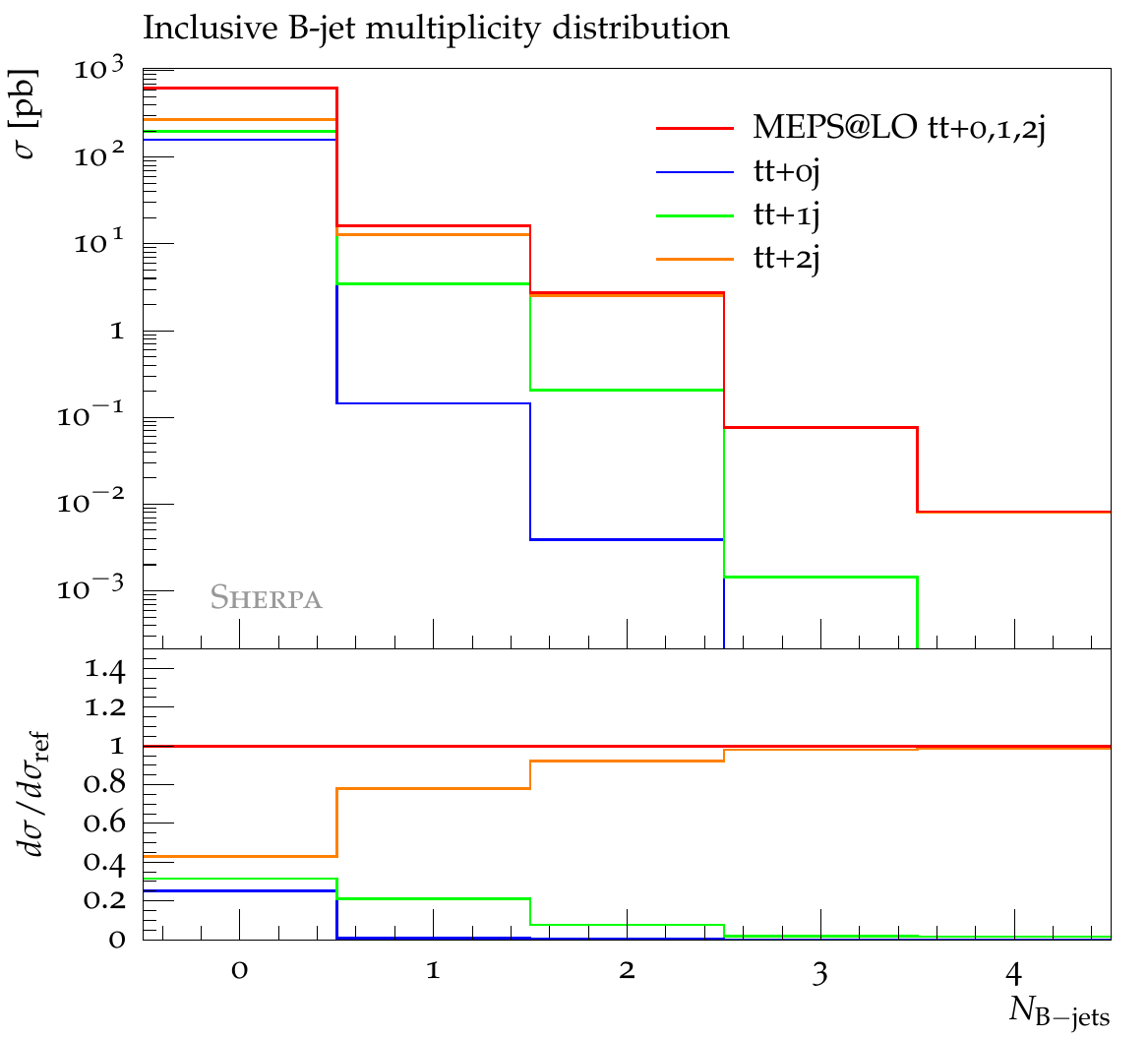}
}
\subfigure[]{
\includegraphics[width=0.4\textwidth]{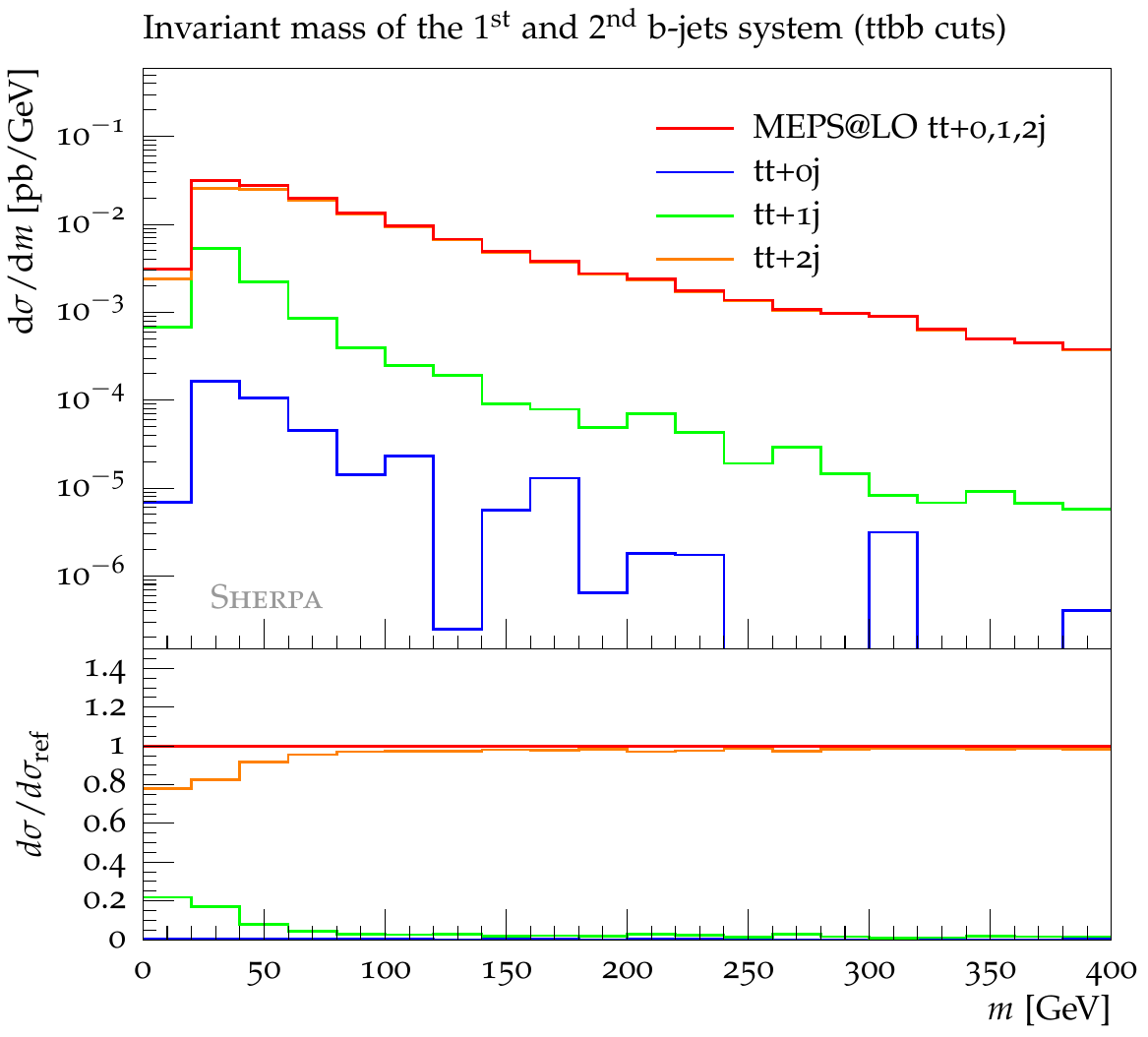}
}
\subfigure[]{
\includegraphics[width=0.4\textwidth]{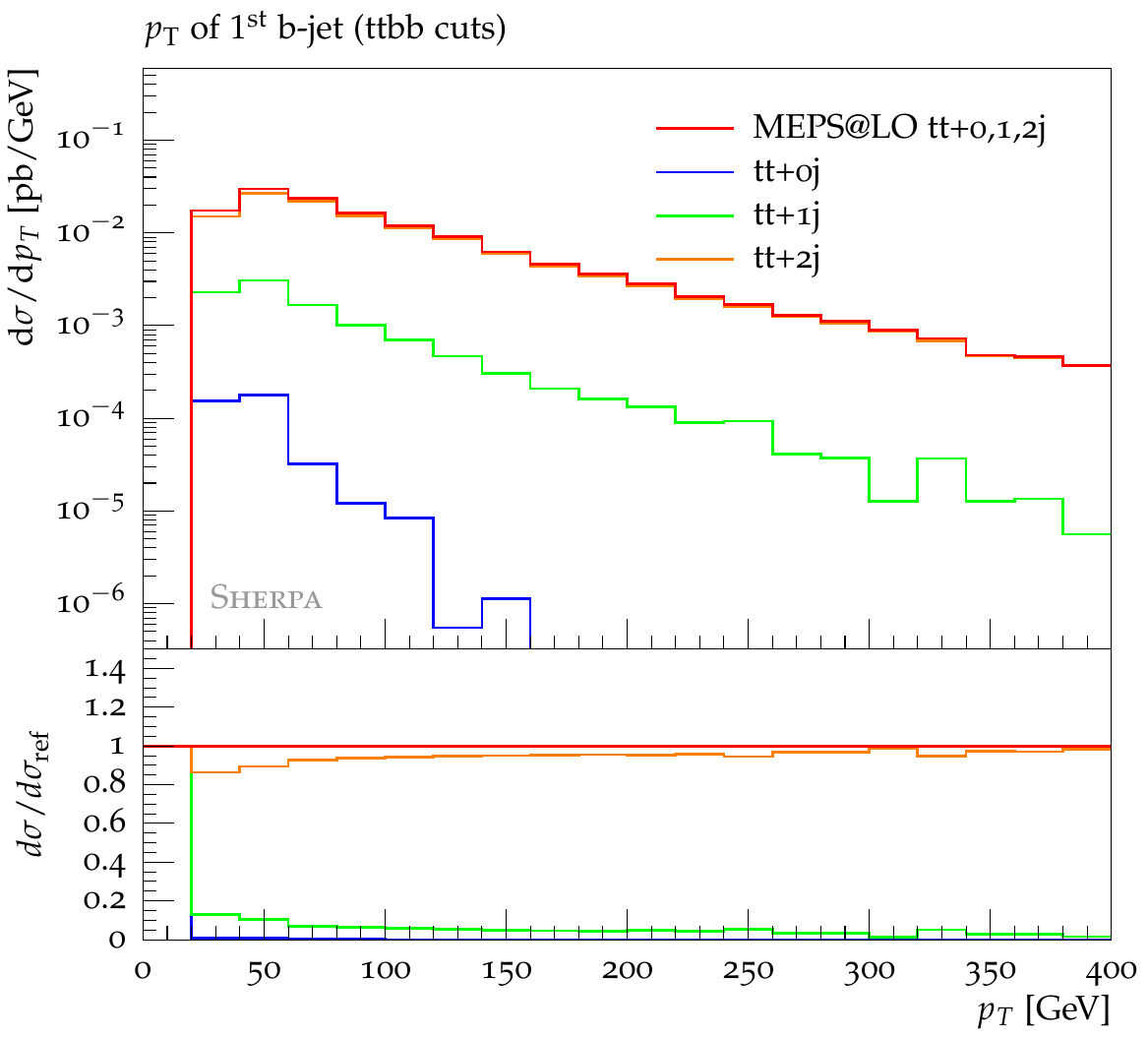}
}%
\caption{Breakdown of merged LOPS simulations of $pp\to \ttbar+0,1,2$\,jets production 
at 13\,TeV into 
the contributions from matrix elements with $\ttbar+$0,1 and 2 generic QCD partons:
distributions in the number of $b$-jets with $p_\rT>25$\,GeV (a),
the invariant mass of the two leading $b$-jets (b), 
and the  $p_\rT$ of the leading $b$-jet (c).
}
\label{fig:MEPSatLONjets}
\end{figure}

\begin{figure}[t!]
\centering
\subfigure[]{
\includegraphics[width=0.4\textwidth]{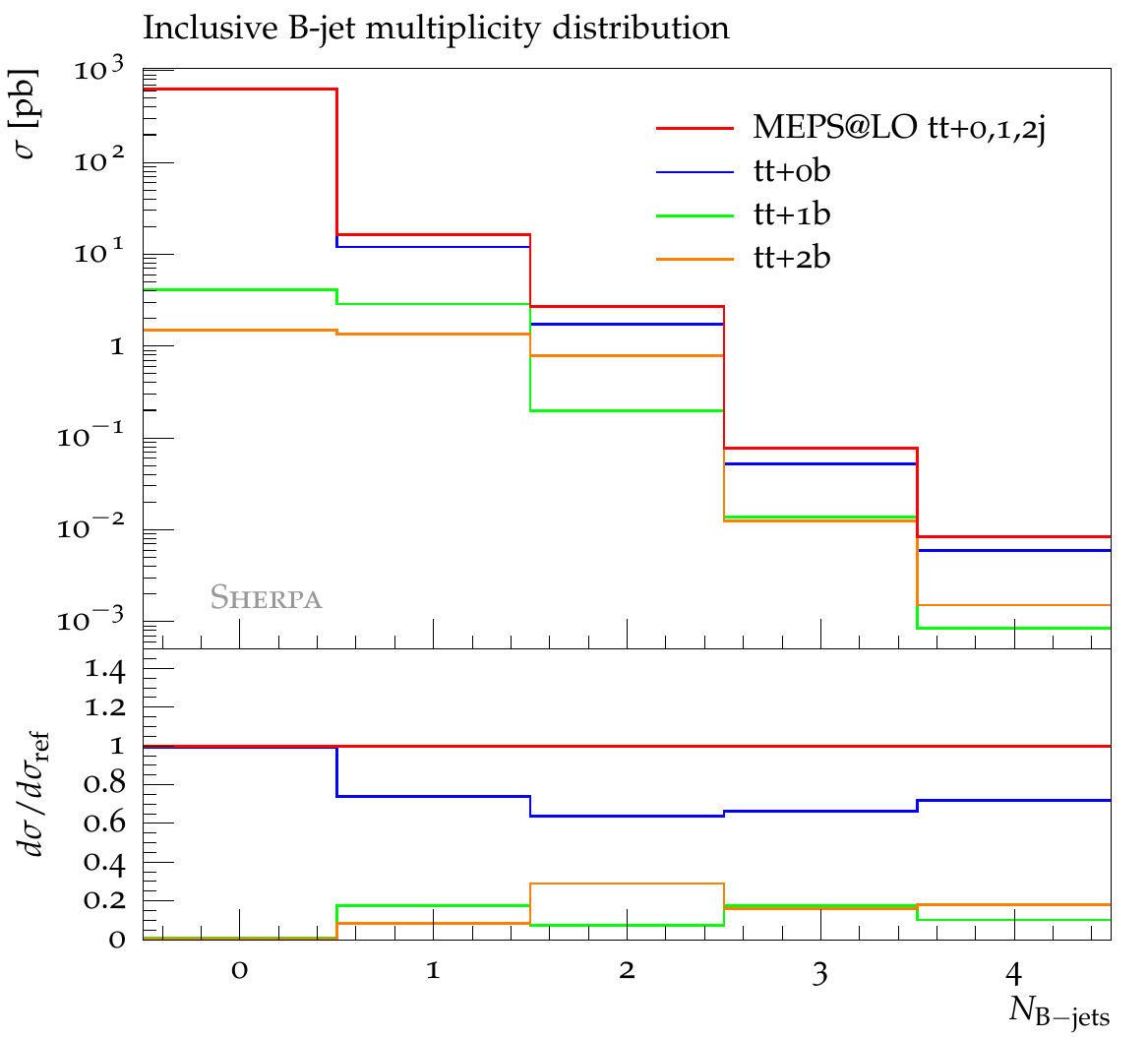}
}
\subfigure[]{
\includegraphics[width=0.4\textwidth]{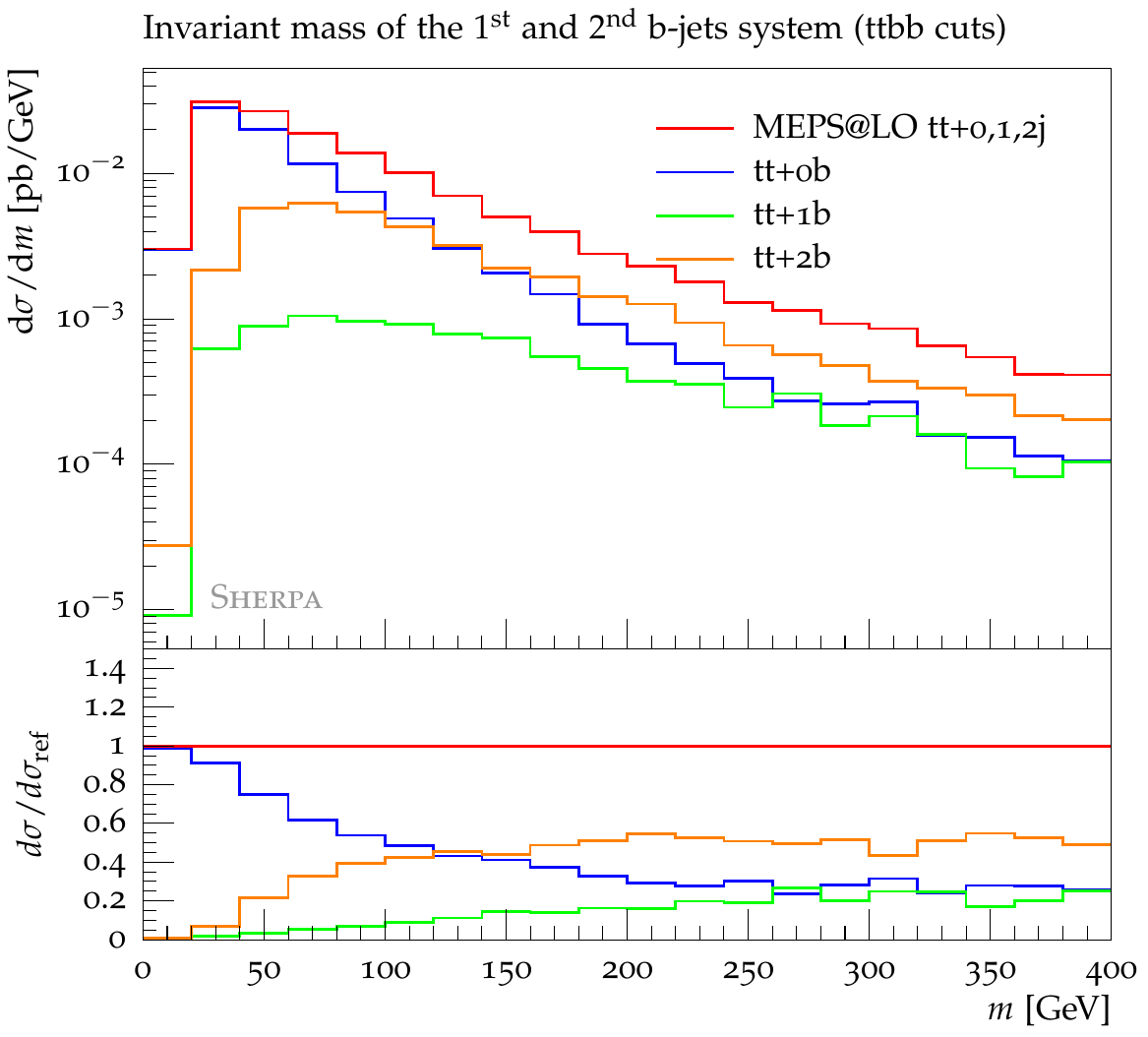}
}
\subfigure[]{
\includegraphics[width=0.4\textwidth]{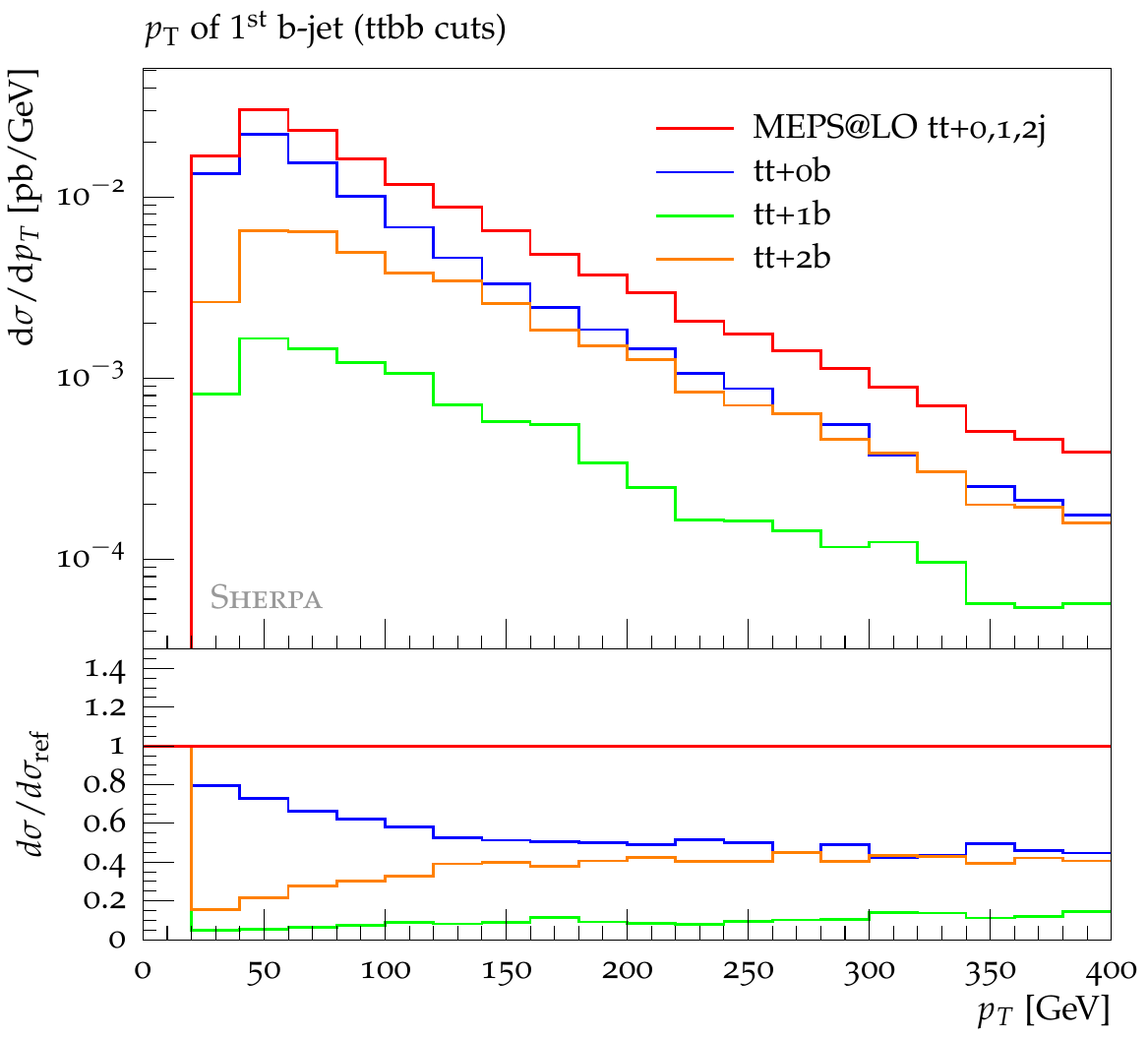}
}
\subfigure[]{
\includegraphics[width=0.4\textwidth]{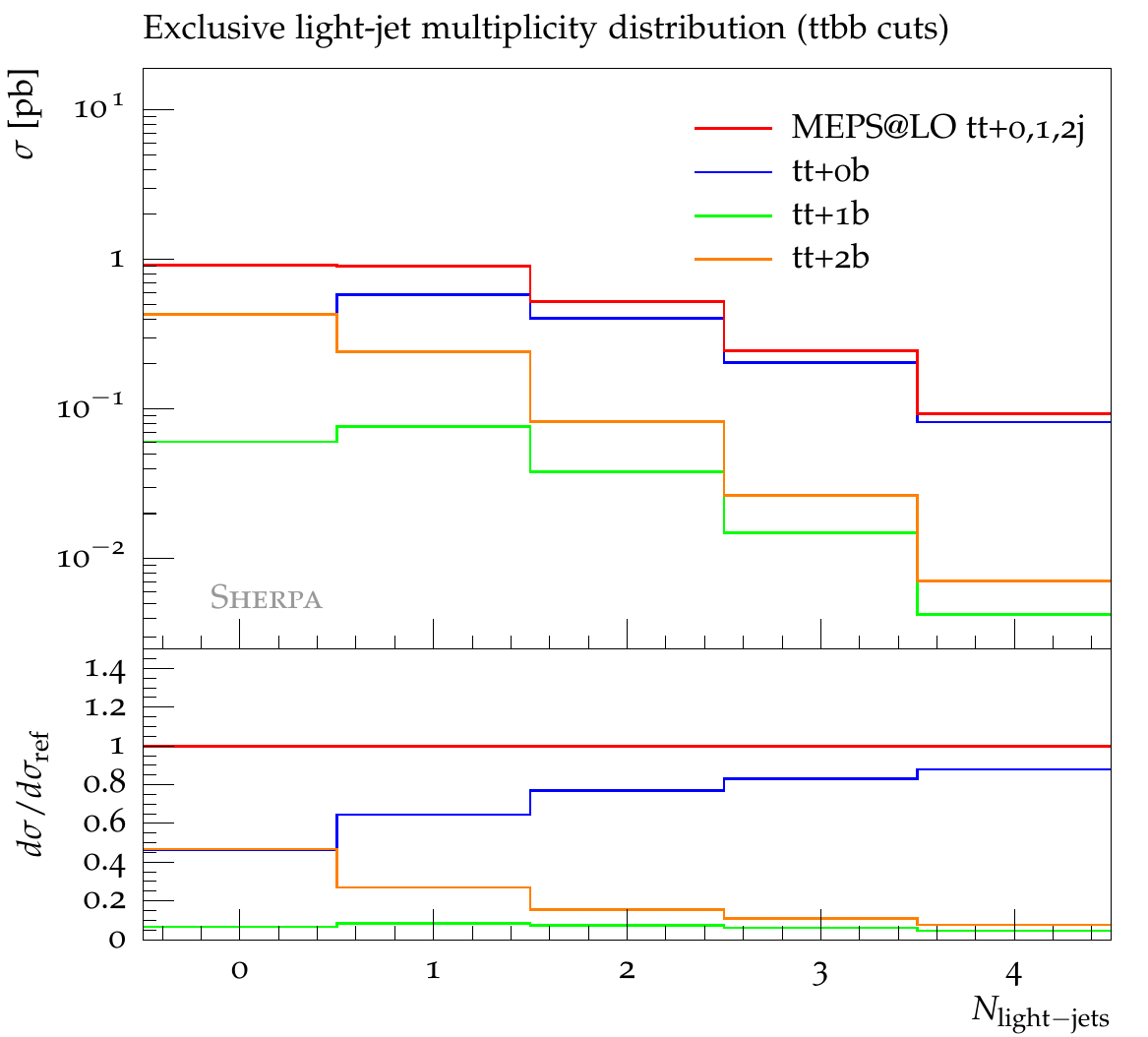}
}
\caption{Breakdown of merged LOPS simulations of $\ttbar+0,1,2$\,jets production at 13\,TeV 
into 
the contributions from matrix elements with $\ttbar+$0,1 and 2 $b$-quarks.
Same observables (a)--(c) as in \reffi{fig:MEPSatLONjets} 
and distribution in the 
exclusive number of light jets with $p_\rT>25$\,GeV (d) in the presence of ttbb cuts.
}
\label{fig:MEPSatLONbjets}
\end{figure}

In \reffis{fig:MEPSatLONjets}{fig:MEPSatLONbjets} we analyse the
matrix-element content of a $\ttbar+0,1,2$-jet merged simulation at LO.
These studies are based on the \MEPSatLO method~\cite{Hoeche:2009rj} implemented in \Sherpa,
but the main findings are expected to hold also for other merging
methods. 
Besides the $b$-jet multiplicity distribution, in
\reffis{fig:MEPSatLONjets}{fig:MEPSatLONbjets} we plot differential
observables in the presence of ttbb cuts, i.e.~requiring $\Nb\ge 2$ $b$-jets
as defined in \refse{se:cuts}.  For jets we apply the acceptance cuts
\refeq{eq:jetcuts} and, in order to maximise the possibility to resolve jets
at matrix-element level, we choose a merging scale lower than the
jet-$p_\rT$ threshold, $\qcut=20$\,GeV.

As expected, in \reffi{fig:MEPSatLONjets} we find that $\ttbar+b$-jet
observables with $\Nb\ge 2$ resolved $b$-jets are largely dominated by
$\ttbar+$2-parton matrix elements. This holds also for $\Nb\ge 1$.
However, the breakdown of the merged sample into contributions from matrix
elements with different $b$-quark multiplicity in \reffi{fig:MEPSatLONbjets}
reveals that, in spite of the low merging scale, the cross section for
producing one or more $b$-jets is dominated by matrix elements with zero
$b$-quarks.  The contribution of $\ttbb$ matrix elements hardly exceeds 50\%
even in the region of large $b$-jet $\pt$ or large invariant mass of the
$b$-jet pair.
This counterintuitive feature can be attributed to the fact that, in $\ttbar+$jet events that
involve $g\to \bbbar$ splittings, the two hardest QCD branchings are
typically associated with the emission of the parent gluon of the $\bbbar$
pairs and/or with the production of other light jets.
As a consequence, in LOPS merged samples with $\Nmax=2$, 
$g\to \bbbar$ splittings are
left to the parton shower.  We have verified that the contribution of
$\ttbb$ matrix elements remains relatively low even when the merging
procedure is extended up to 3 or 4 jets with 
$\qcut=20$\,GeV.
Moreover, we have checked that 
increasing the merging scale leads to a further suppression
of the contribution of $\ttbb$ matrix elements.

\reffi{fig:MEPSatLONbjets}d demonstrates that $\ttbar+2\,b$-jet events 
are indeed accompanied by abundant emission of extra light jets,
and the importance of $\ttbb$ matrix elements decreases with 
increasing light-jets multiplicity. 
Moreover, even in the bin with zero additional light jets it turns out that
the contribution of $\ttbb$ matrix elements remains below 50\%.  This is probably 
due to the fact that $g\to \bbbar$ splittings tend to take place at
branching scales below the jet-$p_\rT$-threshold of 25\,GeV.

In summary, contrary to naive expectations, $\ttbar+$jets samples
based on LOPS merging do not guarantee a matrix-element description of
$b$-jet production but largely rely on the parton shower modelling of $g\to
\bbbar$ splittings.  In the case of multi-jet merging at NLO, to a certain
extent $g\to \bbbar$ shower splittings should be matched to $\ttbb$ and
$\ttbb g$ tree matrix elements.  Nevertheless, based on the above
observations, the theoretical accuracy in the description of $b$-jet
production is expected to remain between the LOPS 
and the pure PS level.\footnote{
This is due to the fact that the jet-resolution criterion employed for LO and NLO merging is exactly the same,
while regions  that are described at pure PS accuracy in LO merging, i.e.~regions with
$N>\Nmax$, 
can be either promoted to LOPS accuracy or remain at pure PS accuracy 
in the case of NLOPS merging (see above).}
Finally, in the light of the presence of abundant light-jet radiation 
with a typical hardness beyond the one of 
$b$-jets, the role of 
hard radiation on top of the $\ttbb$ system should be studied with 
great care also in the context of NLOPS simulations of $\ttbb$ production
(see~\refse{se:methods}).

\subsection[$\ttbb$ production in the four-flavour scheme]{$\boldsymbol{\ttbb}$ production in the four-flavour scheme}
\label{se:ttbbISvsFS}

\begin{figure}[t!]
\centering
\subfigure[]{
\includegraphics[width=0.25\textwidth]{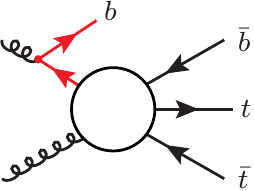}
\label{subfig:ISgen}
}\hspace{.15\textwidth}
\subfigure[]{
\includegraphics[width=0.25\textwidth]{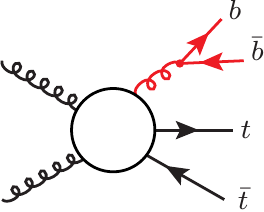}
\label{subfig:FSgen}
}
\\
\subfigure[]{
\includegraphics[width=0.25\textwidth]{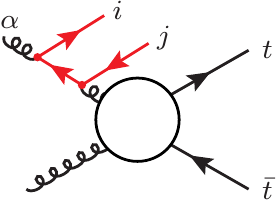}
\label{subfig:ISdom}
}\hspace{.15\textwidth}
\subfigure[]{
\includegraphics[width=0.25\textwidth]{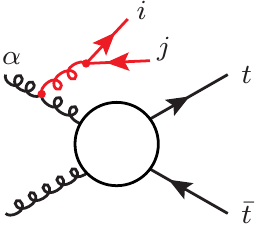}
\label{subfig:FSdom}
}
\caption{Generic leading-order $gg\to \ttbb$ topologies. The first line 
shows the most general form of topologies with 
IS (a) and FS (b) 
$g\to \bbbar$ splittings. The second line shows 
the generic form of those topologies with IS (c) and FS (d)
splittings that turn out to dominate 
$gg\to \ttbb$.
The labels $ij=56, 65$ stand for the $b \bar b$ system,
while $\alpha=1,2$ indicates the initial-state gluon that 
generates the radiation.
}
\label{fig:ttbbtopologies}
\end{figure}

In order to minimise the dependence on parton-shower modelling 
and to maximise the use of higher-order matrix elements, in the following 
we will adopt a description of $\ttbar+b$-jet production based on $ttbb$ matrix
elements in the 4F scheme.
In this scheme, $b$-quarks are treated as massive partons, and $g\to
\bbbar$ splittings are free from collinear singularities.  Thus $\ttbb$
matrix elements can be used in the entire phase space.
Generic $\ttbb$ topologies where $b$-quarks emerge from IS and FS splitting
processes are illustrated in \reffi{fig:ttbbtopologies}.
In the case of FS $g\to \bbbar$ splittings, $\ttbb$ matrix elements with
$m_b>0$ can be extended to the collinear regime, where the $\bbbar$ pair
becomes unresolved within a single $b$-jet.
Similarly, 4F $\ttbb$ matrix elements describe also collinear IS $g\to
\bbbar$ splittings, where the spectator $b$-quark is emitted in the beam
direction and remains unresolved, while the $bg\to \ttbar b$ sub-process
with a single $b$-jet corresponds to the description of $\ttbar+b$-jet
production at LO in the 5F scheme.
Thus, $\ttbb$ matrix elements provide a fully inclusive description of
$\ttbar+b$-jet production, and NLO predictions in the 4F scheme yield NLO
accuracy both for observables with two $b$-jets and for more inclusive
observables with a single resolved $b$-jet.

The inclusion of $m_b$ effects in $g\to \bbbar$ splittings represents a
clear advantage of the 4F scheme with respect to the 5F scheme.  However,
the 4F scheme has the disadvantage that potentially large $\as \ln(m_b/Q)$
terms that arise from IS $g\to \bbbar$ splittings are not resummed through the 
PDF evolution.
In the following, in order to assess the relevance of this limitation, 
we decompose the $pp\to \ttbb$ LO cross section into contributions from 
IS and FS $g\to \bbbar$ splittings.
Since the $q\bar q$ channel involves only FS $g\to \bbbar$ splitting, 
we focus on the $gg$ channel and
we first consider a naive diagrammatic 
splitting of the $gg\to\ttbb$ matrix element,
\begin{equation}
\label{eq:ISFSMEsplitting}
\mathcal{M}_{\ttbb}=
\mathcal{M}_{\mathrm{IS},\ttbb}
+ \mathcal{M}_{\mathrm{FS},\ttbb}+ 
\mathcal{M}_{\mathrm{rem},\ttbb}.
\end{equation}
The terms $\mathcal{M}_{\mathrm{IS},\ttbb}$ and
$\mathcal{M}_{\mathrm{FS},\ttbb}$ correspond, respectively, to 18 diagrams
with IS $g\to \bbbar$ splittings and 16 diagrams with FS $g\to \bbbar$
splittings.  Generic diagrams with IS and FS splittings are depicted in
\reffi{fig:ttbbtopologies}a--b.
The term $\mathcal{M}_{\mathrm{rem},\ttbb}$ corresponds to two remaining
diagrams where an $s$-channel gluon splits into an on-shell $b$-quark and an
off-shell $b$-line coupled to the $\ttbar$ system.  Its numerical impact
turns out to be negligible.
Based on \refeq{eq:ISFSMEsplitting} we split the $\ttbb$ cross section into three terms,
\begin{equation}
\label{eq:ISFSXSsplitting}
\rd\sigma_{\ttbb}= \rd\sigma_{\IS,\ttbb}
+\rd\sigma_{\FS,\ttbb}
+\rd\sigma_{\mathrm{int},\ttbb},
\end{equation}
where the IS and FS parts are defined as
\begin{eqnarray}
\label{eq:ISFSXSdef}
\rd\sigma_{\IS,\ttbb}=\frac{|\mathcal{M}_{\IS,\ttbb}|^2}{|\mathcal{M}_{\ttbb}|^2}\,
\rd\sigma_{\ttbb},
\qquad
\rd\sigma_{\FS,\ttbb}=\frac{|\mathcal{M}_{\FS,\ttbb}|^2}{|\mathcal{M}_{\ttbb}|^2}\,
\rd\sigma_{\ttbb},
\end{eqnarray}
while $\rd\sigma_{\mathrm{int},\ttbb}$ consists of the interference between
$\mathcal{M}_{\mathrm{IS},\ttbb}$ and $\mathcal{M}_{\mathrm{FS},\ttbb}$ plus
a minor contribution from $\mathcal{M}_{\mathrm{rem},\ttbb}$.

In order to check the soundness of the above gauge-dependent 
separation we compare it to an alternative definition of 
IS and FS $g\to \bbbar$ contributions  based on the 
collinear limits of the $gg\to \ttbb$ matrix element.
In this case we define
\begin{eqnarray}
\label{eq:ISFScoll}
\rd\sigma_{\IS\otimes\ttbar}=
\frac{|\mathcal{M}_{\IS\otimes\ttbar}|^2}
{|\mathcal{M}_{\ttbb}|^2}\,\rd\sigma_{\ttbb},
\qquad
\rd\sigma_{\FS\otimes\ttbar}=
\frac{|\mathcal{M}_{\FS\otimes\ttbar}|^2}
{|\mathcal{M}_{\ttbb}|^2}\,\rd\sigma_{\ttbb}.
\end{eqnarray}
Here $|\mathcal{M}_{\IS\otimes\ttbar}|^2$ and
$|\mathcal{M}_{\FS\otimes\ttbar}|^2$ describe the collinear limits of the
topologies depicted in \reffi{fig:ttbbtopologies}c and
\ref{fig:ttbbtopologies}d, respectively.
Note that, for simplicity, we consider only the leading
collinear enhancements where the $\bbbar$ system originates 
either through the combination of 
$g\to \bbbar$ and $b\to gb$ IS splittings ($\mathcal{M}_{\IS\otimes\ttbar}$)
or via IS $g\to gg$ plus FS $g\to \bbbar$ splittings
($\mathcal{M}_{\FS\otimes\ttbar}$).
For events with external momenta 
\begin{equation} 
\label{eq:ggttbbkin}
g(p_1)\,g(p_2) \to t (p_3)\, \bar t(p_4)\, b(p_5)\, \bar b(p_6),
\end{equation} 
the collinear limits take the general form
\begin{eqnarray}
\label{eq:ISFSapprox}
\big|\mathcal{M}_{\mathrm{IS}\otimes\ttbar/\mathrm{FS}\otimes\ttbar}\big|^2
&=&(8\pi\as)^2\,
\max_{\alpha=1,2 \atop ij=56,65}\left\{
\frac{K_{\IS/\FS}(p_\alpha,p_i,p_j)}{(p_\alpha-p_i-p_j)^2}\,
\big|\mathcal{M}_{gg\to\ttbar}\big|^2_{p_\alpha\to z p_\alpha}\right\},
\end{eqnarray}
where $\alpha\in\{1,2\}$ and $ij\in\{56,65\}$ specify, respectively, the
IS gluon emitter and the ordering of the $\bbbar$ pair as
depicted in \reffi{fig:ttbbtopologies}c and \ref{fig:ttbbtopologies}d.
$K_{\IS/\FS}(p_\alpha,p_i,p_j)$ are the corresponding splitting kernels.
The choice of $\alpha$ and $ij$ specifies a particular 
topology, and the maximum in \refeq{eq:ISFSapprox} defines 
$\big|\mathcal{M}_{\mathrm{IS}\otimes\ttbar}\big|^2$
and
$\big|\mathcal{M}_{\mathrm{FS}\otimes\ttbar}\big|^2$
as the collinear limit of the most likely 
topology of IS and FS type.
The splitting kernels read 
\begin{eqnarray}
\label{eq:ISapproxa}
K_{\mathrm{IS}}(p_\alpha,p_i,p_j)
&=&\frac{1}{(p_\alpha-p_i)^2-m_b^2}
\frac{P_{gq}(\xis)}{\xis}
\frac{P_{qg}(\yis)}{\yis},\nonumber\\
K_{\mathrm{FS}}(p_\alpha,p_i,p_j)
&=&\frac{-1}{(p_i+p_j)^2}
\frac{P_{gg}(z)}{z}
P_{gq}(\xfs)
,
\end{eqnarray}
where
\begin{align}
\label{eq:ISapproxc}
P_{gq}(x)&=T_R\left[1-2x(1-x)\right], &  P_{qg}(x)& =C_F\left[\frac{1+(1-x)^2}{x}\right],
\nonumber\\
P_{gg}(x)&= 2C_A\left[\frac{x}{1-x}+\frac{1-x}{x}+x(1-x)\right], &
\end{align}
with $T_R=1/2$, $C_F=4/3$ and $C_A=3$. The various momentum fractions are 
set to
\begin{align}
\label{eq:ISapproxb}
\xis & =\frac{E_\alpha-E_i}{E_\alpha} &  \yis& =\frac{E_\alpha-(E_i+E_j)}{E_\alpha-E_i},
\nonumber\\
\xfs & =\frac{E_j}{E_i+E_j},          &     z& = \xis\,\yis\,. 
\end{align}
Finally, the underlying $gg\to \ttbar$ squared matrix element
in \refeq{eq:ISFSapprox}
reads
\begin{eqnarray}
\label{eq:ttME}
\big|\mathcal{M}_{gg\to\ttbar}\big|^2_{p_\alpha\to z p_\alpha} 
&=& 
\left(4\pi\as\right)^2
\theta\left(z -\frac{m_{\ttbar}}{p_1\cdot p_2}\right)
\Bigg\{
\left[\frac{\pp{1}{2}^2}{6\pp{1}{3}\pp{2}{3}}-\frac{3}{8}\right]
\nonumber\\
&\times&
\left[\frac{\pp{1}{3}^2}{\pp{1}{2}^2}
+\frac{\pp{2}{3}^2}{\pp{1}{2}^2}
-\frac{2m_t^2}{\pp{1}{2}}
+\frac{m_t^4}{\pp{1}{3}\pp{2}{3}}
\right]
\bigg\}_{p_\alpha\to z p_\alpha}\hspace{-4mm},\hspace{4mm}
\end{eqnarray}
where helicity/colour sums and average factors are included, and the
momentum of the IS emitter has to be rescaled by $z$.

\begin{figure}[t!]
\centering
\subfigure[]{
\includegraphics[width=0.4\textwidth]{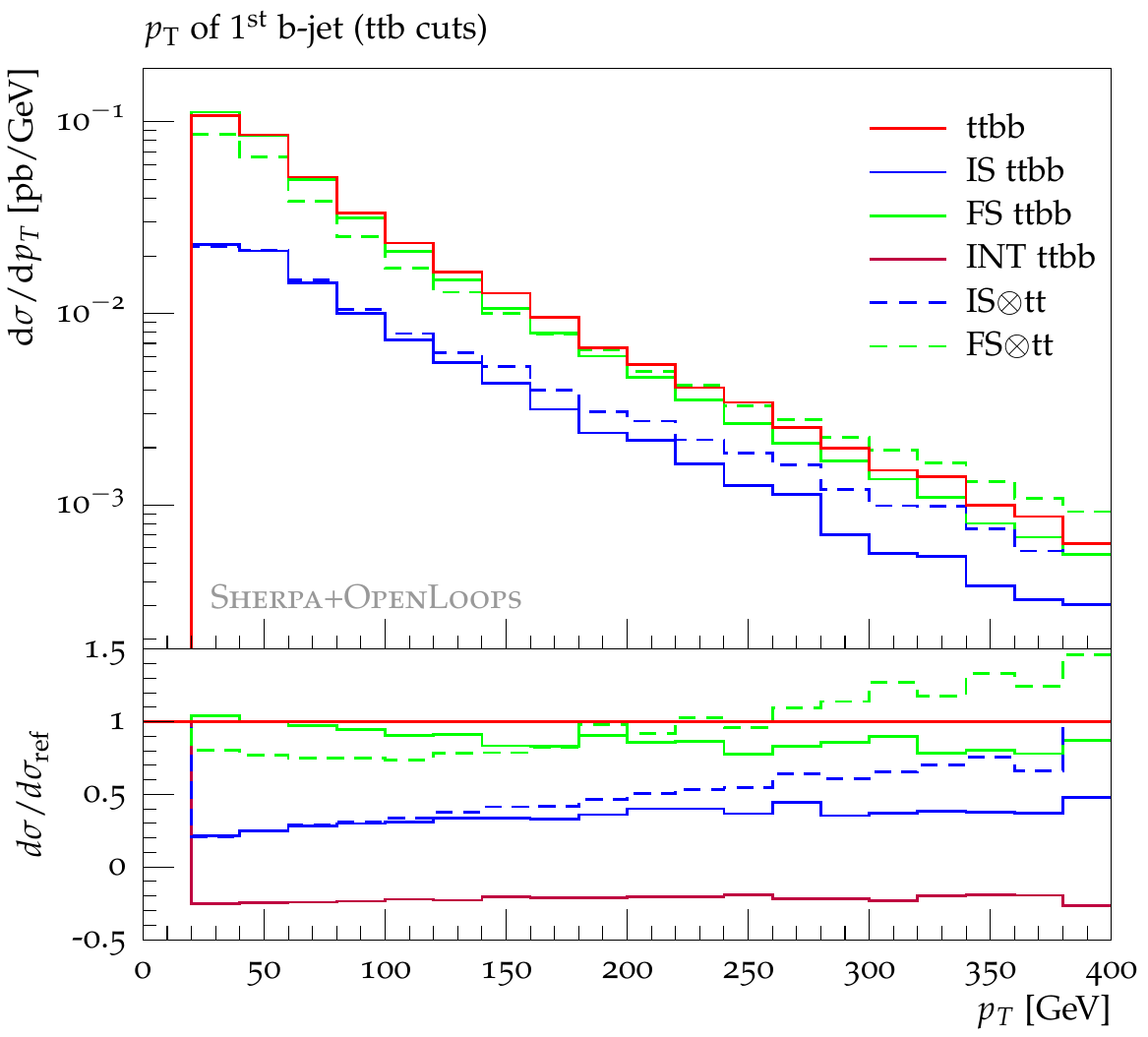}
\label{subfig:ttbbbreakdowna}
}\hspace{.0\textwidth}
\subfigure[]{
\includegraphics[width=0.4\textwidth]{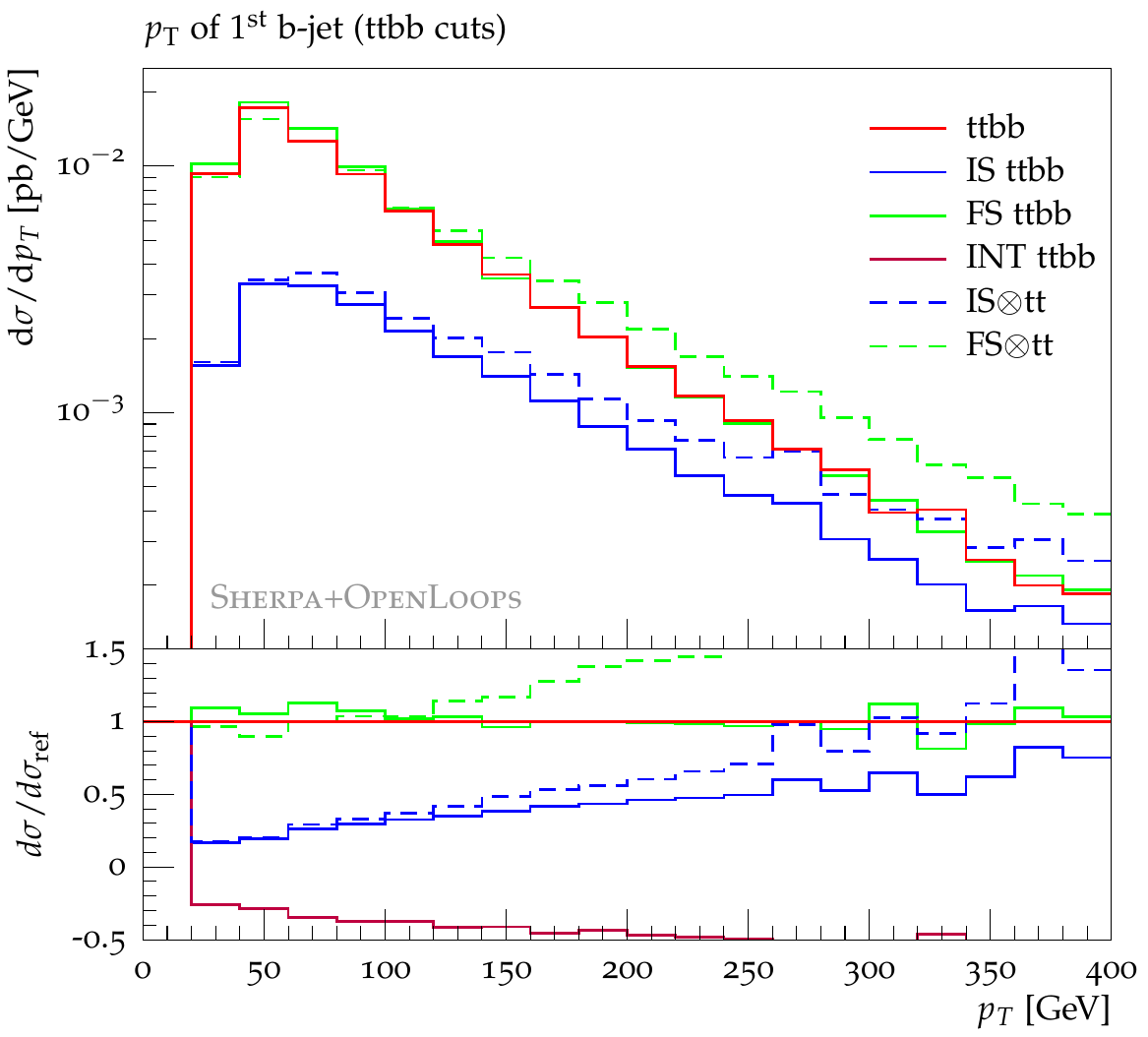}
\label{subfig:ttbbbreakdownb}
}\hspace{.0\textwidth}
\subfigure[]{
\includegraphics[width=0.4\textwidth]{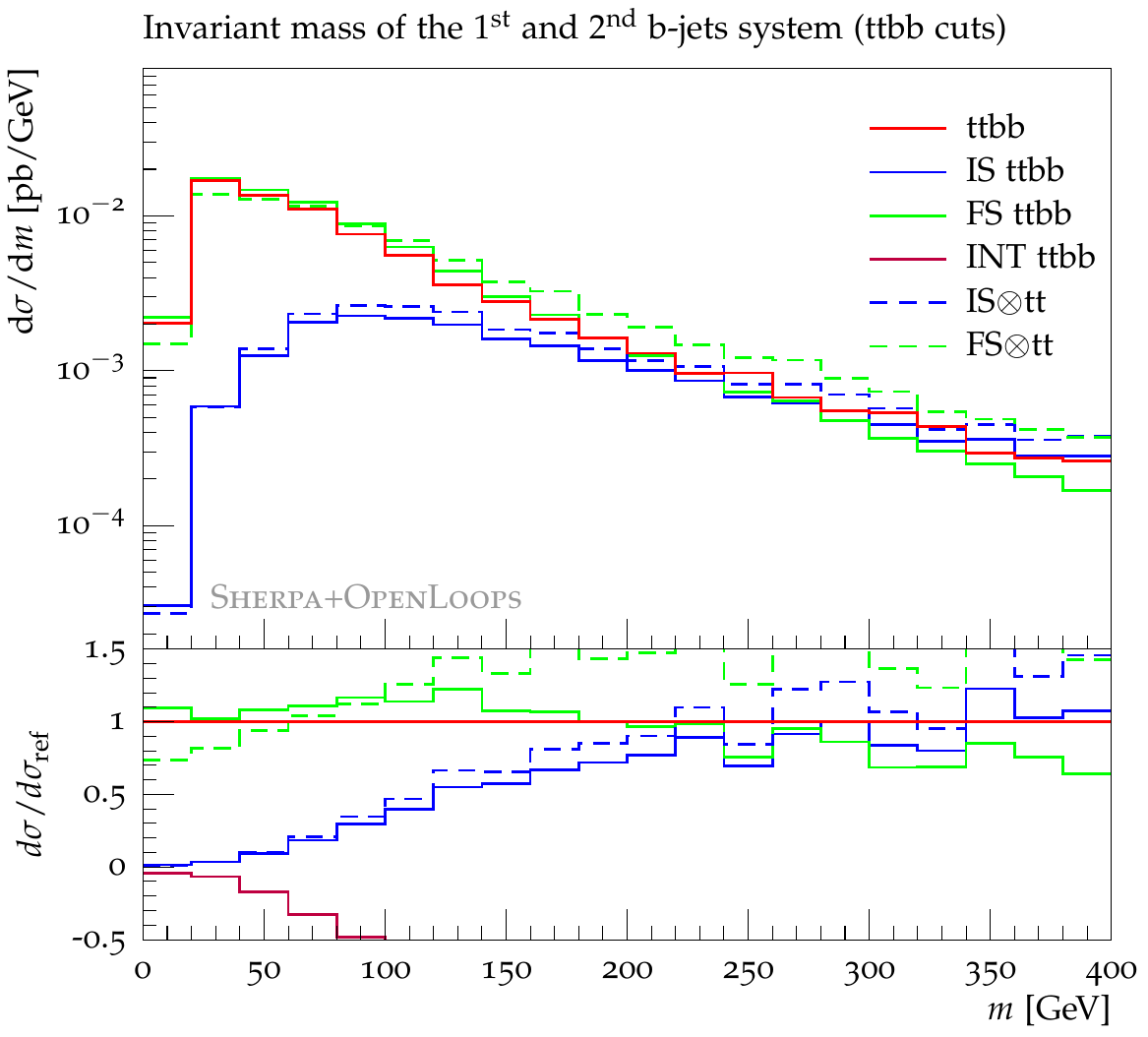}
\label{subfig:ttbbbreakdownc}
}\hspace{.0\textwidth}
\subfigure[]{
\includegraphics[width=0.4\textwidth]{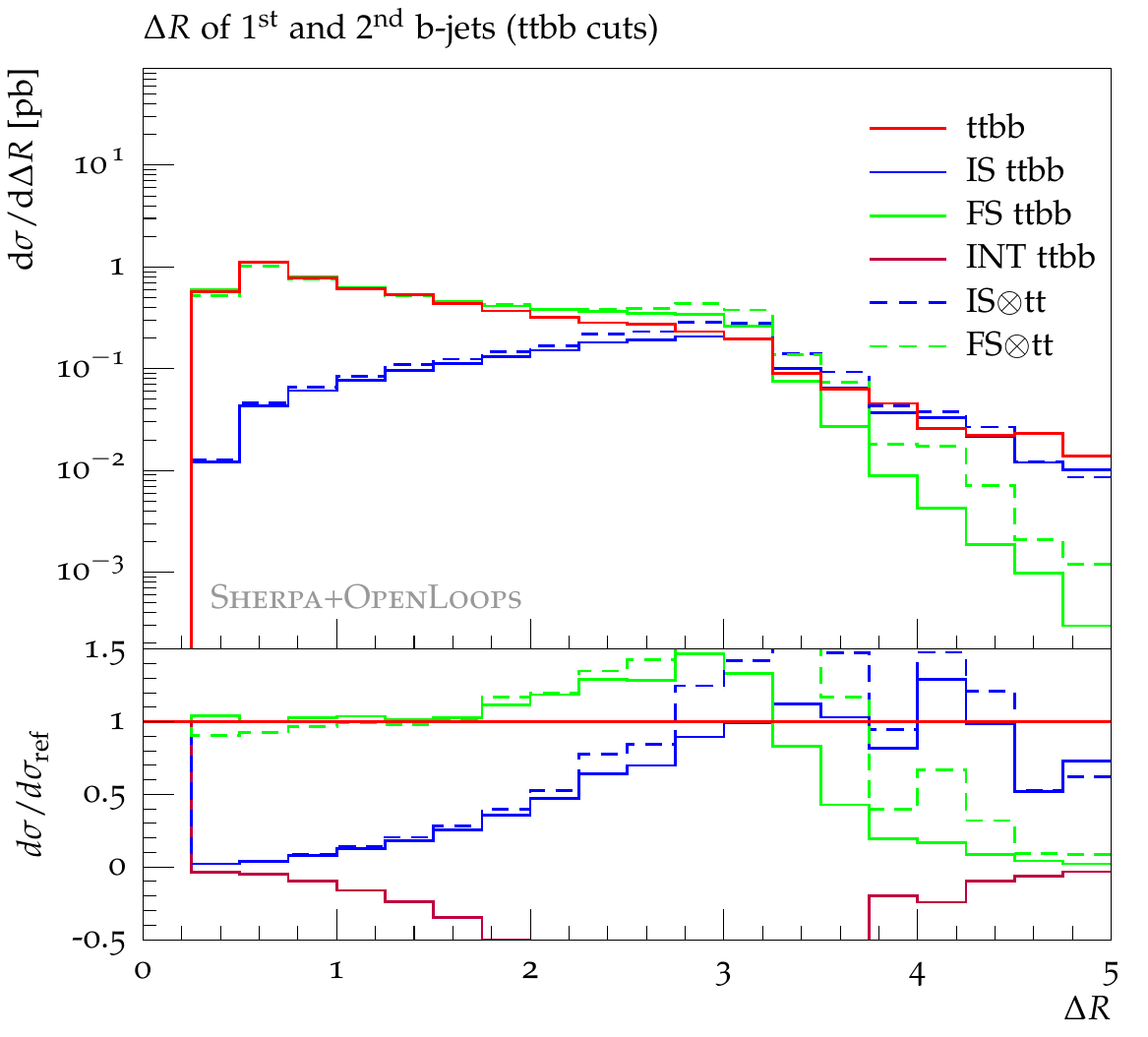}
}
\caption{Breakdown of the $pp\to \ttbb$ cross section into contributions
from topologies with IS and FS splittings at fixed-order LO in the 4F
scheme: distributions in the $p_\rT$ of the leading $b$-jet with ttb (a)
and ttbb (b) cuts, and distributions in the invariant mass (c) and $\Delta R$
separation (d) of the two
leading $b$-jets with ttbb cuts.
The complete $gg/q\bar q\to \ttbb$ matrix-element prediction (solid red) is
split according to \refeq{eq:ISFSXSsplitting} into contributions from
topologies of IS (solid blue) and FS (solid green) type and their
interference (solid purple).  This is compared to the gauge-invariant
breakdown \refeq{eq:ISFScoll} into IS (dashed blue) and FS (dashed green)
parts based on the collinear limits of the $\ttbb$ matrix element.
Note that the $q\bar q$ channel consists solely of FS
$g\to \bbbar$ contributions.}
\label{fig:ttbbbreakdown} 
\end{figure}

Numerical results for the diagrammatic decomposition
\refeq{eq:ISFSXSsplitting} and the collinear decomposition
\refeq{eq:ISFScoll} of $pp\to\ttbb$ at $\sqrt{s}=13$\,TeV
are shown in \reffi{fig:ttbbbreakdown}.
The first two plots display the leading $b$-jet $p_\rT$
distribution in the presence of ttb and ttbb cuts
as defined in \refse{se:cuts}, \ie
requiring $\Nb\ge 1$ and $\Nb\ge 2$ $b$-jets, respectively.
In the ttbb phase space, topologies with $g\to \bbbar$ FS splittings 
turn out to be surprisingly close to the 
full matrix element, with deviations that do not exceed the 10\% level in the entire 
spectrum. This agreement remains remarkably good also in the inclusive 
ttb phase space, where IS splitting processes with one unresolved $b$-quark
are expected to be more pronounced. 
Actually, with ttb and ttbb cuts the pure IS contribution 
ranges between 20--40\% and 
15--75\%, respectively, but is almost entirely cancelled 
by the IS--FS interference.

The fact that the collinear approximations \refeq{eq:ISFScoll} agree rather
well with the corresponding squared Feynman diagrams up to relatively high
$p_\rT$ confirms that $\ttbb$ production is dominated by the topologies in
\reffi{fig:ttbbtopologies}c and \ref{fig:ttbbtopologies}d.
On the other hand, the importance of interference effects provides strong
motivation for using exact $\ttbb$ matrix elements, while collinear
approximations such as those in \refeq{eq:ISFSapprox} or in the parton-shower
modelling of $g\to \bbbar$ splittings should be used with due caution.

The above considerations apply also for the $m_{\bbbar}$ and 
$\Delta R_{\bbbar}$ distribution in
\reffi{fig:ttbbbreakdown}c and d.  In particular, we observe that topologies with
FS $g\to \bbbar$ splittings are very close to the full matrix element in the
whole  $m_{\bbbar}$ spectrum as well as for $\Delta R_{\bbbar}<2$ .  
At the same time, for $50\,\GeV< m_{\bbbar}<
200\,\GeV$ and $1<\Delta R_{\bbbar}<2.5$, i.e.~in the range of interest for 
$\ttbar H(\bbbar)$ analyses,
we observe that IS splitting contributions and negative interference effects
grow fast and tend to become very sizable.  Thus a naive
separation into contributions from IS and FS splittings is not applicable
at large $m_{\bbbar}$  and $\Delta R_{\bbbar}$. 
On the other hand, in the region of moderate invariant mass and $\Delta R$ separation, 
which contains the bulk of the $\ttbb$ cross section,
interference effects are rather small,
and $\bbbar$ pairs turn out to originate almost entirely from 
FS $g\to \bbbar$ splittings.

In summary, given that the 5F scheme is based on the LO process $gb\to
\ttbar b$, where FS $g\to \bbbar$ splittings and interference effects are
entirely neglected, the above observations provide strong motivation for a
description of $\ttbar+b$-jet production based on $\ttbb$ matrix elements in
the 4F scheme.

\section{Technical aspects and setup of NLOPS $\boldsymbol{\ttbb}$ simulations}
\label{se:technical}

In this section we introduce a new \Powheg generator based on $\ttbb$ matrix
elements in the 4F scheme.  Special emphasis is devoted to some technical aspects of
the \Powheg method that turn out to play an important role for 
multi-scale processes like $pp\to \ttbb$.
In addition we describe the setup used for the
$\ttbar+b$-jet simulations presented in 
\refses{se:stabletops}{se:decayedtops}, \ie all 
relevant input parameters, scale choices and parton shower settings, 
as well as the treatment of theoretical uncertainties
and the definitions of physics objects and selection cuts.
Finally we provide details on the treatment of top-quark decays.

The new $\ttbb$ generator is implemented in the \PowhegBoxRes framework~\cite{Jezo:2015aia},
and the relevant LO and NLO matrix elements are 
computed by \OpenLoops~\cite{Cascioli:2011va,OLhepforge,Buccioni:2017yxi} 
through its \PowhegBoxRes interface~\cite{Jezo:2016ujg}.
For the evaluation of one-loop integrals
\OpenLoops employs the \Collier library~\cite{Denner:2014gla,Denner:2002ii,Denner:2005nn,Denner:2010tr}
and, alternatively, \CutTools~\cite{Ossola:2007ax,Ossola:2006us} together with 
the \OneLOop library~\cite{vanHameren:2010cp}.
While we do not apply the resonance-aware method~\cite{Jezo:2015aia},
the \PowhegBoxRes framework allows us to make use of new technical
features, such as the automated implementation of scale variations
, a \Rivet~\cite{Buckley:2010ar} interface and the option to unweight events partially.
This $\ttbb$ generator will be soon publicly available on the \PowhegBox webpage~\cite{PWGBpage}.

\subsection{Powheg methodology}
\label{se:methods}

In the following, we briefly review the \Powheg method~\cite{Nason:2004rx, Frixione:2007vw}
with emphasis on the
separation of radiation into singular and finite parts. In this context we 
discuss technical subtleties that arise in the case of multi-scale processes 
like $pp\to \ttbb$.

The master formula for the description of NLO radiation in the \Powheg approach
consists of two contributions,
\begin{eqnarray}
  \mathd \sigma & = & 
  \mathd \sigma_\sing +
  \mathd \sigma_\fin,
\label{eq:powhegsplit1}
\end{eqnarray}
which arise from the splitting of real emission into singular (s) and 
finite (f) parts,
\begin{equation}
 R(\Phi_{\mathrm{R}}) = R_\sing(\Phi_{\mathrm{R}})+R_\fin(\Phi_{\mathrm{R}}).
\label{eq:powhegsplit2}
\end{equation}
Here $R(\Phi_{\mathrm{R}})$ should be understood as squared real-emission matrix element,
and $\Phi_{\mathrm{R}}$ as the corresponding phase space.
Similarly, Born and virtual contributions in the Born phase space 
are denoted as $B(\Phi_{\mathrm{B}})$ and $V(\Phi_{\mathrm{B}})$.
The splitting \refeq{eq:powhegsplit2} is implemented as
\begin{eqnarray}
R_{\sing}(\Phi_{\mathrm{R}})
=
F(\Phi_{\mathrm{R}})\,
R(\Phi_{\mathrm{R}}),
\qquad
R_{\fin}(\Phi_{\mathrm{R}})
=
\left[1-F(\Phi_{\mathrm{R}})\right]\,
R(\Phi_{\mathrm{R}}),
\qquad
\label{eq:powhegsplit3}
\end{eqnarray}
where $F(\Phi_{\mathrm{R}})\in [0,1]$ is a damping function 
that fulfils $F\to 1$ and $F\to 0$, respectively, in the
infrared and hard regions of phase space (see below).

The singular part of real radiation is resummed according to 
the \Powheg formula
\begin{eqnarray}
  \mathd \sigma_\sing & = & \bar{B} (\Phi_{\mathrm{B}}) \,\mathd \Phi_{\mathrm{B}}  \left[
  \Delta (q_{\tmop{cut}}) + \sum_{\alpha} \Delta (k_{\rT,\alpha}) 
  \frac{R_{\sing,\alpha} (\Phi_{\alpha} 
(\Phi_{\mathrm{B}}, \Phi_{\tmop{rad}})
)}{B
  (\Phi_{\mathrm{B}})} \,\mathd \Phi_{\tmop{rad}} \right],
\label{eq:powheg1}
\end{eqnarray}
where real emission is further split into FKS sectors~\cite{Frixione:1995ms},
\begin{equation}
R_\sing = \sum_{\alpha} R_{\sing,\alpha},\qquad
R_\fin = \sum_{\alpha} R_{\fin,\alpha},
\label{eq:fks1}
\end{equation}
which isolate collinear singularities arising from 
individual emitters. In each sector,  the emission phase space $\Phi_{\mathrm{R}}$
is factorised 
into the  Born phase space $\Phi_{\mathrm{B}}$ and a one-particle radiation phase space 
$\Phi_{\tmop{rad}}$ through an appropriate FKS mapping,
\begin{equation}
(\Phi_{\mathrm{B}}, \Phi_{\tmop{rad}})
\;\longrightarrow\;
\Phi_{\mathrm{R}} = \Phi_{\alpha} (\Phi_{\mathrm{B}}, \Phi_{\tmop{rad}}).
\label{eq:fks2}
\end{equation}
The term within squared brackets in \refeq{eq:powheg1} 
generates the hardest radiation according to an emission probability $R/B$.
The parameter $k_{\rT,\alpha}=k_{\rT,\alpha}(\Phi_\rad)$
stands for the hardness of the radiated parton, and radiation harder than $k_{\rT,\alpha}$
is excluded by means of corresponding Sudakov form factors, 
\begin{eqnarray}
  \Delta (q) & = & 
 \exp \left[ - \sum_\alpha\,\int_{k_{\rT,\alpha}\, >\, q}
  \frac{R_{\sing,\alpha} (\Phi_{\alpha} 
(\Phi_{\mathrm{B}}, \Phi_{\tmop{rad}})
)}{B
  (\Phi_{\mathrm{B}})}\, \mathd \Phi_{\tmop{rad}} \right].
\label{eq:Sudakov}
\end{eqnarray}
The term $\Delta(q_{\tmop{cut}})$ in \refeq{eq:powheg1} represents the
no-emission probability above the infrared cutoff $q_{\tmop{cut}}$,
and Sudakov form factors account for unresolved multiple emissions
in a way that cancels infrared singularities while preserving the
differential NLO cross section $\bar{B} (\Phi_{\mathrm{B}})$ in the Born
phase space.  The latter is defined by integrating out 
the singular part of real radiation,
\begin{eqnarray}
  \bar{B} (\Phi_{\mathrm{B}}) & = & B (\Phi_{\mathrm{B}}) + V (\Phi_{\mathrm{B}}) +
  \sum_{\alpha} \int R_{\sing,\alpha} (\Phi_{\alpha} 
(\Phi_{\mathrm{B}}\,,\Phi_{\tmop{rad}})
) \,\mathd \Phi_{\tmop{rad}}\,.
\label{eq:powheg2}
\end{eqnarray}
Here infrared cancellations between $V$ and $R_\sing$ are controlled via
FKS subtraction.
The
remaining finite part of NLO radiation is treated as in fixed-order
calculations,
\begin{eqnarray}
\mathd \sigma_\fin
&=&
\sum_\alpha
R_{\fin,\alpha} (\Phi_{\alpha} (\Phi_{\mathrm{B}}, \Phi_{\tmop{rad}}))\,
\mathd \Phi_{\mathrm{B}},
\mathd \Phi_{\tmop{rad}}\,.
\label{eq:powheg3}
\end{eqnarray}
Note that flux and symmetry factors as well as the
convolution with PDFs are implicitly understood in \refeq{eq:powheg1} and
\refeq{eq:powheg3}.

Let us now come back to the details of the separation of the singular and
finite parts of real emission in \refeq{eq:powhegsplit3}.  Technically, the
damping function $F$ is implemented based on the kinematics of the actual
FKS sector, \ie
\begin{eqnarray}
R_{\sing,\alpha}(\Phi_{\alpha}) 
= R_{\alpha}(\Phi_{\alpha}) -
R_{\fin,\alpha} (\Phi_{\alpha})
=F_{\alpha} (\Phi_{\alpha}) 
R_{\alpha}(\Phi_{\alpha}).
\end{eqnarray}
The default functional form of $F$ in \PowhegBox~\cite{Alioli:2010xd,Alioli:2008tz}
is
\begin{eqnarray}
F_{\alpha} (\Phi_{\alpha}) 
=
F_{\damp,\alpha} (\Phi_{\alpha})\, 
F_{\bzd, \alpha} (\Phi_{\alpha}), 
\label{eq:dampingdef}
\end{eqnarray}
where 
\begin{eqnarray}
F_{\damp,\alpha} (\Phi_{\alpha}) 
=
\frac{\hdamp^2}{\hdamp^2+k_{\rT,\alpha}^2}
\label{eq:hdampdef}
\end{eqnarray}
is the usual factor that smoothly shifts the weight of real radiation from
$R_\sing$ to $R_\fin$ when the hardness of the emission,
$k_{\rT,\alpha}$, becomes of the order of $\hdamp$ or higher.
Note that the freely adjustable $\hdamp$ parameter in \Powheg plays an
analogous role as the resummation scale $\mu_Q$ in the MC@NLO method. This is because
both parameters act as a $k_{\rT}$-threshold that separates the radiative
phase space into a hard region, which is described by fixed-order matrix
elements, from a singular region, where 
large logarithms of soft and collinear origin are resummed 
to all orders by means of Sudakov form factors.
More precisely, in the MC@NLO approach the factor 
$R_{\sing,\alpha}/B$ and the terms $\Delta$ in \refeq{eq:powheg1} correspond, respectively,
to the parton-shower emission probability and the associated 
Sudakov form factors. 
Thus, in the MC@NLO framework, equation \refeq{eq:powheg1} corresponds
to the weight of so-called soft events supplemented by the probability of the first
parton-shower emission and its no-emission counterpart.
In analogy with \refeqs{eq:dampingdef}{eq:hdampdef}, also the first 
MC@NLO shower emission is modulated by a certain damping function.
The related reference scale $\mu_Q$, i.e.~the MC@NLO counterpart of the scale $\hdamp$ in \Powheg,
corresponds to the upper bound of the first shower emission in MC@NLO.
Thus MC@NLO predictions are sensitive to the choice of the shower 
starting scale.\footnote{Note that in the MC@NLO 
implementation of \cite{Alwall:2014hca} the scale $\mu_Q$ is also 
taken as starting scale for the showering of so-called hard events,
which represent the MC@NLO counterpart of \refeq{eq:powheg3}. Instead,
in \Powheg such events are showered starting from the 
actual $k_\rT$ of the first emission.}
On the contrary, the first \Powheg emission in
\refeq{eq:powhegsplit1}--\refeq{eq:powhegsplit3} is entirely determined by
the matrix element, which also dictates the scale at which the shower starts
emitting further partons.  Thus the \Powheg method has the advantage of being
essentially independent on the shower starting scale.
More generally, thanks to the fact that the first emission is 
completely independent of the 
parton shower, \Powheg predictions are characterised by a rather mild 
sensitivity to systematic uncertainties associated with the parton 
shower.

In addition to the well-known $\hdamp$-dependent damping mechanism~\refeq{eq:hdampdef}, 
the \PowhegBox also implements a theta
function\footnote{%
The damping functions \refeq{eq:dampingdef}--\refeq{eq:bzddef} 
are implemented in the {\tt bornzerodamp} routine of 
\PowhegBox and can be controlled through the flags named
{\tt withdamp} and {\tt bornzerodamp}.
The former activates the overall damping factor, \ie $F_\damp\to 1$ if
{\tt withdamp} is set to 0. The same happens if {\tt hdamp} is not explicitly set by the user.
The remaining {\tt bornzerodamp} flag controls the $\hbzd$-dependent theta function
\refeq{eq:bzddef}. By default {\tt bornzerodamp}={\tt withdamp},
and $F_\bzd\to 1$ if {\tt bornzerodamp} is set to 0.
}
of the form~\cite{Alioli:2008gx,Alioli:2010xd}
\begin{eqnarray}
F_{\bzd, \alpha} (\Phi_{\alpha})
=
\theta\left(\hbzd -\frac{R_{\alpha}(\Phi_{\alpha})}{\calR_{\alpha}(\Phi_{\alpha})}\right),
\label{eq:bzddef}
\end{eqnarray}
where $\calR_{\alpha}$ corresponds to the 
infrared (soft and collinear) approximation 
of the  full matrix element. Schematically it has the factorised form
\begin{eqnarray}
\calR_{\alpha}(\Phi_\alpha) = \calK_{\alpha}(\Phi_{\mathrm{rad}}) B(\Phi_{\mathrm{B}}),
\label{eq:fks3}
\end{eqnarray}
with an FKS kernel $\calK_{\alpha}(\Phi_{\mathrm{rad}})$ and an underlying
Born contribution $B(\Phi_{\mathrm{B}})$, whose kinematics is determined by
the inverse of the mapping \refeq{eq:fks2} in the actual sector $\alpha$.
By default, the cut-off parameter $\hbzd$ in \refeq{eq:bzddef} is set equal
to 5.  In this way, in the vicinity of IR singularities, where
$R_\alpha/\calR_{\alpha}\to 1$, radiative contributions are attributed to
$R_\sing$ and resummed according to~\refeq{eq:powheg1}.
On the contrary, when the real emission matrix element largely exceeds the IR
approximation \refeq{eq:fks3}, the resummation of the full $R/B$ kernel
according to~\refeq{eq:powheg1} is not well justified, and corresponding
events are attributed to the finite remnant~\refeq{eq:powheg3} through the
theta function~\refeq{eq:bzddef}.
In the standard \PowhegBox, 
and in \citere{Bevilacqua:2017cru},
the damping function \refeq{eq:dampingdef}
is applied only to initial-state radiation.  However, in the present $\ttbb$
generator we have extended it to all (massless or massive) final-state
emitters, that have a FKS sector associated with it, 
in order to ensure a consistent resummation of QCD radiation off
$b$-quarks.

The requirement $R_{\alpha}(\Phi_{\alpha}) <
\hbzd\,\calK_{\alpha}(\Phi_{\mathrm{rad}})\,B(\Phi_{\mathrm{B}})$ was
originally introduced in order to avoid possible divergences of
$R(\Phi_\alpha)/B(\Phi_\rB)$ due to so-called Born zeros, i.e.~phase space
regions where $B(\Phi_{\mathrm{B}})\to 0$.
Such divergences cancel in the $\bar B/B$ ratio, i.e.~they are not physical,
and are not related to IR radiation.  Corresponding $\Phi_\rB$ regions
should thus be attributed to the finite remnant.  Otherwise they could lead
to dramatic inefficiencies in the event generation.
More generally, the damping factor \refeq{eq:bzddef} can play an important
role also in case of multi-scale processes where the Born cross section
involves enhancement mechanisms at scales well below the hard energy of the
full process.
Such enhancements can compete with the ones due to soft and collinear QCD
radiation in a way that is somewhat analogous to Born zeros.

In the case of $pp\to \ttbb$, such effects can arise from the interplay of
soft and collinear enhancements due to NLO light-jet radiation and to the
generation of the $\bbbar$ system in regions with $m_{\bbbar}\ll m_{\ttbb}$
and/or $p_{\rT,\bbbar}\ll m_{\ttbb}$.
For example, let us consider a $gg\to \ttbb g$ event with a gluon emission
of ISR type.
Its kinematics is generated starting from a $gg\to \ttbb$ Born event through
a mapping of type \refeq{eq:fks2}, which creates the required gluon recoil
by boosting the final state of the $gg\to\ttbb$ Born event in the transverse
direction.
The relevant boost factor, $\gamma=1/(1-\beta^2)^{1/2}$, is determined by
$p_{\rT,j}=p'_{\rT,\ttbb}=\gamma\beta E_{\ttbb}$, where $E_{\ttbb}$ is the
$\ttbb$ energy of the $\bbbar$ system in the  Born event.
If we assume, for simplicity, that he gluon is emitted in the same azimuthal
direction as the $\bbbar$ system in the Born event, then the $\bbbar$
transverse momentum of the radiative event becomes
$p'_{\rT,\bbbar}=\gamma(p_{\rT,\bbbar}-\beta E_{\bbbar})$, where
$p_{\rT,\bbbar}$ and $E_{\bbbar}$ are the $\bbbar$
transverse momentum and energy
in the Born event.
Thus, the FKS mapping can lead to a very significant reduction of
$p_{\rT,\bbbar}$.  More precisely, for radiative events with 
\begin{equation}
\frac{p_{\rT,j}}{p_{\rT,\bbbar}}=(1+\epsilon)\frac{E_{\ttbb}}{E_{\bbbar}},
\label{eq:pTbbboosta}
\end{equation}
the effect of the FKS boost on the $\bbbar$ system amounts to 
\begin{equation}
\frac{p'_{\rT,\bbbar}}{p_{\rT,\bbbar}}=\tilde \epsilon=\gamma-1-\epsilon.
\label{eq:pTbbboostb}
\end{equation}
Thus, since the bulk of the $\ttbb$ cross section is characterised by $E_{\ttbb}
\gg E_{\bbbar}$ and $\gamma\sim 1$, in the case of hard QCD
radiation with $\tilde \epsilon\ll 1$ the FKS mapping can lead to a drastic reduction of the
$p_{\rT}$ of the $\bbbar$ system.
As a result, in the region of small $m_{\bbbar}$, the ISR boost can enhance the
$R_\alpha(\Phi_\alpha)/\calR_\alpha(\Phi_\alpha)$ ratio by up to a
factor\footnote{The $1/p_{\rT,\bbbar}$ dependence arises from the collinear
singularity associated with the emission of the parent gluon in the
topologies of type (d) in \reffi{fig:ttbbtopologies}.}
$(p_{\rT,\bbbar}/p'_{\rT,\bbbar})^2\sim 1/{\tilde\epsilon}^2$.
This violates the main assumption that justifies the \Powheg
formula~\refeq{eq:powheg1}, namely 
$R_\alpha(\Phi_\alpha)/B(\Phi_B)\sim
\calR_\alpha(\Phi_\alpha)/B(\Phi_B)={\cal K}_\alpha(\Phi_\rad)$,
which requires a sufficiently hard 
$\ttbb$ process as compared to the $k_\rT$ of NLO radiation. 
In particular, due to the sensitivity of the Born amplitude to scales of the order
${p_{\rT,\bbbar}}\sim (E_{\bbbar}/E_{\ttbb})\, p_{\rT,j}\ll p_{\rT,j}$,
the factorisation formula~\refeq{eq:fks3} is not fulfilled.

Fortunately, this problematic behaviour emerges only in relatively hard
regions of the $\Phi_\rad$ phase space.\footnote{This holds also for similar
issues due to final-state radiation.} 
Thus, as a remedy it is natural to
shift such events into the finite remnant by means of the damping factor
\refeq{eq:bzddef}.
In fact, in the case of $\ttbb$ production we have found that the
$\hbzd$-dependent cut plays an important role for the efficiency generation
of Les~Houches events (LHEs) as well as for a consistent scale dependence.  Moreover,
applying a large $\hbzd$ cut we have observed a significant enhancement of
the QCD scale dependence.
This can be attributed to the fact that scale variations in the soft term~\refeq{eq:powheg1}
are restricted to the $\bar B$ factor \refeq{eq:powheg2},
where the unphysical distortions of the $\bbbar$ kinematics induced by the
FKS mappings can jeopardise the natural cancellation of virtual and real
contributions associated with a given Born configuration.

As discussed in \refse{se:stabletops}, \Powheg predictions for ttbb
observables are rather stable with respect to variations of 
$\hbzd$. Thus, in order
to avoid an unphysical enhancement of the scale dependence,
we have reduced $\hbzd$ from its default value of 5 
to $\hbzd=2$.  This guarantees a more reasonable consistency with the
fixed-order scale dependence without shifting an excessive fraction of
the cross section from $\rd\sigma_\sing$ to $\rd\sigma_\fin$.

\subsection{Input parameters, PDFs and scale choices}
\label{se:input}

The predictions in \refses{se:stabletops}{se:decayedtops} are based on the following
input parameters, scale choices and PDFs.
Heavy-quark mass effects are included throughout using
\beq
 m_{t} = 172.5 \;\GeV\,,\qquad     m_{b} = 4.75 \;\GeV\,.
\eeq
All other quarks are treated as massless in the perturbative part of the
calculations.  Since we use massive $b$-quarks, for the PDF evolution and
the running of $\as$ we adopt the 4F scheme.
Thus, for consistency, we renormalise $\as$ in the decoupling scheme, where
top- and bottom-quark loops are subtracted at zero momentum transfer.  In
this way, heavy-quark loop contributions to the evolution of the strong coupling
are effectively described at first order in $\as$ through the virtual corrections.

For the calculation of hard cross sections at LO and NLO, as well as for the
generation of the first \Powheg emission, we use the {\tt
NNPDF30\_nlo\_as\_0118\_nf\_4} parton distributions~\cite{Ball:2014uwa} as
implemented in the LHAPDFs~\cite{Buckley:2014ana} and the corresponding
$\asfour$.%
\footnote{More precisely, $\asfour$ is taken from the PDFs everywhere except
for the evaluation of the Sudakov form factors in~\refeq{eq:powheg1}, 
where the corresponding
\Powheg implementation of $\asfour$ is used.  In both implementations
$\asfour(M_Z)=0.112$, which corresponds to $\asfive(M_Z)=0.118$.}
To assess PDF uncertainties we re-evaluate the weights of LHEs
with 100 different PDF replicas, while using the nominal PDF set
for parton showering.

Since it scales with $\as^4$, the $\ttbb$ cross section is highly sensitive
to the choice of the renormalisation scale $\mur$, and this choice 
plays a critical role for the stability of perturbative predictions.
Following~\cite{Cascioli:2013era,deFlorian:2016spz}, we adopt a scale choice of the form 
\begin{equation}
\label{eq:mur}
	\mur=\xir \sqrt{\mu_{\ttbar}\,\mu_{\bbbar}},
\end{equation}
with the scale-variation factor $\xir\in[0.5,2]$. This dynamic scale
choice accounts for the fact that $\ttbb$ production is characterised by two 
widely separated scales,
which are related to the $\ttbar$ and $\bbbar$ systems
and are chosen as the geometric average of the respective transverse energies,
\begin{equation}
\label{eq:mubbtt}
\mu_{\bbbar}=\sqrt{E_{\rT,b}E_{\rT,\bar b}},
\qquad
\mu_{\ttbar}=\sqrt{E_{\rT,t}E_{\rT,\bar t}}\;.
\end{equation}
The transverse energies $E_{\rT,i}=\sqrt{m_i^2+p^2_{\rT,i}}$ are defined in terms of 
the rest masses $m_i$ and the transverse momenta $p_{\rT,i}$ 
of the bare heavy quarks.
The scales \refeq{eq:mubbtt} are computed according to physical kinematics,
i.e.~without projecting real emission events to the underlying Born phase
space.  The choice~(\ref{eq:mur}) is applied to all (N)LO matrix elements
apart from the $\as$ factor that results from the $R/B$ ratio
in~\refeq{eq:powheg1}.  In that case $\as$ is evaluated at the transverse
momentum of the hardest \Powheg emission, 
and that $\as(k_{\rT,\alpha})$ factor is not subject to scale variations.

For the factorisation scale $\muf$ we use\footnote{This choice does not 
coincide with the scale
$\muf=\frac{1}{2}\sum_{i=t, \bar t} E_{T,i}$
adopted in~\cite{Cascioli:2013era}. 
However, this difference has a rather minor impact on 
our predictions.}
\begin{equation}
\label{eq:muf}
	\muf=\xif\,\frac{H_\rT}{2}
= \frac{\xif}{2} \sum_{i=t,\bar t, b, \bar b,j} E_{\rT,i}\;,
\end{equation}
where $\xif\in [0.5,2]$, and  the total transverse energy of the $\ttbb$ system, $H_\rT$, is computed in terms of 
bare-quark transverse momenta including also QCD radiation at NLO.
Our nominal predictions correspond to $\xir=\xif=1$, and to quantify scale
uncertainties we take the envelope of the seven-point variation
$(\xir,\xif)=(0.5,0.5)$, $(0.5,1)$, $(1,0.5)$, $(1,1)$, $(1,2)$, $(2,1)$, $(2,2)$.

For the \PowhegBox parameters $\hbzd$ and $\hdamp$, which 
control the resummation of NLO radiation according
to \refeq{eq:dampingdef}--\refeq{eq:bzddef} as discussed in 
\refse{se:methods}, we set
\begin{equation}
\label{eq:bzd}
	\hbzd=2
\end{equation}
and
\begin{equation}
	\hdamp=\frac{H_\rT}{2}
= \frac{1}{2} \sum_{i=t,\bar t, b, \bar b} E_{\rT,i}\;.
\label{eq:hdampchoice}
\end{equation}
Here the various $E_{\rT,i}$ are defined in the underlying Born phase space.
To account for the uncertainties associated with these choice we apply the 
independent variations $\hbzd=2$, $5$, $10$ and
$\hdamp={H_\rT}/{4},{H_\rT}/{2},H_\rT,1.5\,m_t$, varying both parameters one at a time.\footnote{The choice
$\hdamp=1.5\,m_t$ corresponds to the default setting used for inclusive
$\ttbar$ production in ATLAS.}

The above choices for $\mur,\muf$ and $\hdamp$, as well as the employed PDFs
 correspond to the setup recommended in~\cite{deFlorian:2016spz}.

\subsection{Parton shower settings and variations}
\label{se:shower}

By default, LHEs are showered with~\Pythia8.2 using the A14
tune%
\footnote{More precisely we use the \Pythia8.219 version and
the specific A14 tune for NNPDFs, which is based on the NNPDF2.3 LO PDFs.},
where ISR and FSR parameters as well as the MPI activity have been tuned in
a single step using most of the available 
$\ttbar$ ATLAS data
from Run\,1~\cite{ATL-PHYS-PUB-2014-021}.
In the A14 tune, $m_b=4.75$\,GeV and $\asfive(M_Z)=0.127$, both for ISR and FSR, while in 
the default Monash tune
$\asfive(M_Z)=0.13650$.
Since the shower evolution is implemented in the 5F scheme 
we shower events using 5F PDFs. Specifically, we choose the {\tt NNPDF30\_nlo\_as\_0118} 5F 
PDFs.\footnote{We have checked that showering LHEs with the
{\tt NNPDF30\_nlo\_as\_0118\_nf\_4} (used for the NLO hard cross section)
or the {\tt NNPDF30\_lo\_as\_0130\_nf\_4} 
instead of the 5F NNPDFs
does not induce any significant change in our results.
}

The interplay between \Powheg and \Pythia is controlled by the
{\tt scalup} parameter, which describes the hardness of radiation 
in LHEs and may be taken as starting scale for \Pythia.
However, in order to avoid inconsistencies due to the fact that the \Pythia
evolution variable does not coincide with the definition of hardness in
\Powheg, we apply the following two-step procedure
based on the \PowhegHooks class.
Instead of starting below {\tt scalup},
we instruct \Pythia to generate radiation up to 
the kinematic limit 
by setting
\begin{verbatim}
    pythia.readString("SpaceShower:pTmaxMatch = 2");
    pythia.readString("TimeShower:pTmaxMatch = 2");
\end{verbatim}
Then, to guarantee the correct ordering of emissions in \Powheg and 
\Pythia,
we apply a veto on each \Pythia emission that is harder than 
{\tt scalup} according to the \PowhegBox definition of hardness. 
This is achieved by setting
\begin{verbatim}
    pythia.readString("POWHEG:nFinal = 4");
    pythia.readString("POWHEG:veto = 1");
\end{verbatim}
The remaining \PowhegHooks settings are left to their default values.

At LOPS level we set the shower starting scale equal to $H_{\rT}/2$ and  
vary it up and down by a factor two in order to assess the related uncertainty.
At NLOPS, the shower starting scale 
is dictated by the kinematics of real emission matrix elements 
in the \Powheg method.
Thus, at variance with NLOPS predictions based on the MC@NLO method,
\Powheg predictions are free from uncertainties related to the choice of
the shower starting scale.

In order to assess uncertainties 
due to the parton-shower modelling of $g\to \bbbar$ 
splittings we vary the parameter
{\tt TimeShower:weightGluonToQuark}, which permits to select various 
optional forms of the $g\to Q\bar Q$ splitting kernel in \Pythia8.
The default is option 4, which corresponds to the splitting 
probability~\cite{pythianote}
\begin{equation}
\label{eq:gbbmesplitting}
\rd P_{g\to \bbbar} = \frac{\as(p_T^2)}{2\pi}\frac{\rd m_{\bbbar}^2}{m_{\bbbar}^2}
\frac{\beta_b}{2}\left[z^2+(1-z)^2+8r_b z(1-z)\right](1-\delta)^3,
\end{equation}
where $r_b=m_b^2/m_{\bbbar}^2$,  $\beta_b=\sqrt{1-4r_b}$ and
$\delta=m^2_{\bbbar}/m^2_{\mathrm{dipole}}$.
The factor $(1-\delta)^3$, which suppresses the production of
high-mass $\bbbar$ pairs, is derived from
the $H\to g\bbbar$ matrix element by interpreting $m_H$ as the mass of a gluon 
dipole, $m_{\mathrm{dipole}}$. 
Omitting the factor $(1-\delta)^3$ in
\refeq{eq:gbbmesplitting} corresponds to option 2
and results in a DGLAP splitting probability of 
type $\gamma^*\to
\bbbar$ with mass effects.
More precisely, in option 2, $g\to \bbbar$ splittings are generated
based on massless kinematics,
and the $r_b$ mass correction is implemented through 
reweighting, while 
massive kinematics is restored through momentum reshuffling.
Option 3, which implements massive DGLAP splittings
in a more realistic way, involves an additional 
$(1+\delta)/(1-\delta)$ factor
that leads to a significant enhancement of the $g\to \bbbar$ rate.
This option is 
excluded by LEP/SLC data and also by direct 
measurements of $\ttbar+b$-jet production~\cite{Aad:2015yja}.
Finally, option 1 corresponds to option 2 with $r_b=0$ and 
yields very similar results. 
Thus, for the assessment of $g\to \bbbar$ shower uncertainties we will
compare options 4 and 2.

In addition to the functional form of the heavy-quark splitting kernel
we also vary the scale of $\as$ in the parton shower.
To this end,  we 
set {\tt TimeShower:weightGluonToQuark} to 6 and 8,
which corresponds to options 2 and 4
with 
$\as(p_T^2)$ replaced by $\as(m_{\bbbar}^2)$
in the heavy-quark splitting kernel \refeq{eq:gbbmesplitting}.
Moreover, using {\tt TimeShower:renormMultFac},
we vary $\as(p_\rT^2)\to \as(\xi p_\rT^2)$ 
with prefactors $\xi=0.1,1,10$
both for options 2 and 4. This latter variation is applied to 
all final-state QCD splittings, i.e.~also splittings of type
$g\to gg$, $q\to q g$, etc.

\subsection{Comparisons against alternative generators}
\label{se:gencompsettings}

In order to assess systematic uncertainties related to the 
parton shower and the matching scheme, 
in \refses{se:PSplots}{se:gencomp} we compare
\PowhegPythia predictions of $\ttbb$ production
against corresponding predictions generated with \PowhegHerwig 
and with \Sherpa~\cite{Cascioli:2013era}.
The \PowhegPythia and \Sherpa generators of 
$\ttbb$ production are also compared against 
corresponding generators of inclusive $\ttbar$ production in the 5F
scheme\footnote{In the case of \Powheg, the well known {\tt hvq} generator~\cite{Frixione:2007nw} is used.}.

In the case of \Herwig~\cite{Bellm:2015jjp}
we apply the angular ordered shower
using version 7.1, setting $m_b = 4.75$\,GeV,
and leaving the strong coupling to its 
default value, 
$\alpha_S(m_Z) = 0.126234$.
To restrict the hardness of  \Herwig emissions 
according to the value of {\tt scalup} in the LHEs
we set
\begin{verbatim}
    set /Herwig/Shower/ShowerHandler:RestrictPhasespace Yes
    set /Herwig/Shower/ShowerHandler:MaxPtIsMuF Yes
\end{verbatim}

In the case of \Sherpa we use version 2.2.4 
with its default tune\footnote{More precisely, 
in order to 
be consistent with the \Sherpa2.1 benchmarks
presented in~\citere{deFlorian:2016spz}
we have used the shower recoil scheme proposed in \citere{Hoeche:2009xc},
which  was the default in  \Sherpa2.1.
This corresponds to setting {\tt CSS\_KIN\_SCHEME=0},
while {\tt CSS\_KIN\_SCHEME=1}, which became the new 
default in \Sherpa2.2, 
leads to slightly more significant 
differences with respect to the 
$\ttbar+b$-jet predictions of~\citere{deFlorian:2016spz}.
More precisely, comparing \PowhegPythia 
against \Sherpa2.2 with the new recoil scheme 
we observe differences at the level of 10\% 
in the ttbb cross section and up to about 40\%
in the light-jet $p_\rT$ spectrum.
Such differences are well consistent with QCD scale variations.
For comparison they are three times smaller 
with respect to the differences 
between \SherpaOpenLoops and
\MadgraphaMC in \citere{deFlorian:2016spz}.
Note that {\tt CSS\_KIN\_SCHEME} acts only on the second and 
subsequent shower emissions, \ie it does not affect the 
SMC@NLO matching procedure.
}
for \Sherpa's dipole shower~\cite{Schumann:2007mg}. 
The relevant one-loop matrix elements are computed with \OpenLoops,
and matching to the parton shower is based on the \Sherpa implementation~\cite{Hoeche:2011fd}
of the MC@NLO method~\cite{Frixione:2002ik}, dubbed SMC@NLO.
As for the hard cross section we use
the same input parameters, PDFs and scale settings as specified in
\refse{se:input} for the case of \Powheg.
Moreover, as motivated in~\refse{se:methods} 
we identify the resummation scale $\mu_Q$ in \Sherpa with the $\hdamp$ parameter in \Powheg, \ie
we set $\mu_Q=H_\rT/2$.
In the \Sherpa simulation the {\tt
NNPDF30\_nlo\_as\_0118\_nf\_4} PDF set is used throughout, \ie also for paton showering.

For \Powheg and \Sherpa 
simulations of inclusive $\ttbar$ production we use the same
setup as for the corresponding $\ttbb$ generators, 
with the only exceptions being 
the QCD scales, $\mur=\muf=0.5\sqrt{E_{\rT,t}E_{\rT,\bar t}}$,
and the choice of the {\tt NNPDF30\_nlo\_as\_0118\_nf\_5} PDF set.
In this setup the inclusive NLO cross section amounts to
$\sigma_{\ttbar}=815$\,pb, which is only 2\% below the 
NNLO prediction of $832^{+45}_{-50}$\,pb~\cite{Czakon:2013goa}.

\subsection{Simulations with stable or decayed top quarks}
\label{se:topdecaytreatment}

In \refses{se:stabletops}{se:decayedtops}
predictions for $\ttbb$ production are presented both for the case of
stable top
quarks  and with spin-correlated top decays.  
Simulations with stable top
quarks permit to avoid the combinatorial complexity that results from the
presence of four $b$-quarks in decayed $\ttbb$ events.  In this way one can
focus on the production of the $\bbbar$ pair that is governed by QCD
dynamics and which represents the main source of theoretical uncertainty in
$pp\to \ttbb$.  Moreover, results with stable top quarks can be compared to
the benchmarks of~\citeres{Cascioli:2013era, deFlorian:2016spz}.
Since top quarks do not hadronise, when we switch off 
top decays we disable hadronisation
and, following 
\citere{Cascioli:2013era, deFlorian:2016spz}, we also deactivate multi-parton interactions (MPIs) and QED radiation in the 
parton shower.
This is achieved by setting
\begin{verbatim}
   pythia.readString("6:mayDecay = off");
   pythia.readString("-6:mayDecay = off");
   pythia.readString("SpaceShower:QEDshowerByQ = off");
   pythia.readString("SpaceShower:QEDshowerByL = off");
   pythia.readString("TimeShower:QEDshowerByQ = off");
   pythia.readString("TimeShower:QEDshowerByL = off");
   pythia.readString("PartonLevel:MPI = off")
   pythia.readString("HadronLevel:All = off");
\end{verbatim}

For the case of decaying top quarks we show results both with 
hadronisation and MPI switched off or on, while 
the QED shower is always activated and hadrons are kept stable throughout.
For the implementation of spin-correlated decays in the \Powheg framework
we follow the approach of~\citere{Frixione:2007zp}, which has already been employed in the \PowhegBox framework
in~\citeres{Frixione:2007nw,Alioli:2011as,Hartanto:2015uka}.
More precisely, we use
resonant tree matrix elements  for the full 
$2\to 8$ Born processes 
$q\bar q/gg\to t (\to b ij)\bar{t}(\to \bar b kl)\bbbar$,
where $ij$ and $kl$ stand for the
leptons or quarks from $W$ decays,
and corresponding $2\to 9$ 
processes with an additional external gluon at the level of the 
$pp\to\ttbb$ sub-process.
In the $2\to 8(9)$ matrix elements we include only 
topologies with two intermediate top resonances.
This accounts for spin correlations as well as for off-shell effects associated with
the top and the $W$ propagators.
Technically, top decays are generated starting from on-shell 
$\ttbb$ events with a veto algorithm based on the
ratio between $2\to 8(9)$  matrix elements
and corresponding $2\to 4(5)$ matrix elements for the underlying
$pp\to\ttbb$(+jet) process.

As additional input parameters for top decays 
we use~\cite{Patrignani:2016xqp}
\begin{equation}
\label{eq:ewinputs}
M_{W} = 80.385 \;\GeV,\qquad G_{\rF} = 1.1663787\cdot 10^{-5}\; \GeV^{-2},
\end{equation}
the total widths
\begin{equation}
\label{eq:inputwidths}
\Gamma_t=1.329\;\GeV,\qquad
\Gamma_W=2.089\;\GeV\,,
\end{equation}
and the branching ratios
\begin{eqnarray}
\label{eq:GammaWa}
\mathrm{BR}_{t\to b\ell_i\nu_j}&=&
\mathrm{BR}_{W\to \ell_i\nu_j}=
\frac{\delta_{ij}}{3}\mathrm{BR}_{W\to \mathrm{lept}},
\\
\mathrm{BR}_{t\to b u_i d_j}&=&
\mathrm{BR}_{W\to u_i d_j}=
\frac{|V_{ij}|^2}{2}\mathrm{BR}_{W\to \mathrm{had}},
\end{eqnarray}
where we assume a 100\% branching ratio for 
$t\to bW$ decays.
For the total $W$-boson branching ratios into leptons and hadrons
we use the values~\cite{Patrignani:2016xqp}
\begin{eqnarray}
\label{eq:GammaWb}
\mathrm{BR}_{W\to \mathrm{had}}=0.675,\qquad
\mathrm{BR}_{W\to \mathrm{lept}}=0.325\,,
\end{eqnarray}
which include state-of-the-art higher-order corrections.

\subsection{Jet observables and acceptance cuts}
\label{se:cuts}
For the reconstruction of jets we use the anti-$k_\rT$~\cite{Cacciari:2008gp}
algorithm with $R=0.4$. We select jets that fulfil
\begin{equation}
\label{eq:jetcuts}
p_{\rT}>25\,\GeV,\qquad
|\eta|<2.5,
\end{equation}
both for the case of light jets and $b$-jets.  
At parton level, we define as $b$-jet a jet that contains at least a $b$-quark, \ie jets that contain a $\bbbar$ pair
arising from a collinear
$g\to \bbbar$ splitting are also tagged as $b$-jets. At particle level, i.e.~when hadronisation is switched on,
we tag as $b$-jets those jets that are matched to a $B$-hadron
using the ghost method 
as implemented in \FastJet~\cite{Cacciari:2011ma}.

When studying $\ttbb$ production with stable top quarks, in
Sections~\ref{se:gbbstudies} and \ref{se:stabletops}, we categorise events
according to the number $\Nb$ of $b$-jets that do not arise from top decays
and fulfil the acceptance cuts~\refeq{eq:jetcuts}.
For the analysis of cross sections and distributions
we consider an inclusive selection with $N_b\ge 1$
and a more exclusive one with $\Nb\ge 2$. We refer to them
as ttb and ttbb selections, respectively.

In \refse{se:decayedtops} we present
predictions for $\ttbb$ production with 
top-quark decays in the dilepton channel.
In this case we require two oppositely charged leptons,
$\ell=e$ or $\mu$, with
\begin{equation}
p_{\rT,\ell}>20\,\GeV,
\qquad
|\eta_{\ell}|< 2.5.
\label{eq:leptoncuts}
\end{equation}
Charged leptons are dressed with collinear photon radiation 
within a cone of radius $0.1$.
We do not apply any cut on missing transverse energy.
Jets are defined  as for the case of
stable top quarks, and  we select events with at least four
$b$-jets that fulfill the acceptance cuts 
\refeq{eq:jetcuts}.

\section{Predictions for $\boldsymbol{\ttbb}$ production with stable top quarks}
\label{se:stabletops}

In this section we present numerical predictions for $pp\to \ttbb$ at
$\sqrt{s}=13$\,TeV in the 4F scheme.  The presented results have been
obtained with \PowhegOpenLoops using the setup of \refse{se:technical}.
Top quarks are kept stable throughout as specified in
\refse{se:topdecaytreatment}, and we study cross sections and distributions
in the inclusive ttb phase space with $\Nb\ge 1$ $b$-jets, as well as in the
ttbb phase space with $\Nb\ge 2$.

\subsection{NLOPS predictions with perturbative uncertainties}
\label{se:pertplots}

{\renewcommand{\arraystretch}{1.5}
\begin{table*}[t]
  \vspace*{0.3ex}
  \begin{center}
\resizebox{\textwidth}{!}{\begin{tabular}{c|c|cc|cc|cc|cc}
& 
  LO    &     NLO   &   $\frac{\mathrm{NLO}}{\mathrm{LO}}$   &
  LOPS  &   $\frac{\mathrm{LOPS}}{\mathrm{LO}}$ &
  NLOPS &   $\frac{\mathrm{NLOPS}}{\mathrm{NLO}}$ &
  LHE   &   $\frac{\mathrm{LHE}}{\mathrm{NLO}}$ \\
\hline
$\sigma_{\mathrm{ttb}}[\fb]$  & 
  $6545^{+74\%}_{-39\%}$  &   $12813^{+34\%}_{-27\%}$      &  1.96  &
  $7006^{+75\%}_{-39\%}$  &    1.07                                 &
  $13090^{+39\%}_{-29\%}$  &   1.02                                 &
  $13029^{+36\%}_{-28\%}$  &  1.02
\\ 
$\sigma_{\mathrm{ttbb}}[\fb]$  & 
  $1209^{+70\%}_{-38\%}$  &   $2261^{+30\%}_{-26\%}$      &  1.87  &
  $1562^{+73\%}_{-39\%}$  &   1.29                                 &
  $2537^{+40\%}_{-29\%}$  &   1.12                                 &
  $2392^{+34\%}_{-27\%}$  &   1.06
\\ 
$\sigma_{\mathrm{ttbb}_{100}}[\fb]$ & 
  $358^{+70\%}_{-38\%}$   &   $640^{+26\%}_{-25\%}$       &  1.79  &
  $584^{+73\%}_{-39\%}$   &   1.63                                 &
  $810^{+41\%}_{-29\%}$  &   1.27                                  &
  $678^{+31\%}_{-26\%}$  &   1.06
\\ \hline
$ \frac{\sigma_{\mathrm{ttb}}}{\sigma_{\mathrm{ttbb}}}$ & 
   5.41 &   5.67  &  1.05  &
   4.48 &     0.83  &
   5.16 &     0.91 &
   5.45 &     0.96
\\ 
$ \frac{\sigma_{\mathrm{ttbb}}}{\sigma_{\mathrm{ttbb}_{100}}}$ & 
   3.38 &   3.53  &  1.05  &
   2.67 &     0.79  &
   3.13 &     0.88  &
   3.53 &     1.00
    \end{tabular}}
  \end{center}
  \caption{
Cross sections for $pp\to\ttbb$ at $\sqrt{s}$=13\,TeV
and their ratios in the phase space regions with $\Nb\ge 1$ (ttb)
and $\Nb\ge 2$ (ttbb) $b$-jets as well in the 
region $m_{b_1b_2}>100\,$GeV of the ttbb phase space
(ttbb$_{100}$).
Nominal fixed-order predictions at LO and NLO
accuracy are compared to corresponding LOPS and NLOPS predictions
of \PowhegPythia. 
Also cross sections at LHE level are reported.
Uncertainties correspond to the envelope of
the 7-point factor-two variations of $\mur$ and $\muf$. 
}
\label{tab_pert:XS}
\end{table*}
}

In this section we compare (N)LO and (N)LOPS predictions focusing on 
NLO and matching effects as well as perturbative and PDF uncertainties.
\refta{tab_pert:XS} presents  cross sections in the ttb and ttbb phase
space, as well as in the presence of an additional cut,
$m_{b_1b_2}>100\,$GeV, on the invariant mass of the two hardest $b$-jets.
At fixed order we find perfect agreement with the NLO results 
of~\citere{deFlorian:2016spz}. 
The various phase space regions feature similar NLO uncertainties, around
25--30\%, while corresponding LO scale variations are roughly a factor two
larger.  Both at LO and NLO, scale uncertainties are strongly dominated by
$\mur$ variations.
The large $\sigma_{\mathrm{ttb}}/\sigma_{\mathrm{ttbb}}$ ratio, which
exceeds a factor 5, reflects the appearance of large logarithms of $m_b$
when a $b$-quark becomes unresolved.
As shown in \refse{se:ttbbISvsFS}, such logarithms are mainly due to FS
$g\to \bbbar$ splittings.  Thus the use of 4F PDFs, where $\ln(m_b/Q)$ effects
of IS origin are not resummed in the PDF evolution is well justified.
Note also that $\ln(m_b/Q)$ effects in  $\sigma_{\mathrm{ttb}}$ are present
already at LO.  Thus they do not jeopardise the convergence of the
perturbative expansion.  In fact,
$\sigma_{\mathrm{ttb}}/\sigma_{\mathrm{ttbb}}$ turns out to be very stable
with respect to NLO corrections. The same hold for 
$\sigma_{\mathrm{ttbb}}/\sigma_{\mathrm{ttbb}_{100}}$.

At variance with \cite{Cascioli:2013era}, where LO calculations were
performed using LO PDFS and the corresponding value of $\as$, here,
in order to obtain a more realistic picture of the convergence of the
$\as$-expansion, we use NLO inputs throughout.%
\footnote{For
processes like $\ttbb$ production, whose LO cross section scales with
$\as^4$, evaluating the $K$-factor with LO inputs for the LO cross section
results is a very strong dependence on the LO value of $\as$.  The latter
can depend very strongly on the employed PDF set.  In particular, the two
existing NNPDF 4F LO sets, which correspond to $\as(M_Z)=0.118$ and
$\as(M_Z)=0.130$,
can result in factor 1.5 ambiguity in the $\ttbb$
$K$-factor.  We also note that the local $K$-factor in the \Powheg matching
formula, i.e.~the $\bar B/B$ ratio in \refeq{eq:powheg1}, is computed using
NLO input throughout.}
This approach increases the 
NLO $K$-factors from 1.15--1.25 \cite{Cascioli:2013era} to 
1.80--1.95. This observation raises some concerns regarding the possible
presence of significant higher-order corrections beyond NLO and calls for 
a better understanding of the origin of the large $K$-factor at NLO.
This question as well as the search for possible improvements 
is deferred to future studies.

Comparing fixed-order (N)LO cross sections against (N)LOPS ones we find that
matching and showering effects are almost negligible in
$\sigma_{\mathrm{ttb}}$, while in the case of $\sigma_{\mathrm{ttbb}}$ they
slightly exceed 10\%, and in the Higgs-signal region,
$m_{b_1b_2}>100\,\GeV$, they approach 30\% .
As pointed out in~\citere{Cascioli:2013era}, such effects can be understood
in terms of $\ttbar+2b$-jet production via double $g\to \bbbar$ splittings. 
In practice, one of the $b$-jets results from a $g\to \bbbar$ splitting in
the $\ttbb$ matrix element, while the second one is created by the parton
shower via a further $g\to \bbbar$ collinear splitting.
This interpretation is confirmed by the fact that 
the enhancement at hand is not present in the LHE-level 
cross sections presented in~\refta{tab_pert:XS}.
In fact, double splittings are generated only at NLOPS level through parton showering.
Double-splitting enhancements in~\refta{tab_pert:XS} behave in a
qualitatively similar way as in \citeres{Cascioli:2013era,
deFlorian:2016spz}, but their size turns out to depend on the employed NLOPS
generator.
As compared to~\citere{deFlorian:2016spz}, we observe that the NLOPS/NLO
correction to ${\sigma_\mathrm{ttbb}}$ in \refta{tab_pert:XS} (+12\%) is
twice as large as in \Sherpa (+6\%), very close to \Powhel\footnote{We note in
passing that for other observables, especially in the ttb phase space, 
results in
this paper can deviate more significantly from the \Powhel predictions
of~\citere{deFlorian:2016spz}.
This can be attributed to differences in the \Pythia settings and, most
importantly, to the fact that the \Powhel generator used
in~\citere{deFlorian:2016spz} was based on 5F $\ttbb$ matrix elements with
$m_b=0$ and made use of technical generation cuts in order to avoid
collinear singularities from $g\to \bbbar$ splittings.
This limitation has now been overcome by upgrading the \Powhel
$\ttbb$ generator to the 4F scheme~\cite{Bevilacqua:2017cru}.
}
(+13\%) and well
below the prediction of \MadgraphaMC (+41\%).

For what concerns scale variations,  in~\refta{tab_pert:XS} we see that their 
impact at NLOPS tends to be 5--10\% higher as compared to fixed-order NLO. 
This is consistent with the behaviour of \MadgraphaMC and \Powhel
in~\citere{deFlorian:2016spz}, while \Sherpa features a significantly lower
scale uncertainty.  Such differences may be an artefact of the incomplete
implementation of scale variations in the various NLOPS tools.
In the case of \Powheg, as anticipated in \refse{se:methods}
we have found that increasing $\hbzd$  can lead to unphysical enhancements of the
scale uncertainty. This effect is mostly visible in the ttbb phase space, 
where the maximum scale variation amounts to $+40\%$ 
for $\hbzd=2$ and grows up to $+45\%$ and $+54\%$
when setting $\hbzd=5$ and 50, respectively.
Based on these observations, as default for our $\ttbb$ simulations
we have set $\hbzd=2$. This choice
guarantees a decent consistency with fixed-order scale variations
without altering the matching procedure in a drastic way.  
In particular, when $\hbzd$ is reduced from its standard
\PowhegBox value of $5$ down to $2$,
we have checked that the fraction of the $\ttbar+b$-jet cross
section that is shifted from the singular part~\refeq{eq:powheg1} to the
finite remnant~\refeq{eq:powheg3} amounts to only 10--20\%. This holds for all
considered distributions in the ttb and ttbb phase space.


\newcommand{\pertplot}[1]{
\begin{minipage}{.45\textwidth}
\begin{center}
\includegraphics[width=\textwidth,trim={2 0 2 0},clip]{pert/#1}\\
\end{center}
\end{minipage}
}

\newcommand{\pertplotL}[1]{\hspace{-0mm}\pertplot{#1}}
\newcommand{\pertplotR}[1]{\pertplot{#1}\hspace{-0mm}}

\begin{figure}[t!]
\centering
\subfigure[]{
\pertplotL{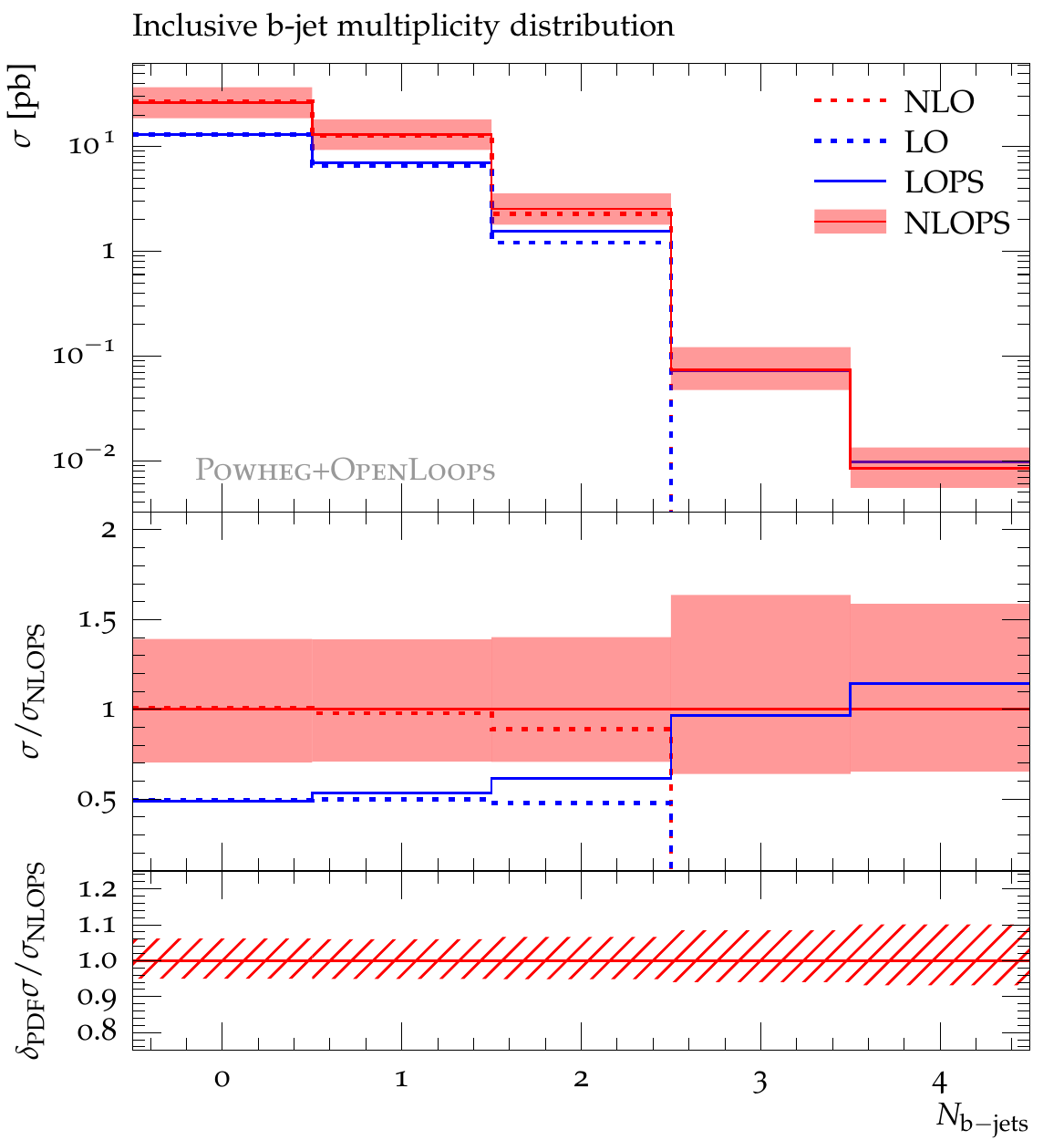}
\label{fig_pert:NBj_XS}}
\subfigure[]{
\pertplotR{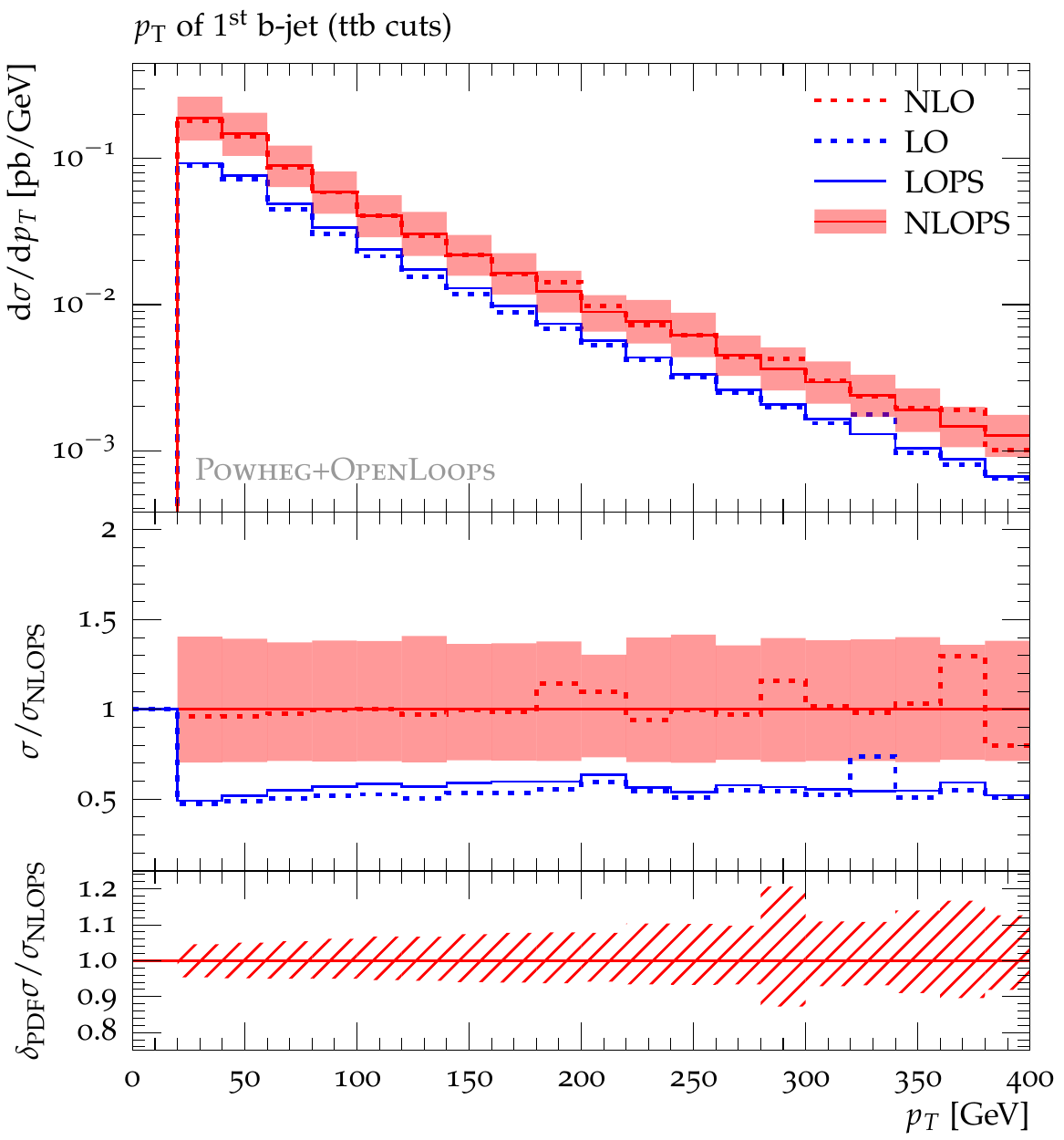}
\label{fig_pert:1_PT_B1}}

\subfigure[]{
\pertplotL{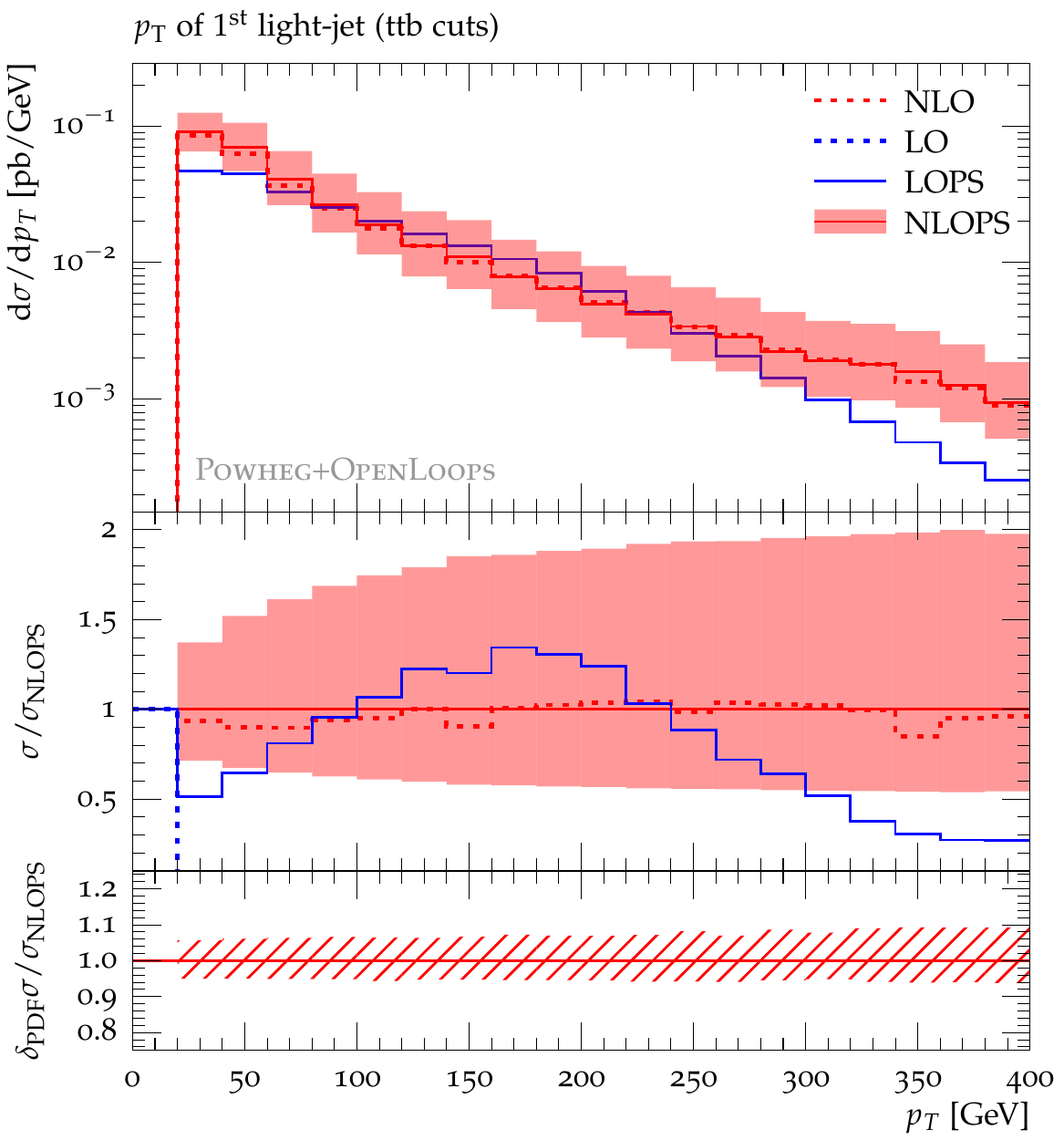}
\label{fig_pert:1_PT_J1}}
\subfigure[]{
\pertplotR{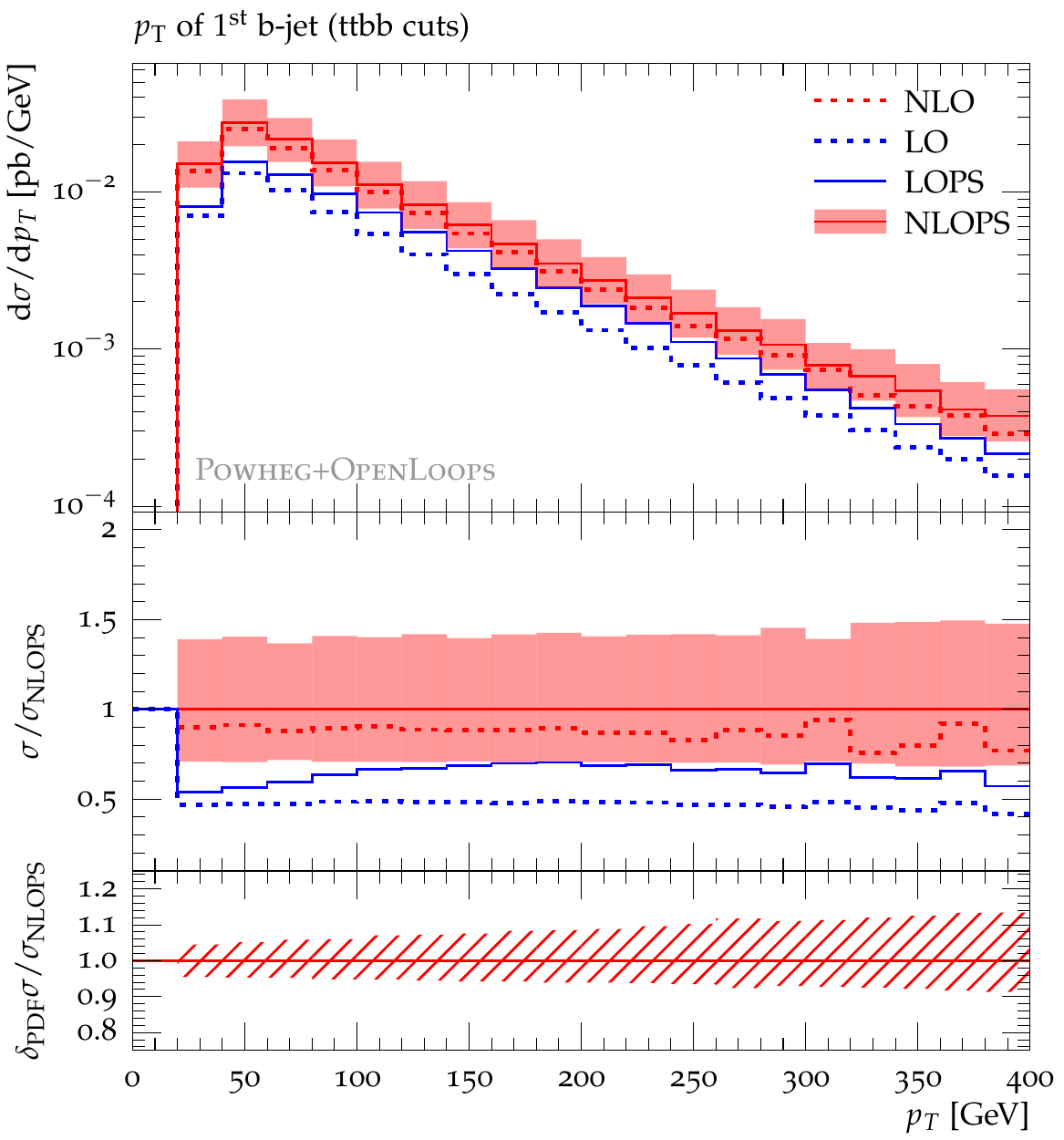}
\label{fig_pert:2_PT_B1}}
\caption{Predictions for $pp\to\ttbb$ at $\sqrt{s}$=13\,TeV: distributions
in the inclusive number of additional $b$-jets (a), the $\pt$ of the first
$b$-jet (b) and the first light jet (c) with ttb cuts, and in the $\pt$ of
the second $b$-jet with ttbb cuts (d).
Results at LO and NLO are in blue and red, respectively, and dashed lines
correspond to fixed-order (N)LO predictions, while solid curves represent
(N)LOPS predictions.  The bands illustrate the envelope of 7-point
$\muR,\muF$ variations.
Absolute predictions are shown in the main frame.  The first ratio plot
shows LO, LOPS and NLOPS predictions normalised to fixed-order NLO.  The
second ratio plot displays the relative effect of PDF uncertainties applied
to NLOPS predictions
Top quarks are kept stable throughout.
}
\label{fig:pert1}
\end{figure}

\begin{figure}[t!]
\centering
\subfigure[]{
\pertplotL{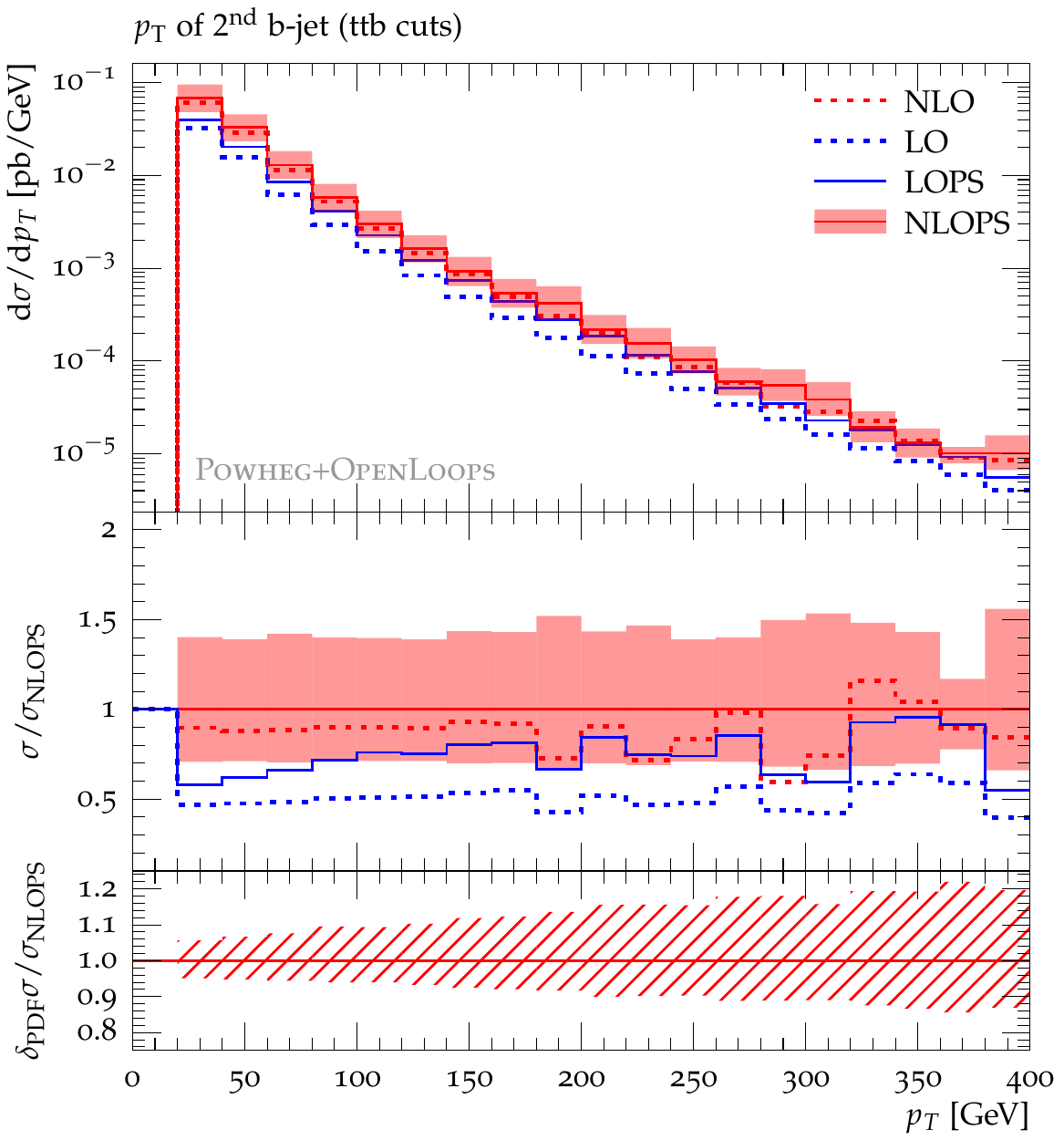}
\label{fig_pert:2_PT_B2}}
\subfigure[]{
\pertplotR{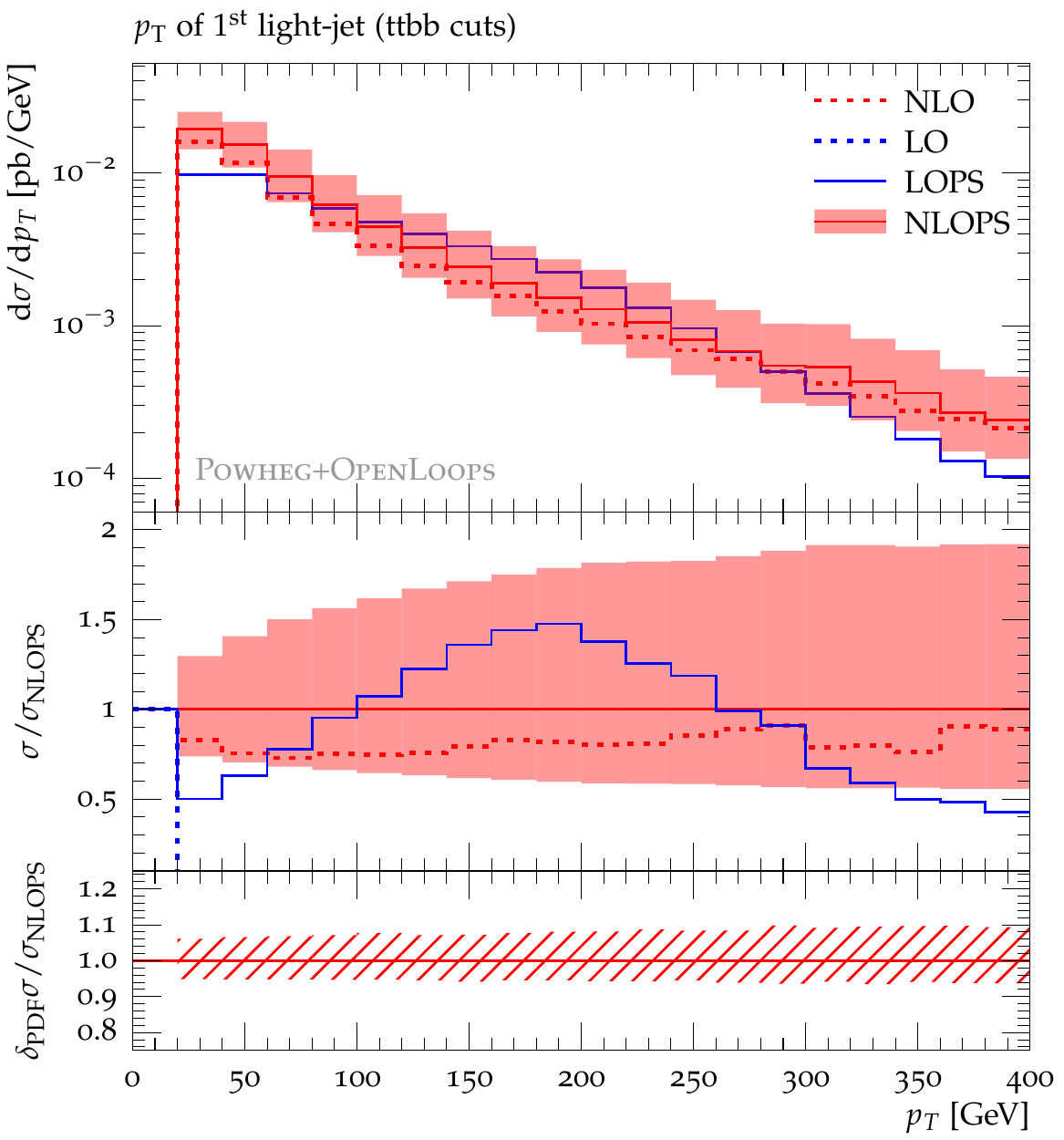}
\label{fig_pert:2_PT_J1}}
\subfigure[]{
\pertplotL{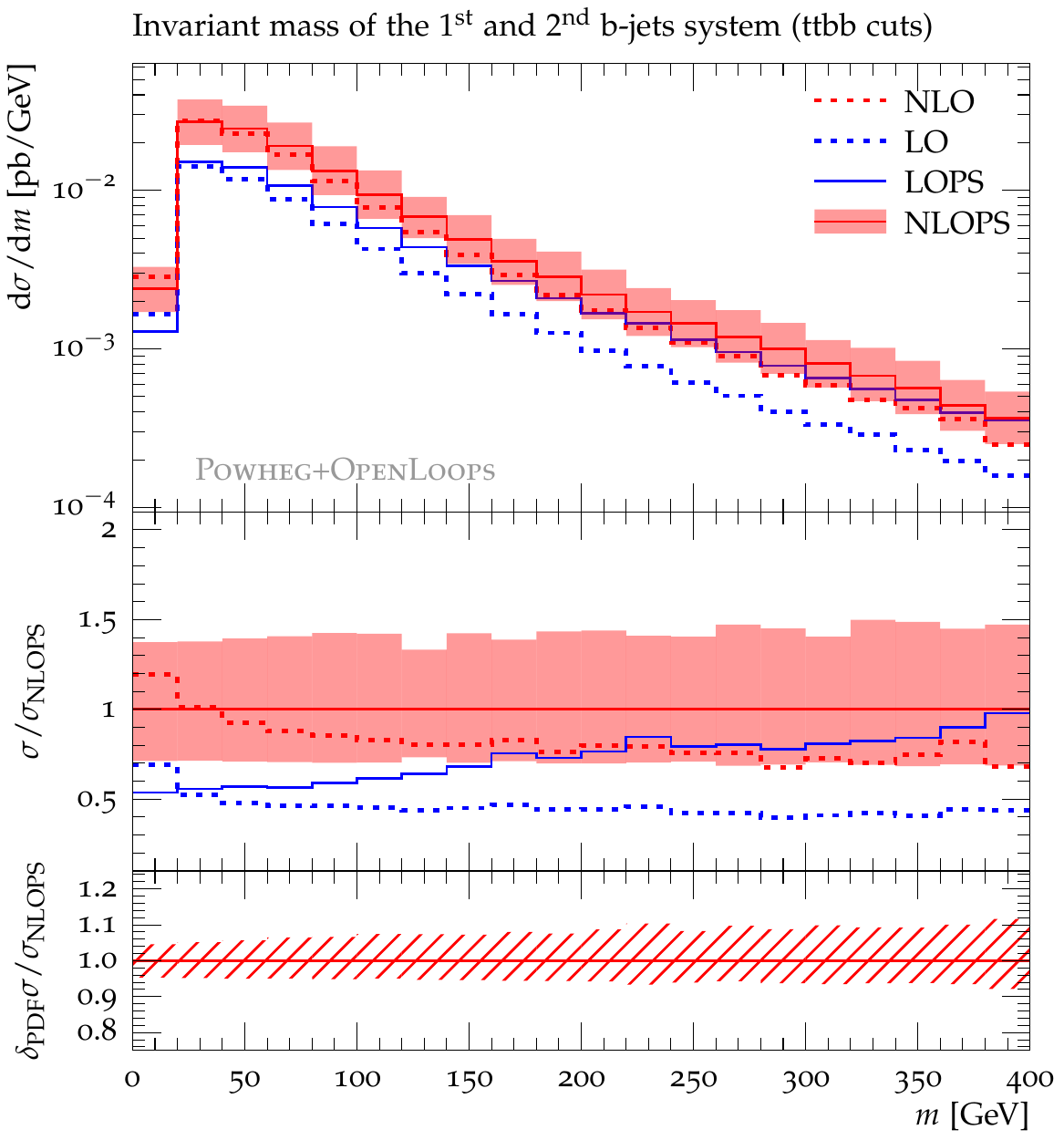}
\label{fig_pert:2_M_B1B2}}
\subfigure[]{
\pertplotR{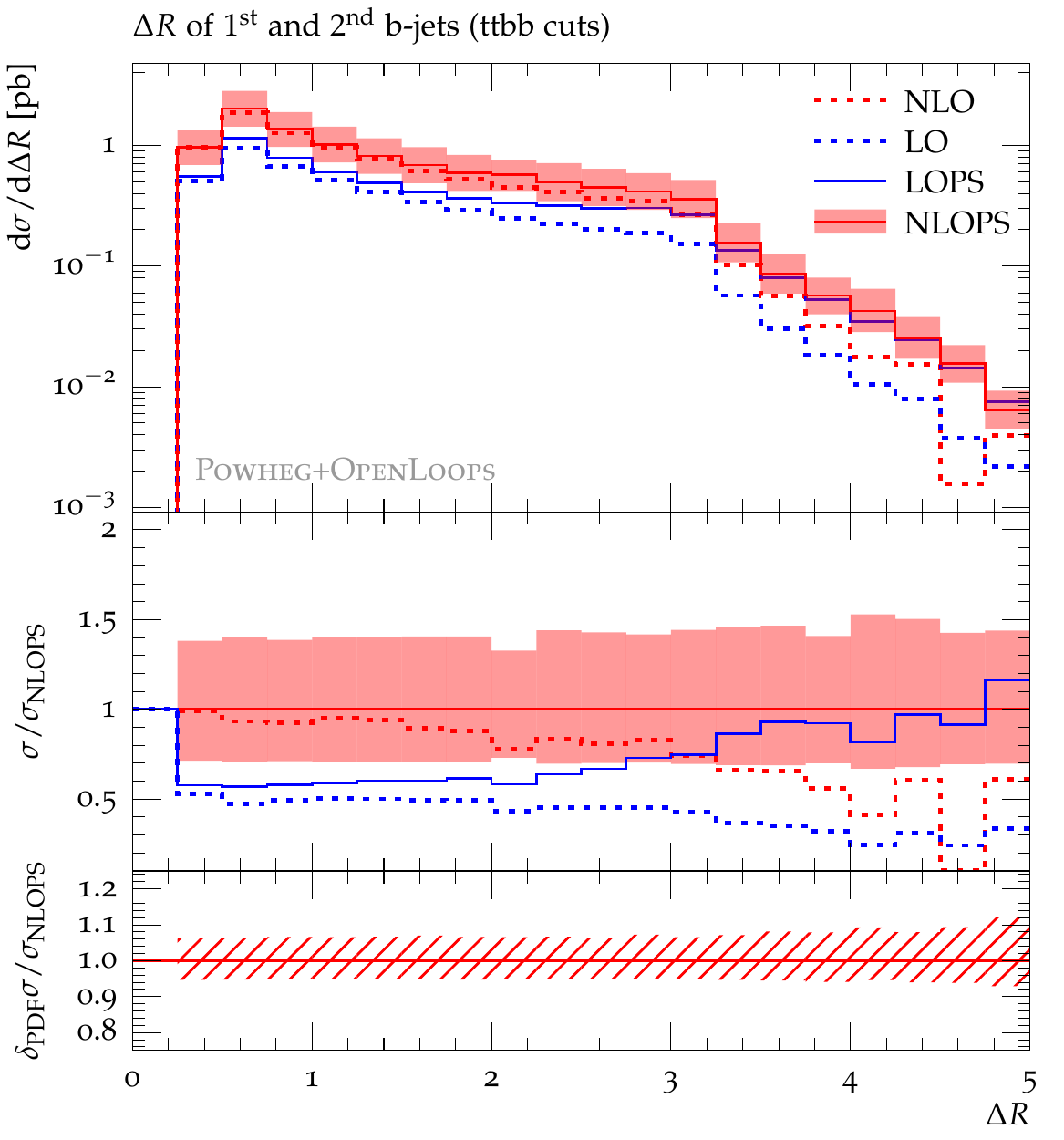}
\label{fig_pert:2_DR_B1B2}}
\caption{Distributions in the $\pt$ of the second $b$-jet (a) in the $p_\rT$
of the first light jet (b), and in the invariant mass (c) and the $\Delta R$
separation (d) of the first two $b$-jets with ttbb cuts throughout.
Predictions and uncertainties as in \reffi{fig:pert1}.
}
\label{fig:pert2}
\end{figure}

Differential observables with ttb and ttbb cuts are presented in
\reffis{fig:pert1}{fig:pert2}.  The inclusive $b$-jet multiplicity
distribution in \reffi{fig:pert1}a extends the results of
\refta{tab_pert:XS}, which correspond to $\Nb\ge 1,2$, to the bins
with $\Nb\ge 3,4$.
The latter are populated by events that result from the interplay of
real-emission matrix elements and $g\to \bbbar$ parton-shower splittings. 
Thus they feature an enhanced scale dependence.

For kinematic distributions that are inclusive with respect to NLO QCD radiation, NLOPS
scale variations have a minor impact on shapes and amount essentially to a
normalisation shift, similar to what observed at the level of the ttb and ttbb
cross sections.
In contrast, in the case of the light-jet $p_\rT$ spectra, scale variations
increase from about 30\% in the soft region up to 100\% in the hard tails. 
This is consistent with the fact that such observables are only LOPS
accurate and depend on $\as^5(\mur)$.
The effect of PDF variations is clearly 
subleading as compared to scale uncertainties and 
has little impact on shapes.

Comparing (N)LOPS predictions to the respective fixed-order (N)LO results,
we observe that matching and shower effects remain almost negligible also at
the level of distributions in the ttb phase space.
As for the ttbb region, the NLOPS effects of order 10\%
observed in $\sigma_{\mathrm{ttbb}}$ turn out to be 
quite sensitive to the kinematics of $b$-jets.
In particular, as expected from the QCD dynamics 
of double $g\to \bbbar$ splittings~\cite{Cascioli:2013era},
the most pronounced effects are observed in the tails of the
$m_{b_1b_2}$ and $\Delta R_{b_1b_2}$ distributions, where the NLOPS/NLO ratio
approaches a factor two.
In the Higgs signal region, $m_{b_1b_2}\sim 125$\,GeV,
the NLOPS enhancement is around 1.25
and well consistent with~\citere{Cascioli:2013era}.

Comparing fixed-order NLO predictions to LO ones we find that, in spite
of the fairly large $K$-factors observed in \refta{tab_pert:XS},
the shapes of distributions turn out to be quite stable with respect
to higher-order QCD corrections.  In the case of (N)LOPS predictions,
the situation is different, especially for the shape of the light-jet $p_\rT$ spectra, 
which receives significant NLO distortions.
This is not surprising, since at LOPS
the light-jet $p_\rT$ is entirely generated by the parton shower. 
Thus the NLOPS/LOPS ratio should be regarded as a LO matrix-element
correction to the parton-shower approximation, rather than a NLO correction
in the perturbative sense.

Significant differences between NLOPS and LOPS shapes are observed also in
the $m_{b_1b_2}$ and $\Delta R_{b_1b_2}$ distributions.  Since the
respective NLO and LO shapes are very similar, this behaviour can be attributed
to the parton shower.  More precisely, it can be understood as a
side effect of the above-mentioned NLOPS/LOPS correction 
to the light-jet $p_\rT$ spectra, which is converted into 
a  double-splitting effect by 
$g\to \bbbar$ 
splittings 
inside the light jet.


\subsection{Shower uncertainties}
\label{se:PSplots}

\newcommand{\psplot}[1]{
\renewcommand{\arraystretch}{0}
\begin{minipage}{.45\textwidth}
\begin{center}
\includegraphics[width=\textwidth,trim={2 9.9mm 2 0},clip]{PS/#1_top}\\[0.mm]
\includegraphics[width=\textwidth,trim={2 9.9mm 2 6.mm},clip]{PS/#1_bottom} \\[0.mm]
\includegraphics[width=\textwidth,trim={2 0 2 6.mm},clip]{PS/#1_center}
\end{center}
\end{minipage}
}

\newcommand{\psplotL}[1]{\hspace{0mm}\psplot{#1}}
\newcommand{\psplotR}[1]{\psplot{#1}\hspace{0mm}}

\begin{figure}[t!]
\centering
\subfigure[]{
\hspace{-5mm}
\psplot{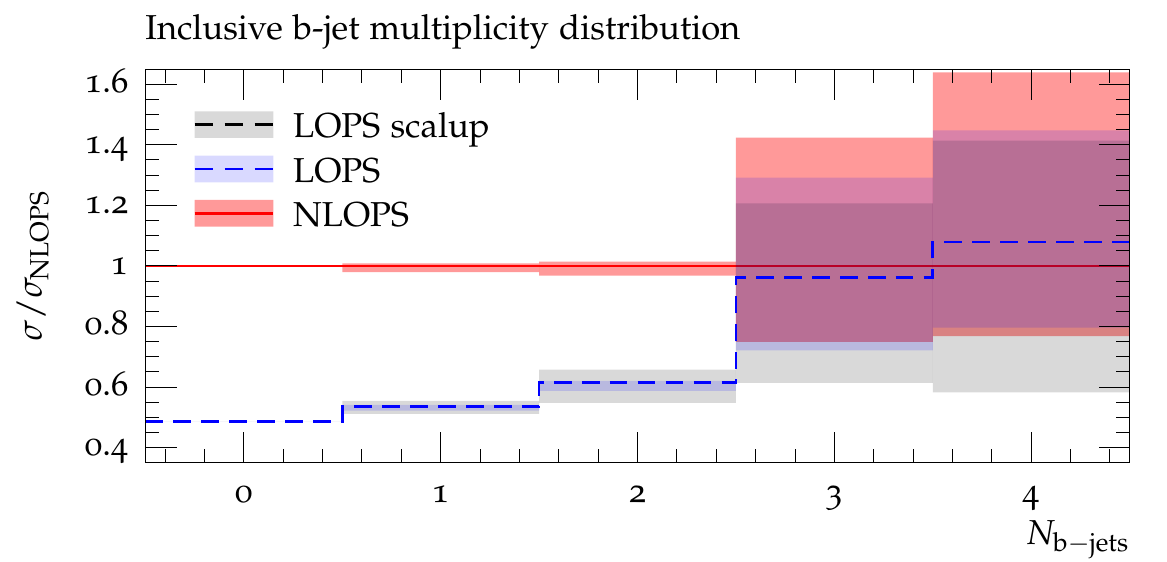}
\label{fig_PS:NBj_XS}}
\subfigure[]{
\psplot{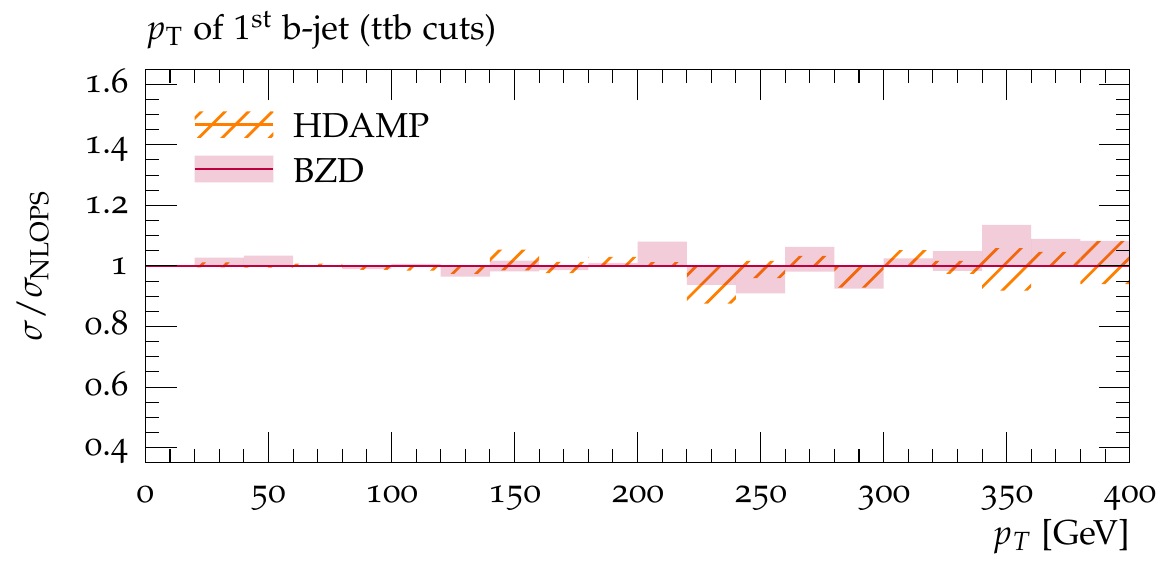}
\label{fig_PS:1_PT_B1}
\hspace{-5mm}}
\subfigure[]{
\hspace{-5mm}
\psplot{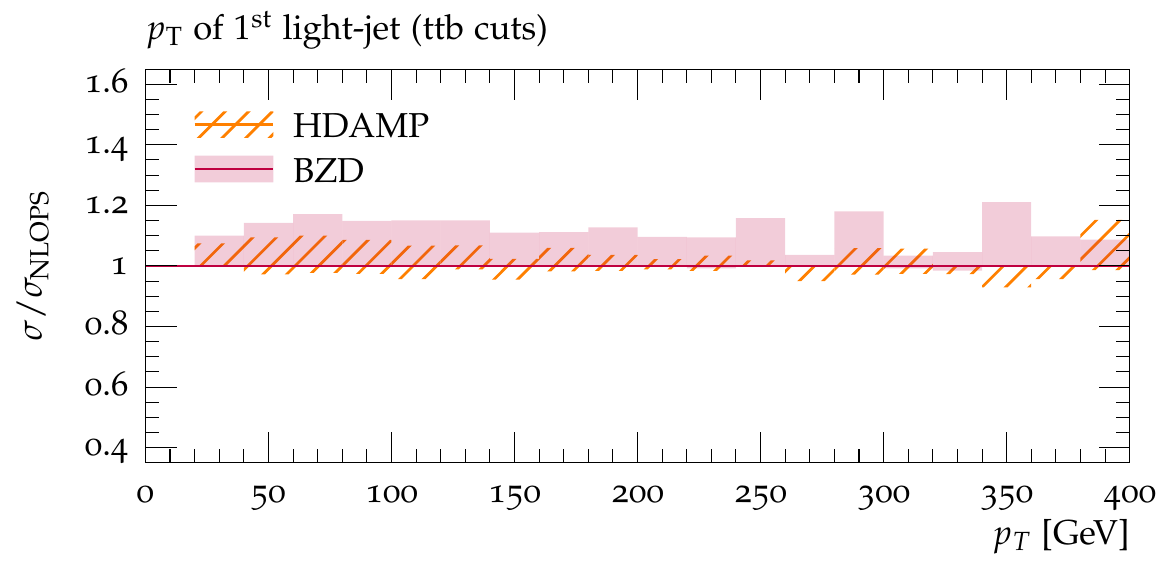}
\label{fig_PS:1_PT_J1}}
\subfigure[]{
\psplot{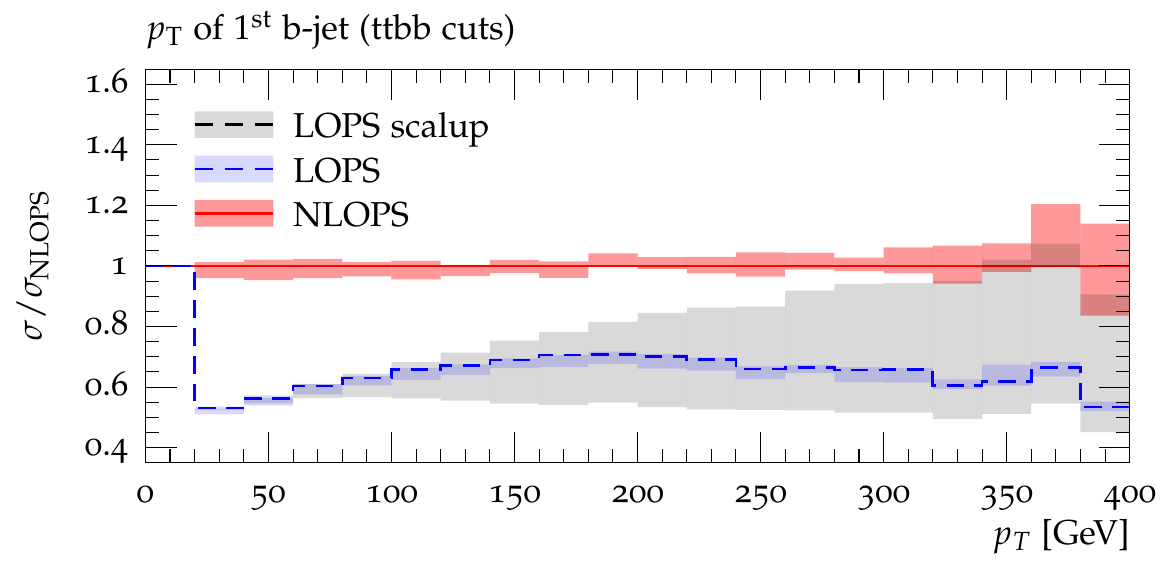}
\label{fig_PS:2_PT_B1}
\hspace{-5mm}}
\caption{Relative impact of shower effects and uncertainties
in (N)LOPS simulations of $pp\to\ttbb$ at $\sqrt{s}$=13\,TeV: distributions
in the inclusive number of additional $b$-jets (a), the $\pt$ of the first
$b$-jet (b) and the first light jet (c) with ttb cuts, and in the $\pt$ of
the second $b$-jet with ttbb cuts (d).
All results are normalised to 
nominal NLOPS predictions with \Pythia8.
The upper frame compares NLOPS result 
based on \Pythia8 (PY8) 
or \Herwig7 (HW7) against LHE results. 
The central frame compares NLOPS (red) and LOPS (blue) predictions
with uncertainties related to $\as$ variations and 
to the 
modelling of $g\to \bbbar$ splittings in \Pythia
(red NLO band and blue LO
band).  At LOPS, also variations of the shower starting scale
{\tt scalup}=$H_\rT/4,H_\rT/2,H_\rT$ are shown (grey band).
The lower frame illustrates the relative effect of $\hdamp=H_\rT,H_\rT/2,H_\rT,1.5
m_t$ variations (HDAMP) and $\hbzd=2,5,10$ variations (BZD).
Top quarks are kept stable throughout.
}
\label{fig:PS1}
\end{figure}

\begin{figure}[t!]
\centering
\subfigure[]{
\hspace{-5mm}
\psplot{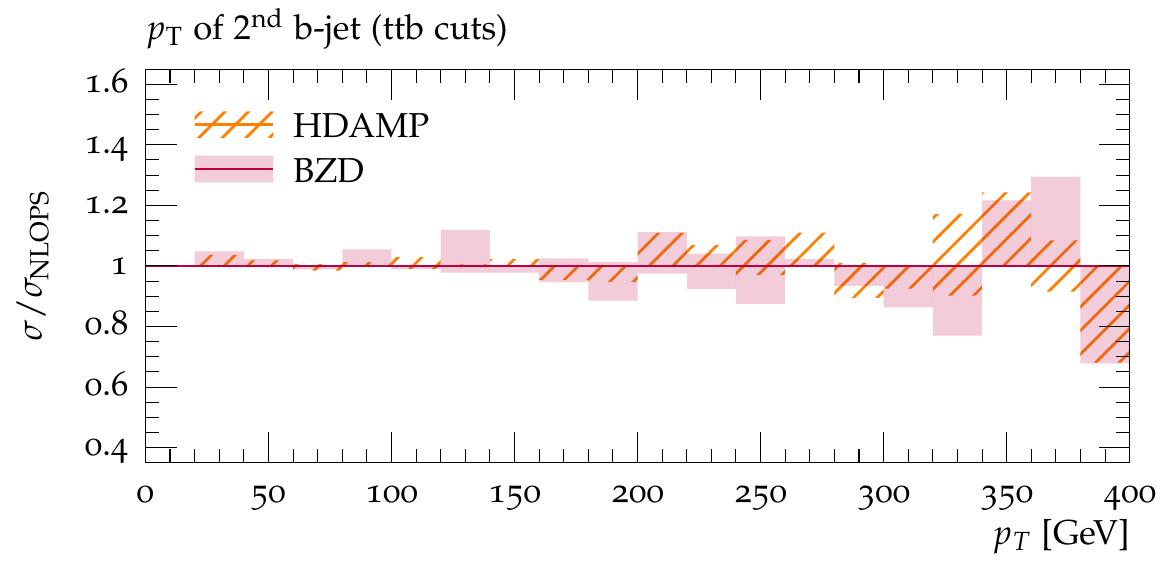}
\label{fig_PS:2_PT_B2}}
\subfigure[]{
\psplot{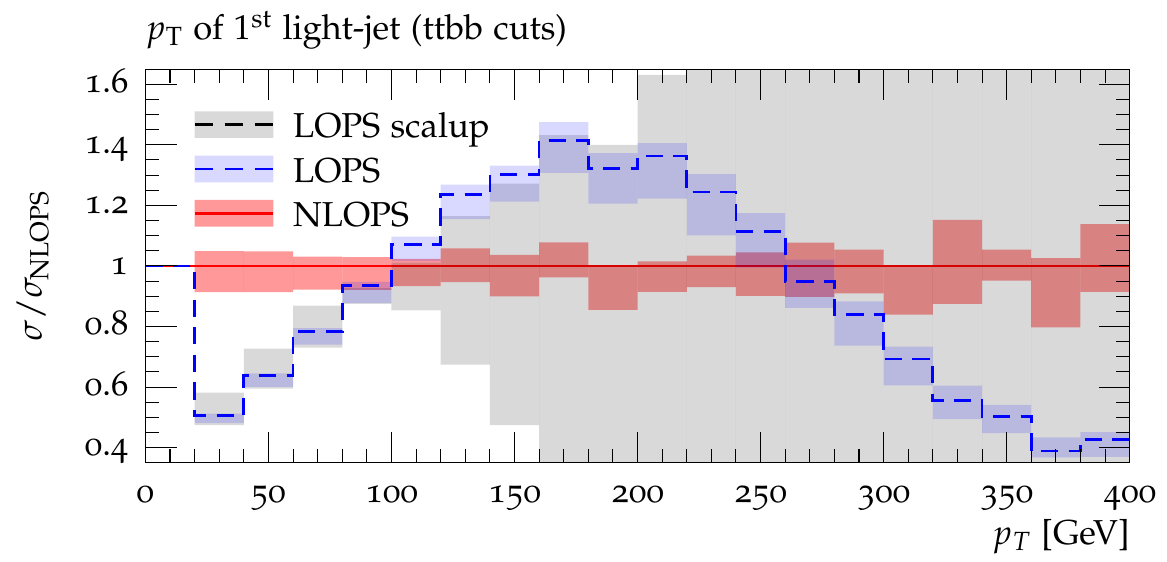}
\label{fig_PS:2_PT_J1}
\hspace{-5mm}}
\subfigure[]{
\hspace{-5mm}
\psplot{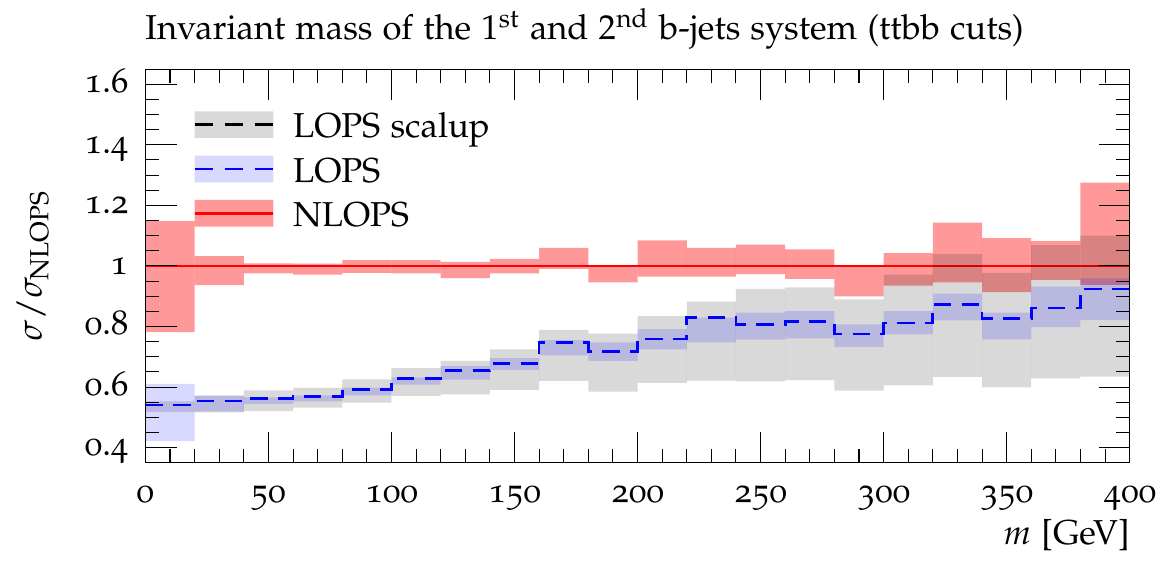}
\label{fig_PS:2_M_B1B2}}
\subfigure[]{
\psplot{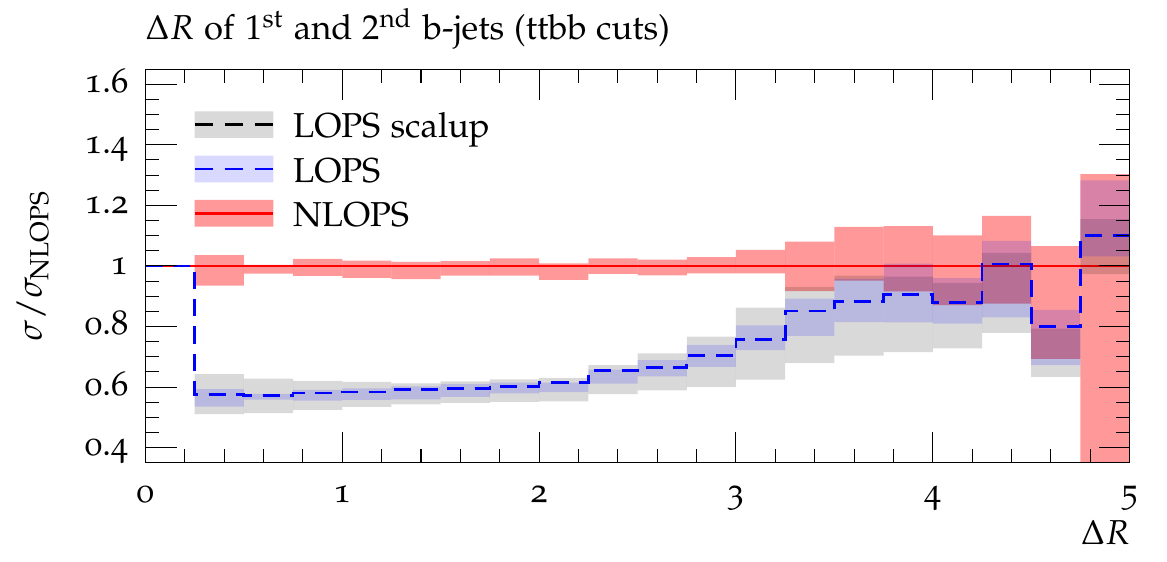}
\label{fig_PS:2_DR_B1B2}
\hspace{-5mm}}
\caption{Distributions in the $\pt$ of the second $b$-jet (a) in the $p_\rT$
of the first light jet (b), and in the invariant mass (c) and the $\Delta R$
separation (d) of the first two $b$-jets with ttbb cuts throughout.
Predictions and uncertainties as in \reffi{fig:PS1}.
}
\label{fig:PS2}
\end{figure}

In \reffis{fig:PS1}{fig:PS2} we study the sensitivity of (N)LOPS predictions
to parton-shower and matching uncertainties for the same observables
considered in \refse{se:pertplots}.

The ratios displayed in the upper frames illustrate the net effect of parton
showering by comparing full NLOPS predictions against results at LHE level. 
In addition, to assess parton-shower uncertainties,
NLOPS predictions based on \Pythia are compared to the corresponding results
obtained with \Herwig.

In the ttb phase space, apart from a mild distortion of the light-jet
spectrum, the net effect of parton showering is essentially negligible.
In contrast, in the ttbb phase space it increases the cross section 
by about 5\% and tends to grow in the tails of distributions.
The most sizable shower effects are observed in the $m_{b_1b_2}$ and $\Delta
R_{b_1b_2}$ distributions, where they reach up to 50--100\%.  This behaviour
is well consistent with the enhancement of the NLOPS/NLO ratio observed in
\reffi{fig:pert2}, and the fact that it is driven by the parton shower
provides further support to its interpretation in terms of double $g\to
\bbbar$ splittings.

\clearpage

In spite of the important role of parton showering, it is 
reassuring to observe that the sensitivity
of NLOPS predictions to the choice of parton shower is very small. In fact,
the typical agreement between results based on \Pythia and
\Herwig is at the level of a few percent both in the ttb and ttbb
selections.  Sizeable deviations at the level of 20\% are observed only
when requiring more than three $b$-jets.
As discussed in \refse{se:methods}, the very mild sensitivity 
of \Powheg predictions to the choice of parton shower is due 
to the fact that the first emission 
is completely independent of the parton shower in the \Powheg approach.
 
The ratios shown in the central frames of \reffis{fig:PS1}{fig:PS2}
illustrate (N)LOPS uncertainties
related to the modelling of $g\to \bbbar$ splittings and variations of 
$\as$ in \Pythia (see  \refse{se:shower}).
At LOPS also variations of the shower starting scale ({\tt scalup}) are shown.

The fact that $\ttbb$ 4F matrix elements populate the whole $\bbbar$ phase
space restricts the effect of $g\to \bbbar$ shower splittings to events with
four or more $b$-quarks. Thus, only the cross sections with $\Nb\ge 3,4$
$b$-jets suffer from sizable shower uncertainties.
Vice versa, all considered observables with ttb or ttbb cuts turn out to be
very stable, with typical shower uncertainties of a few percent
at NLOPS.
This holds also for the observables that are most sensitive to double
splittings, i.e.~$m_{b_1b_2}$ and $\Delta R_{b_1b_2}$, the only exception
being the tail of the $\Delta R_{\bbbar}$ distribution, where
double-splitting effects can reach 50\% of the NLOPS cross section, while
$g\to \bbbar$ shower uncertainties can reach 15\%.

Predictions at LOPS depend also on the choice of the shower starting scale. 
This uncertainty is especially sizeable in the case of the light-jet spectrum,
where {\tt scalup} acts as a cutoff.  A sizeable {\tt scalup}
dependence is visible also in the LOPS predictions for the 
$p_\rT$-distributions of $b$-jets, which indicates that such observables are
rather sensitive to QCD radiation.  
Let us recall that the {\tt scalup} dependence disappears
completely in NLOPS simulations based on the \Powheg approach.

Ratios plotted in the lower frames of \reffis{fig:PS1}{fig:PS2} 
show the dependence of NLOPS predictions
with respect to the choice of the $\hdamp$ and $\hbzd$ parameters, 
which control the separation of the first emission into events of
soft and hard type in the \PowhegBox framework (see \refse{se:shower}).  
The $\hdamp$ band is obtained by varying $\hdamp=H_\rT/4$, $H_\rT/2$, $H_\rT$, $1.5m_t$
with the value of $\hbzd$ fixed to 2, while the $\hbzd$ band is obtained by
varying $\hbzd=2$, $5$, $10$ with fixed $\hdamp=H_\rT/2$.
Observables that are inclusive with respect to light-jet radiation
reveal a remarkably small
dependence, typically of the order of a few percent, on the choice of
$\hdamp$ and $\hbzd$.
Non-negligible but moderate uncertainties are found only in 
the light-jet spectra, which
are enhanced by up to 20\% when $\hbzd$ is
increased from 2 to 10.
Investigating simultaneous variations of $\hdamp$ and
$\hbzd$ (not plotted) we have found that the size of the $\hdamp$ variation band is
fairly stable with respect to the value of $\hbzd$ within the considered range.


\subsection[Comparisons against other $\ttbb$ and $\ttbar$ generators]{Comparisons against other $\boldsymbol{\ttbb}$ and $\boldsymbol{\ttbar}$ generators}
\label{se:gencomp}

In \reffis{fig:tools1}{fig:tools2} we compare $\ttbar+b$-jet predictions
based on \PowhegPythia and \Sherpa.  This comparison is done both for 
(N)LOPS $pp\to \ttbb$ generators in the 4F scheme 
and for corresponding generators of inclusive  $\ttbar$ production in the 5F scheme.
Specifically, in the case of \Powheg we use {\tt hvq}~\cite{Frixione:2007nw}.
As detailed in \refse{se:gencompsettings},
input parameters, QCD scales and matching parameters 
are chosen as coherently as possible across all generators.
In this spirit, the parameter $\hdamp=H_\rT/2$ in \Powheg
is identified with the resummation scale $\mu_Q$ in the SMC@NLO framework of 
\Sherpa.
Instead, for what concerns the parton showers
we simply use standard settings, \ie we do not try to 
improve the agreement between generators by
tuning the \Pythia and \Sherpa showers.

\newcommand{\toolsplot}[1]{
\renewcommand{\arraystretch}{0}
\begin{minipage}{.45\textwidth}
\begin{center}
\includegraphics[width=\textwidth,trim={2 9.9mm 2 0},clip]{tools/#1_top}\\
\includegraphics[width=\textwidth,trim={2 9.9mm 2 7.0mm},clip]{tools/#1_center} \\
\includegraphics[width=\textwidth,trim={2 0 2 7.0mm},clip]{tools/#1_bottom}
\end{center}
\end{minipage}
}

\newcommand{\toolsplotL}[1]{\hspace{0mm}\toolsplot{#1}}
\newcommand{\toolsplotR}[1]{\toolsplot{#1}\hspace{0mm}}

\begin{figure}[t!]
\centering
\subfigure[]{
\hspace{-5mm}
\toolsplot{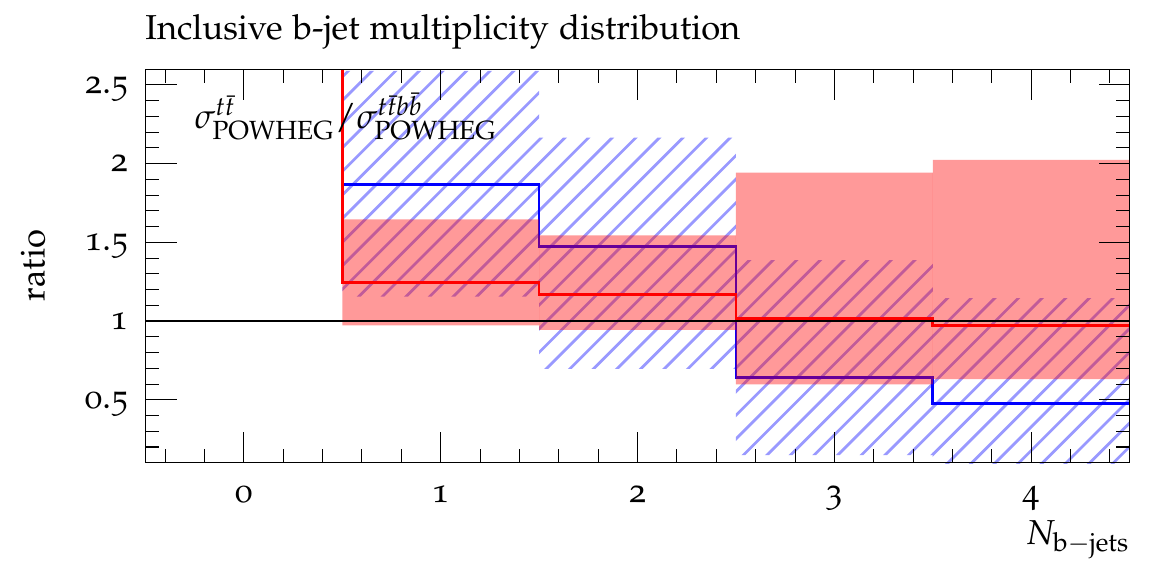}
\label{fig_tools:NBj_XS}}
\subfigure[]{
\toolsplot{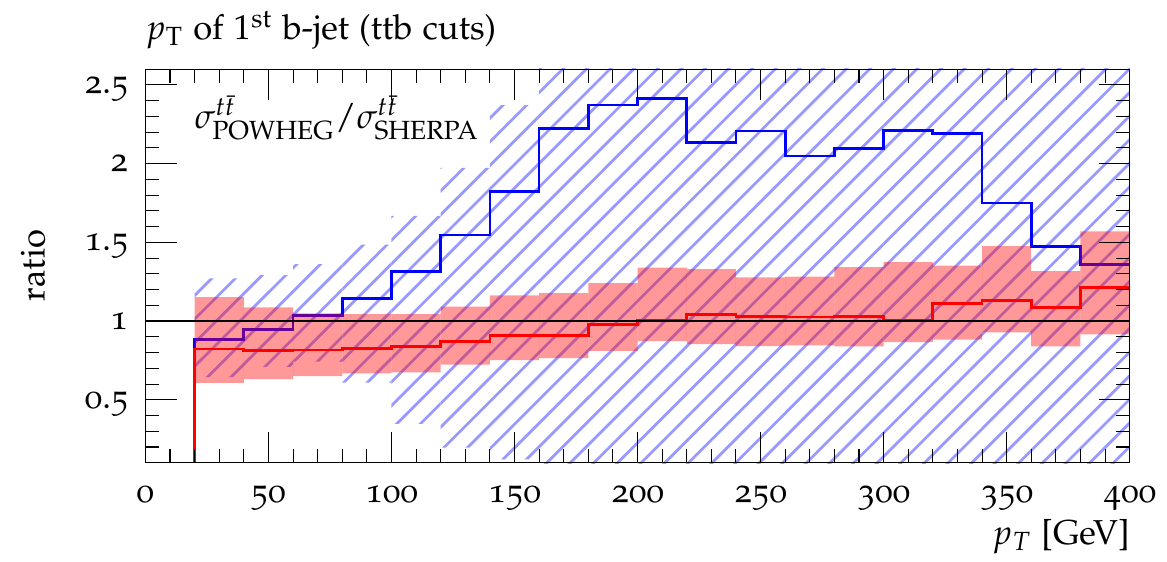}
\label{fig_tools:1_PT_B1}
\hspace{-5mm}}
\subfigure[]{
\hspace{-5mm}
\toolsplot{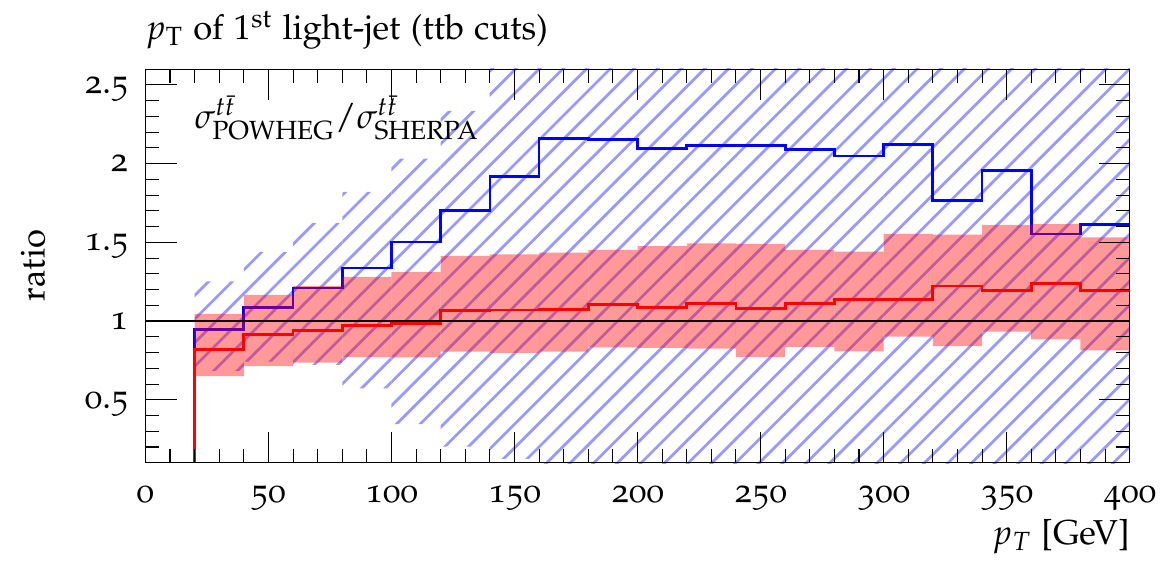}
\label{fig_tools:1_PT_J1}}
\subfigure[]{
\toolsplot{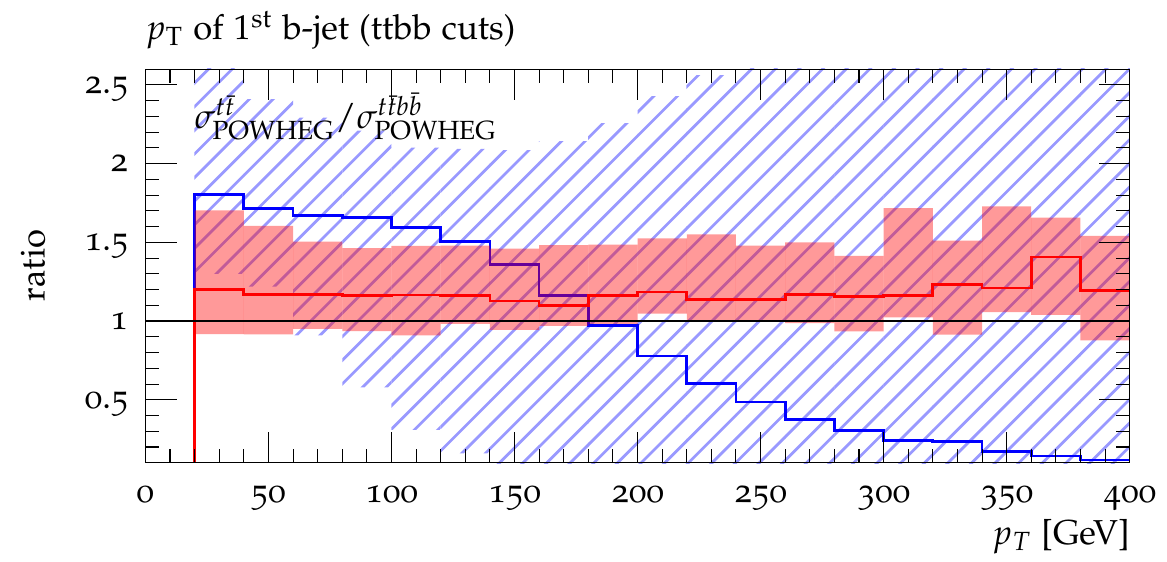}
\label{fig_tools:2_PT_B1}
\hspace{-5mm}}
\caption{Predictions for $pp\to \ttbar+b$-jets at $\sqrt{s}$=13\,TeV: 
distributions in the inclusive number of additional $b$-jets (a), the $\pt$ of the first
$b$-jet (b) and the first light jet (c) with ttb cuts, and in the $\pt$ of
the second $b$-jet with ttbb cuts (d).
The various ratio plots compare $\ttbar+b$-jet observables 
as described in LOPS (blue) and NLOPS (red) simulations based on 
$pp\to \ttbb$ or $pp\to \ttbar$ matrix elements
in \PowhegPythia or \Sherpa.
In the ratios shown in the upper and middle frame 
\Powheg predictions are normalised to \Sherpa ones for the case
of $pp\to \ttbb$ and $pp\to \ttbar$ simulations, respectively.
The third frame displays the ratio of \ttbar
to $\ttbb$ \Powheg predictions.
For all ratios the numerator and
denominator are evaluated at the same order, and
uncertainties are applied only to the 
numerator.  They correspond to the combination in quadrature of 
$\hdamp$ and $\hbzd$ variations with the uncertainties 
due to the  modelling of 
$g\to \bbbar$ splittings and the choice of $\as$ and
{\tt scalup} in \Pythia (see \refses{se:input}{se:shower}).
Top quarks are kept stable throughout.
}
\label{fig:tools1}
\end{figure}

\begin{figure}[t!]
\centering
\subfigure[]{
\hspace{-5mm}
\toolsplot{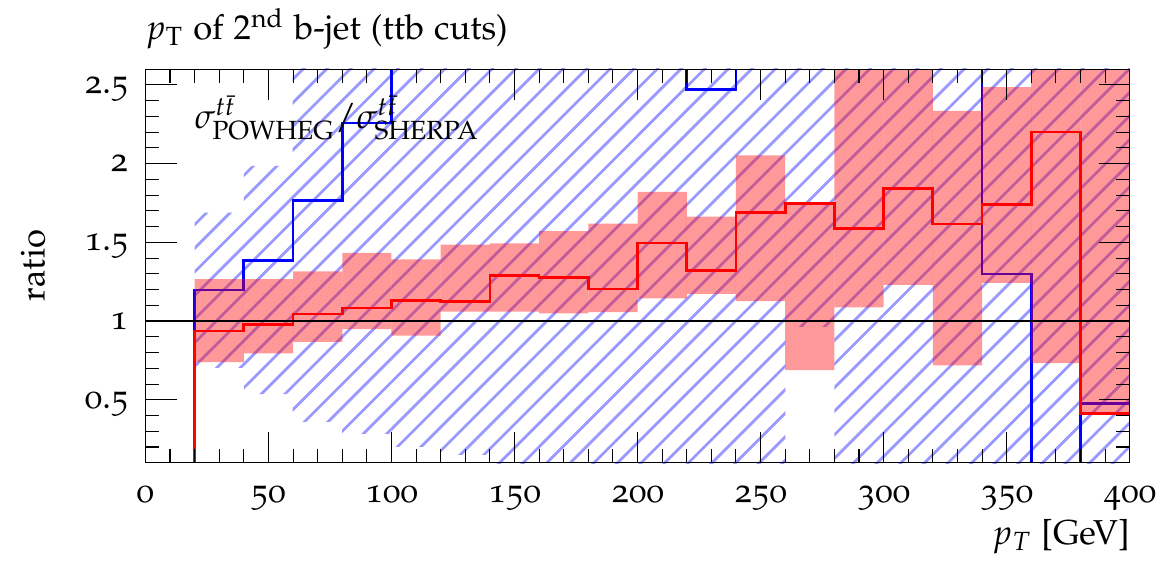}
\label{fig_tools:2_PT_B2}}
\subfigure[]{
\toolsplot{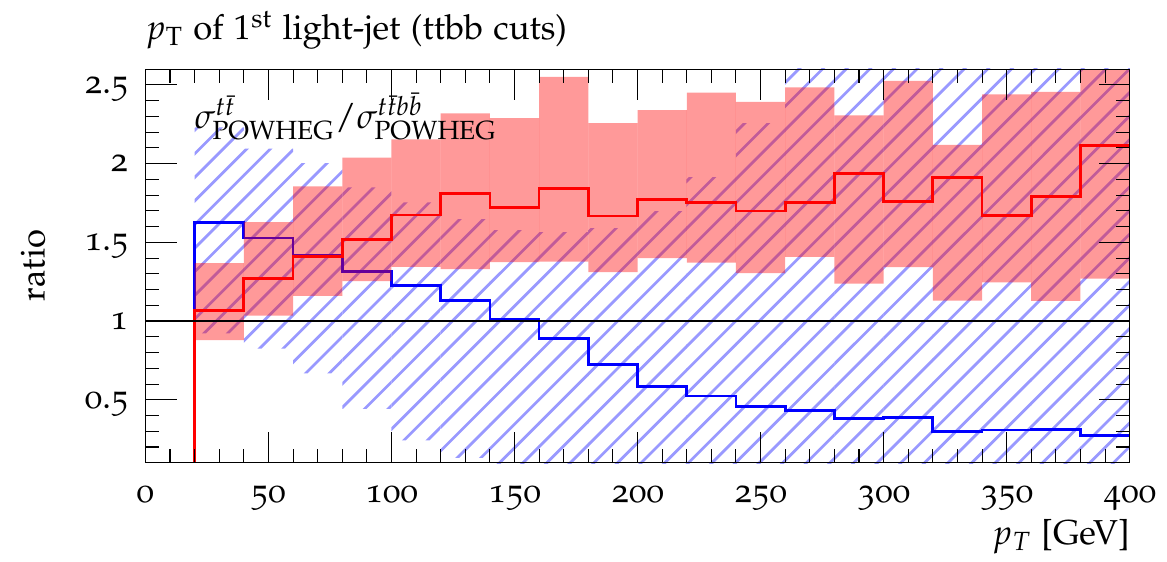}
\label{fig_tools:2_PT_J1}
\hspace{-5mm}}
\subfigure[]{
\hspace{-5mm}
\toolsplot{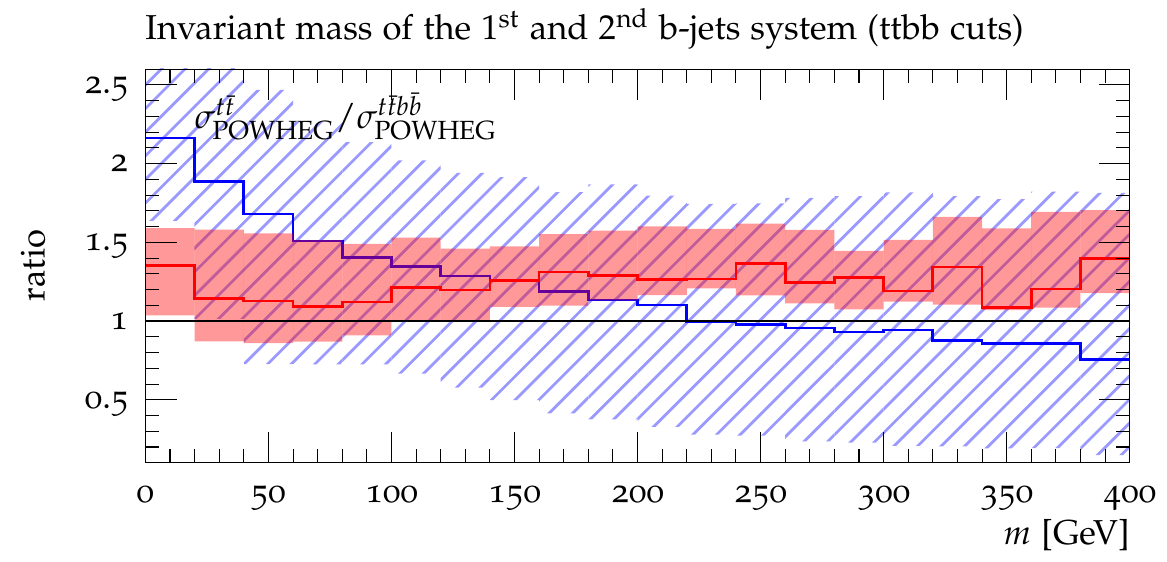}
\label{fig_tools:2_M_B1B2}}
\subfigure[]{
\toolsplot{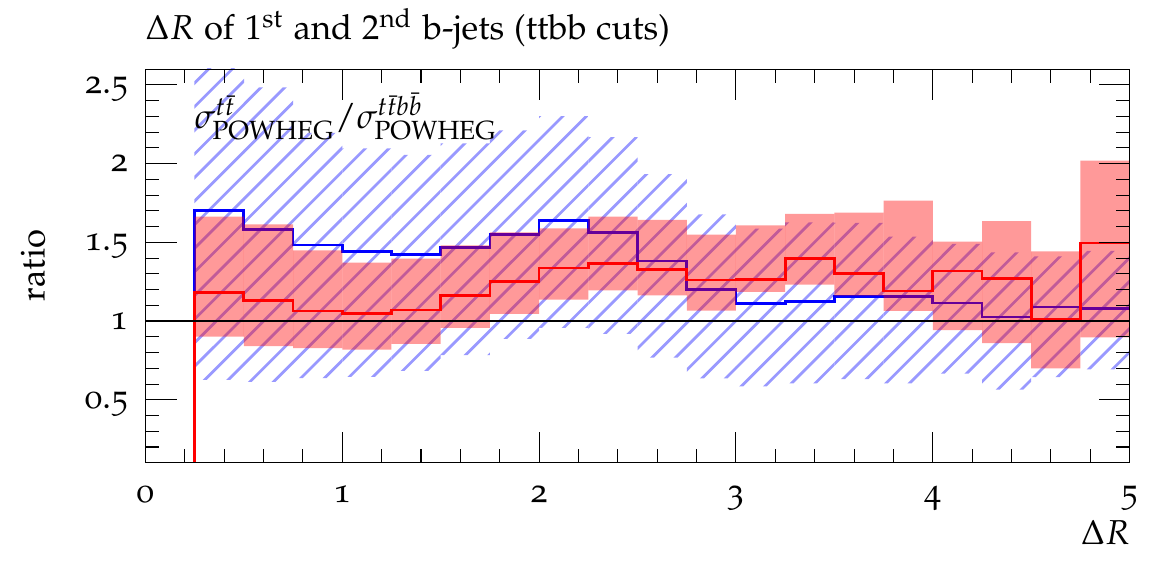}
\label{fig_tools:2_DR_B1B2}
\hspace{-5mm}}
\caption{Distributions in the $\pt$ of the second $b$-jet (a) in the $p_\rT$
of the first light jet (b), and in the invariant mass (c) and the $\Delta R$
separation (d) of the first two $b$-jets with ttbb cuts throughout.
Predictions and uncertainties as in \reffi{fig:tools1}.
}
\label{fig:tools2}
\end{figure}

The ratios in the upper frames of \reffis{fig:tools1}{fig:tools2} show
\Powheg $pp\to \ttbb$  predictions normalised to corresponding \Sherpa
predictions at LOPS and NLOPS accuracy.
The bands describe the combination in quadrature of all 
matching and shower uncertainties\footnote{Note that 
QCD scale uncertainties are not shown here.} in
\PowhegPythia (referred to shower uncertainties in the following), 
while only nominal \Sherpa predictions are considered in the ratios.
Comparing LOPS predictions gives direct insights into the different modelling of radiation 
in \Pythia and \Sherpa.
For observables that are inclusive with respect to jet radiation
we find deviations between 10--40\%
and comparably large shower
uncertainties.
In contrast, in the 
jet-$p_\rT$ distributions the LOPS predictions of \Pythia 
are far above the ones by \Sherpa,
with differences that can reach a factor 2.5
in the tails.
These differences are perfectly consistent with LOPS shower
uncertainties, which are dominated by variations of the \Pythia starting
scale.

Moving to NLOPS reduces the direct dependence on the parton shower. At
the same time, differences between the \Powheg and SMC@NLO matching
methods  come into play.
In practice, at NLOPS we observe a drastic reduction
of shower uncertainties, especially in the light-jet and
$b$-jet $p_\rT$-distributions.
Also the differences between \Powheg and \Sherpa 
become very small at NLOPS. The  
ttb and ttbb cross sections agree at the percent level,
and differential $b$-jet observables deviate by more than 
5\% only in the tails of the $m_{b_1b_2}$ and 
$\Delta R_{b_1b_2}$ distributions. 
Even the light-jet spectra in the ttb and ttbb phase space deviate by less
than 10--20\% up to high $p_\rT$, in spite of the limited formal accuracy (LOPS)
of such observables.
In the light of these results,
NLOPS theoretical uncertainties related to the matching scheme and the parton
shower seem to be well under control in $pp\to \ttbb$.  
In particular, their impact appears to be
clearly subleading as compared to QCD scale uncertainties.

In the central frames of \reffis{fig:tools1}{fig:tools2} 
we compare (N)LOPS generators of 
inclusive $\ttbar$ production based on 
\PowhegPythia and \Sherpa.
In this case, the $g\to \bbbar$ final-state splittings that 
give rise to $\ttbar+b$-jet signatures are 
entirely controlled by the parton shower.
At LOPS, also the parent gluon that splits into $\bbbar$ is generated by the
parton shower. Nevertheless, the ttb and ttbb
LOPS cross sections predicted by \Powheg and \Sherpa 
deviate by less than 30\%--40\%. 
Instead, as expected, the shapes of $\ttbar+b$-jet observables vary very strongly, and
in all considered light-jet and $b$-jet distributions 
\Pythia results exceed \Sherpa ones by a factor of two 
and even more. This excess is well consistent with the estimated LOPS shower uncertainties.
At NLOPS, only $g\to \bbbar$ splittings are controlled by the parton shower, 
while the emission of their parent gluon is dictated by LO matrix elements.
Consequently,  we observe 
a drastic reduction of shower uncertainties as compared to LOPS.
Also the differences between \Powheg and \Sherpa
are largely reduced at NLO, nevertheless they remain quite significant in 
various distributions.

\clearpage

To provide a more complete picture of the uncertainties of inclusive $\ttbar$ simulations,
in the lower frames of \reffis{fig:tools1}{fig:tools2} we
compare \PowhegPythia generators of
inclusive $\ttbar$ production
and $\ttbb$ production.
Shower uncertainties are shown only for the $\ttbar$ generator.
At LOPS, the $\ttbar$ generator 
is strongly sensitive to the modelling of
$pp\to \ttbar g$ through initial-state gluon radiation in \Pythia. 
As a result, the $\ttbar$ generator overestimates 
the ttb and ttbb cross sections by about 90\% and 50\%, respectively.
This excess is strongly sensitive to {\tt scalup}, and in the $p_\rT$-distributions
it is confined to the regions below 100--200\,GeV, while the tails 
are strongly suppressed.
Also  the
$m_{b_1b_2}$ and $\Delta R_{b_1b_2}$ distributions feature 
strong shape differences as compared to LOPS $\ttbb$ predictions.

Such differences go down significantly at NLOPS.
The ttb and ttbb cross sections predicted by the
$\ttbar$ generator overshoot $\ttbb$ results
by only 15--20\%, and also $b$-jet observables 
feature an improved agreement with $\ttbb$ predictions.
Nevertheless, in $b$-jet observables we find 
quite significant shape differences,
especially for the $m_{b_1b_2}$ and $\Delta R_{b_1b_2}$ distributions,
and shower uncertainties remain far above the ones 
of the  $\ttbb$ generator (see upper frame).
As for the light-jet spectra, 
$\ttbar$ predictions turn out to lie above $\ttbb$ ones by about a factor of two 
in the tails.
In principle, with the help of parton shower tuning
NLOPS $\ttbar$ generators may be amenable to a
reasonable description of inclusive $\ttbar+b$-jet observables.
However, in the light of the above results it should be clear that 
NLOPS $\ttbb$ generators are mandatory 
in order to achieve an acceptable level of shower systematics.


\section{$\boldsymbol{\ttbb}$ production with top-quark decays}
\label{se:decayedtops}

In this section we present NLOPS results of the \PowhegPythia
$\ttbb$ generator with leptonic top-quark decays. More precisely 
we consider final states with oppositely charged leptons and/or muons.
By default hadronisation and MPI are deactivated in \Pythia, and 
their effect is shown separately.
As detailed in \refse{se:topdecaytreatment}, our implementation of top
decays is based on resonant $pp\to \ttbb\to 2\ell2\nu \bbbar$(+j) matrix
elements, where spin correlations are consistently taken into account.

Top-quark decays are due to weak interactions and, up to small corrections
of $\mathcal{O}(\Gamma_t/m_t)$, their effect factorises with respect to 
$\ttbb$ production. Thus, while they strongly increase the complexity of
$\ttbar+b$-jet events, top decays are not expected to interfere
with the QCD dynamics of  $pp\to \ttbb$ in a significant way.
In order to verify this hypothesis, in \reffi{fig:decayed1} we compare NLOPS
$pp\to
\ttbb$ simulations with stable and decayed top quarks.
To this end, based on Monte Carlo truth, all reconstructed jets are split 
into two subsets associated with $\ttbb$ production and top decays. Specifically, 
jets that contain a parton originating from showered top-decay products are 
attributed to top decays, otherwise  to $\ttbb$ production.\footnote{Since 
top quarks carry colour charge, a separation of production and decay is
only possible at parton level, while at hadron level QCD radiation 
from production and decay is merged via colour reconnection.}
At the level of top decays we require two $b$-jets and two charged leptons within 
the acceptance cuts \refeq{eq:jetcuts}--\refeq{eq:leptoncuts}, while
for the ``reconstructed'' $\ttbb$ system we consider the same cuts and 
observables as for the case of stable top quarks.
In order to mimic the leptonic branching ratio and the
efficiency of acceptance cuts on top-decay products,
the normalisation of the $\ttbb$ simulation
with stable top quarks is adapted to the predictions with decayed top quarks.
This is done at the level of the ttbb cross section through a constant
normalisation factor.

\newcommand{\decayedplot}[1]{
\renewcommand{\arraystretch}{0}
\begin{minipage}{.45\textwidth}
\begin{center}
\includegraphics[width=\textwidth]{decayed/#1_top}
\end{center}
\end{minipage}
}

\begin{figure}[t!]
\centering
\subfigure[]{
\hspace{-5mm}
\decayedplot{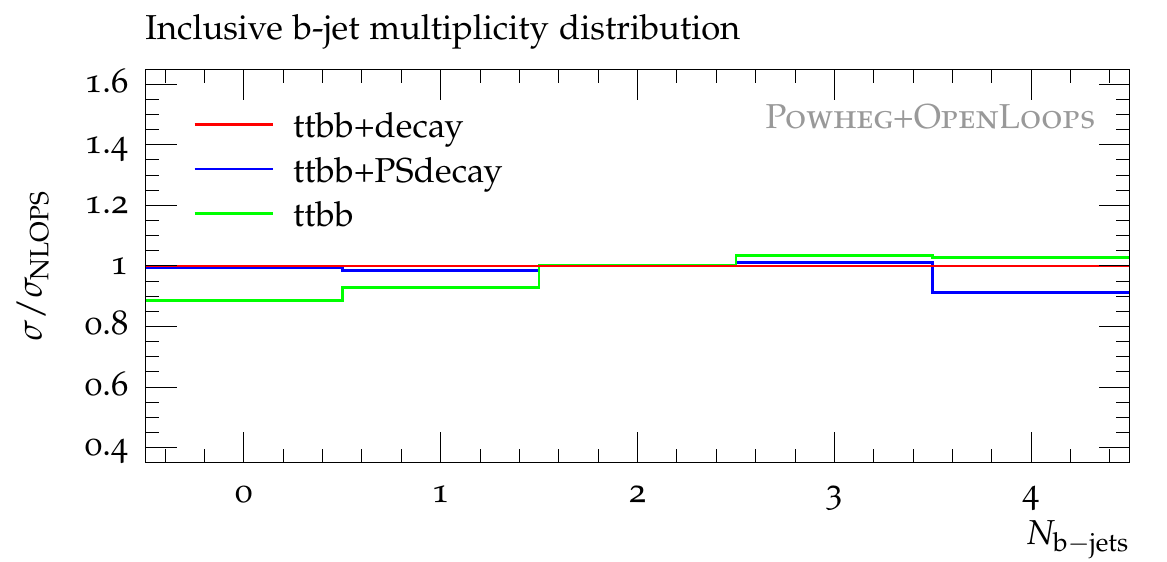}
\label{fig_decayed:NBj_XS}}
\subfigure[]{
\decayedplot{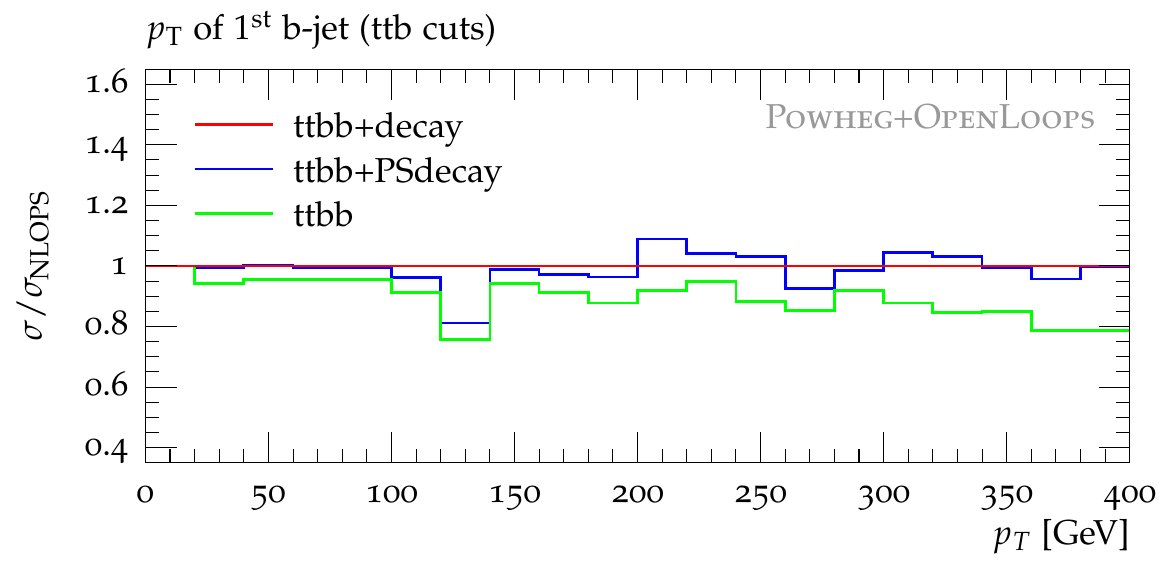}
\label{fig_decayed:1_PT_B1}
\hspace{-5mm}}
\subfigure[]{
\hspace{-5mm}
\decayedplot{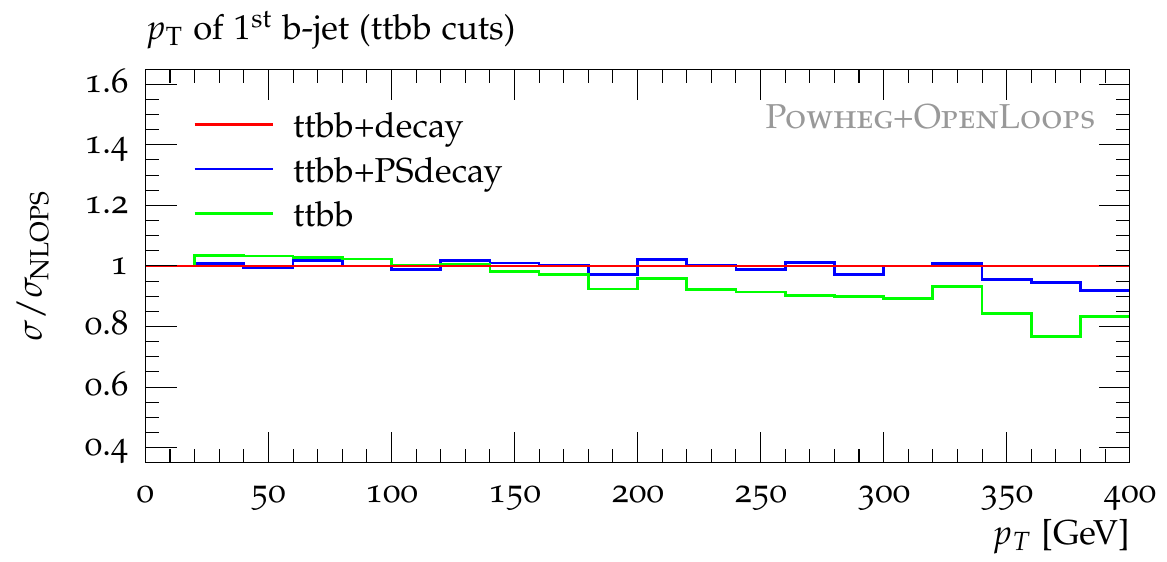}
\label{fig_decayed:2_PT_B1}}
\subfigure[]{
\decayedplot{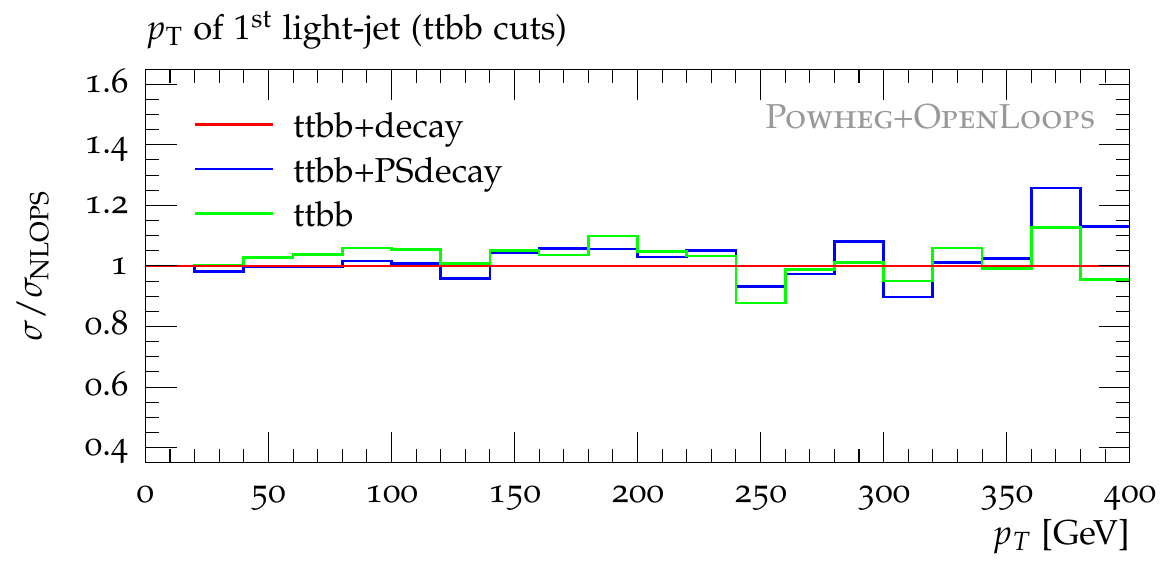}
\label{fig_decayed:2_PT_J1}
\hspace{-5mm}}
\subfigure[]{
\hspace{-5mm}
\decayedplot{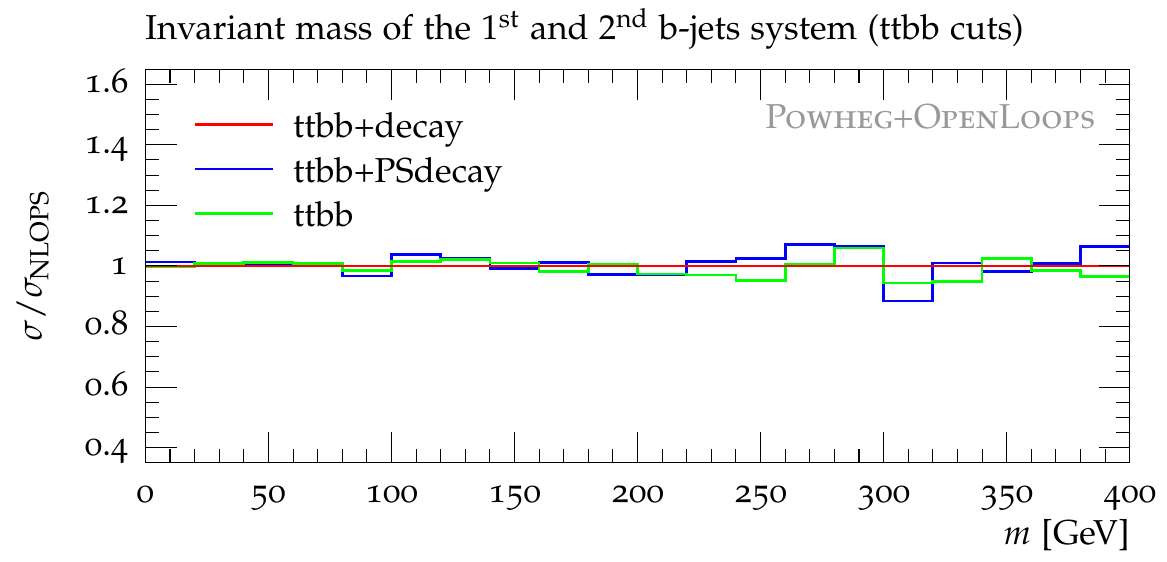}
\label{fig_decayed:2_M_B1B2}}
\subfigure[]{
\decayedplot{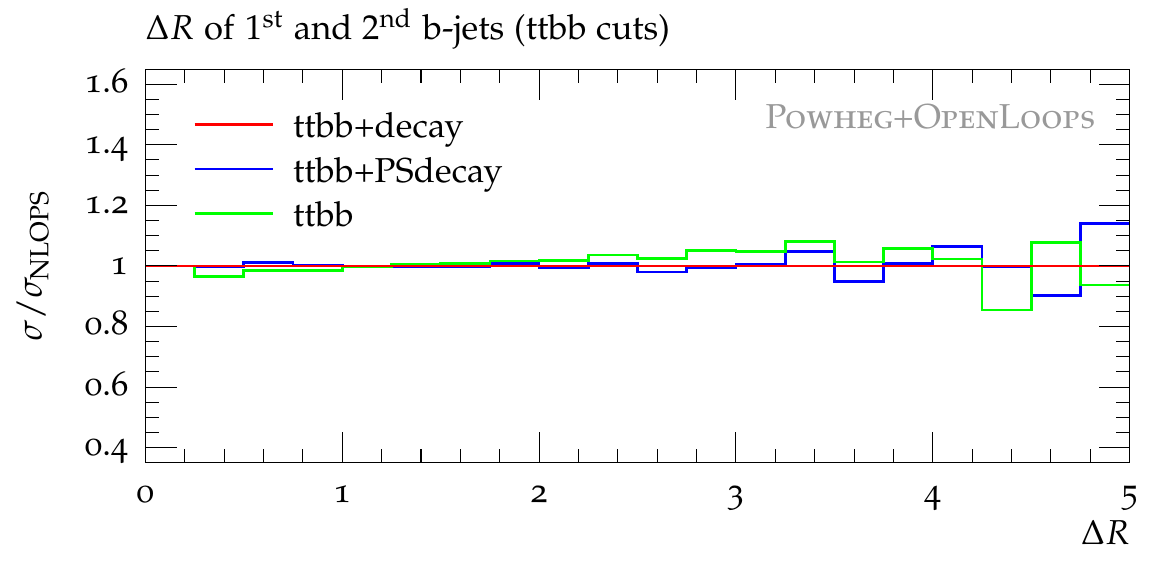}
\label{fig_decayed:2_DR_B1B2}
\hspace{-5mm}}
\caption{
\label{fig:decayed1}
Distributions in the $b$-jets of the reconstructed $\ttbb$ system
for $pp\to \ttbar+b$-jets with dileptonic top decays at $\sqrt{s}$=13\,TeV.
Inclusive number of additional $b$-jets (a), distribution with ttb cuts in the 
$\pt$ of the first
$b$-jet (b) and distributions with ttbb cuts in the 
$\pt$ of the first $b$-jet (c) and light-jet (d)
as well as in the invariant mass (e) and $\Delta R$ (f) of the first and second $b$-jet.
All results are based on \PowhegPythia with
hadronisation and MPI switched off. The ratio corresponds to 
NLOPS predictions with \Pythia decays (ttbb+PSdecay) or stable top quarks (ttbb)
normalised to corresponding ones with spin-correlated decays (ttbb+decay). 
Top-decay products are subject to acceptance cuts, while
predictions with stable top quarks are normalised to ttbb+decay ones at the level of the
ttbb cross section.
}

\end{figure}

\newcommand{\agnosticplot}[1]{
\begin{minipage}{.45\textwidth}
\begin{center}
\includegraphics[width=\textwidth,trim={2 0 2 0},clip]{agnostic/#1}\\
\end{center}
\end{minipage}
}

\newcommand{\agnosticplotL}[1]{\hspace{-0mm}\agnosticplot{#1}}
\newcommand{\agnosticplotR}[1]{\agnosticplot{#1}\hspace{-0mm}}

\begin{figure}[t!]
\centering
\subfigure[]{
\agnosticplotL{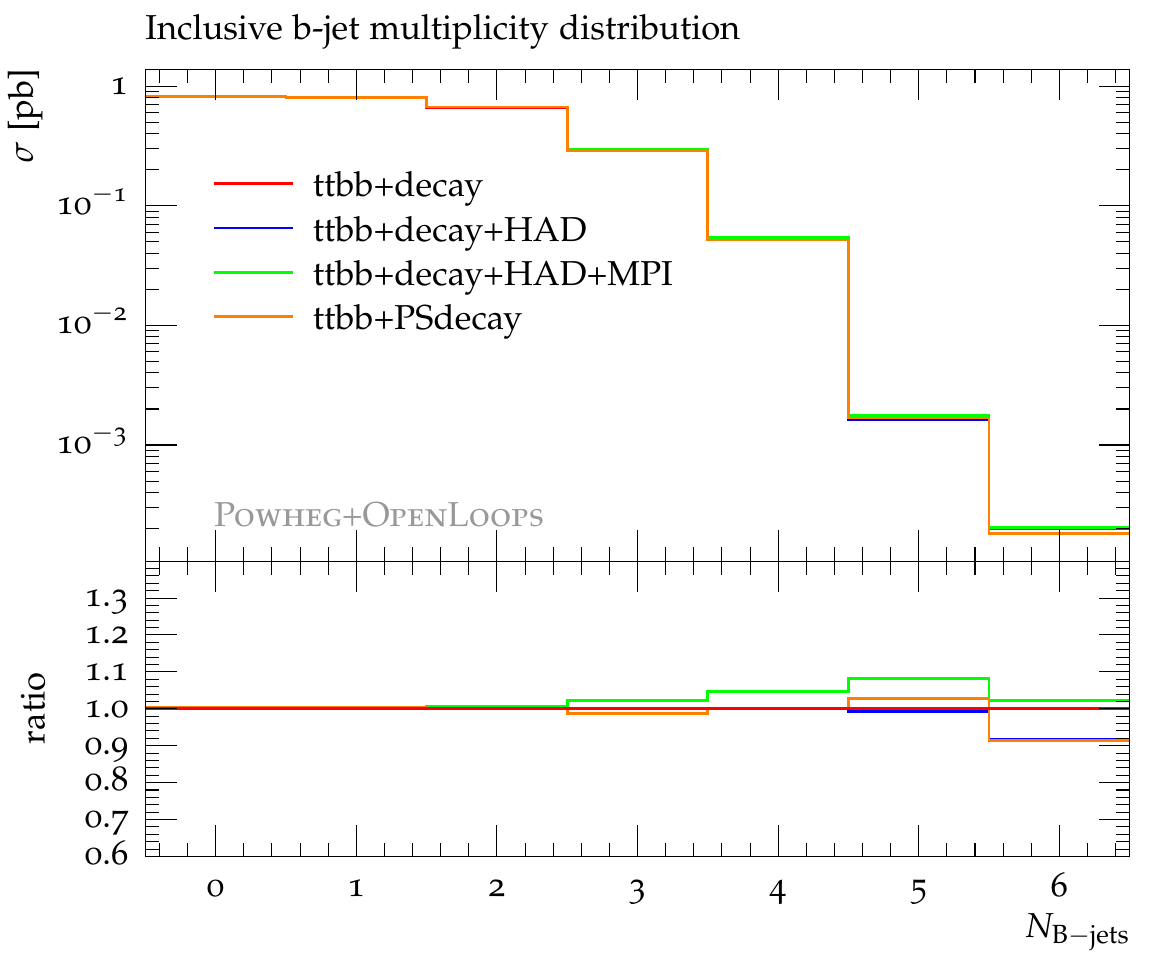}
\label{fig_agnostic:4dec_NBj_XS}}
\subfigure[]{
\agnosticplotR{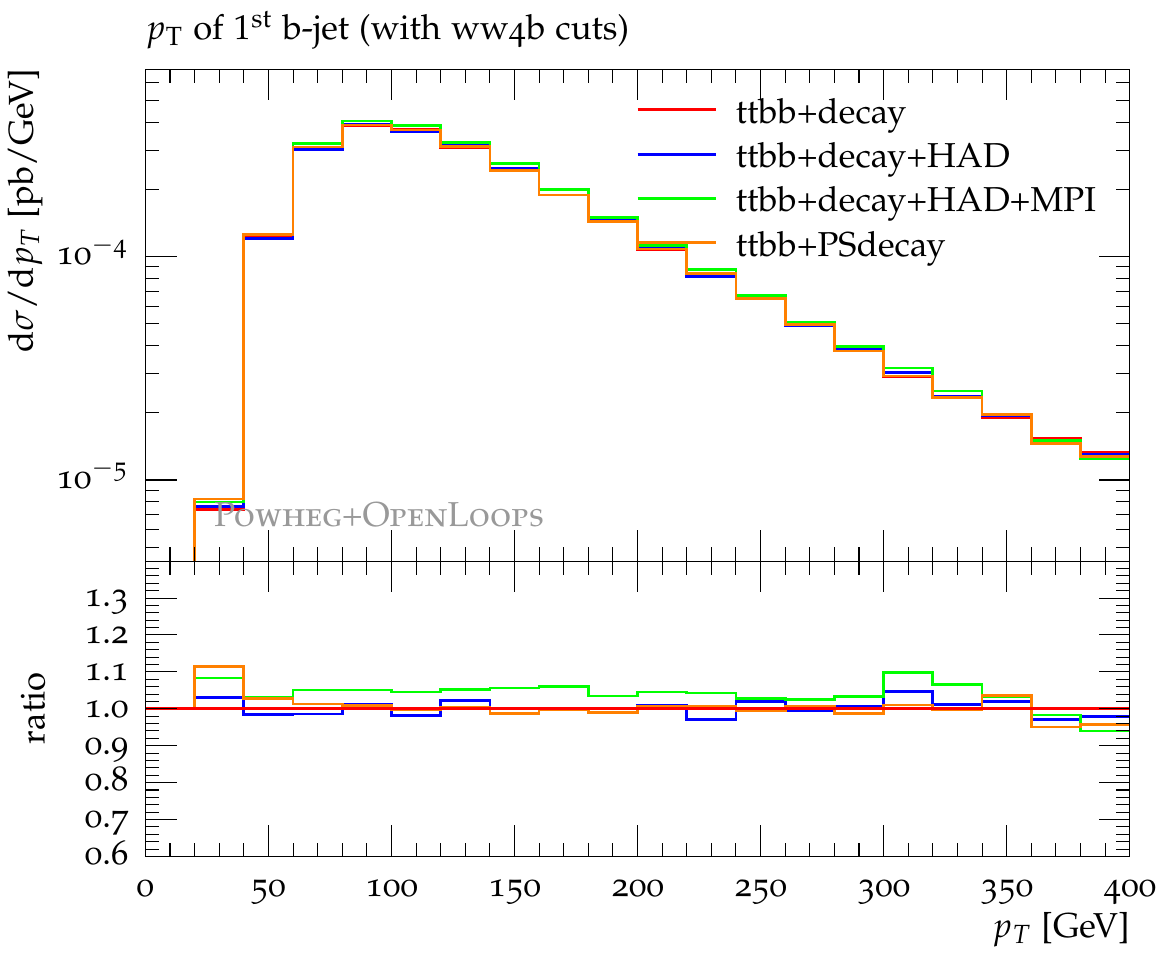}
\label{fig_agnostic:4dec_PT_B1}}

\subfigure[]{
\agnosticplotL{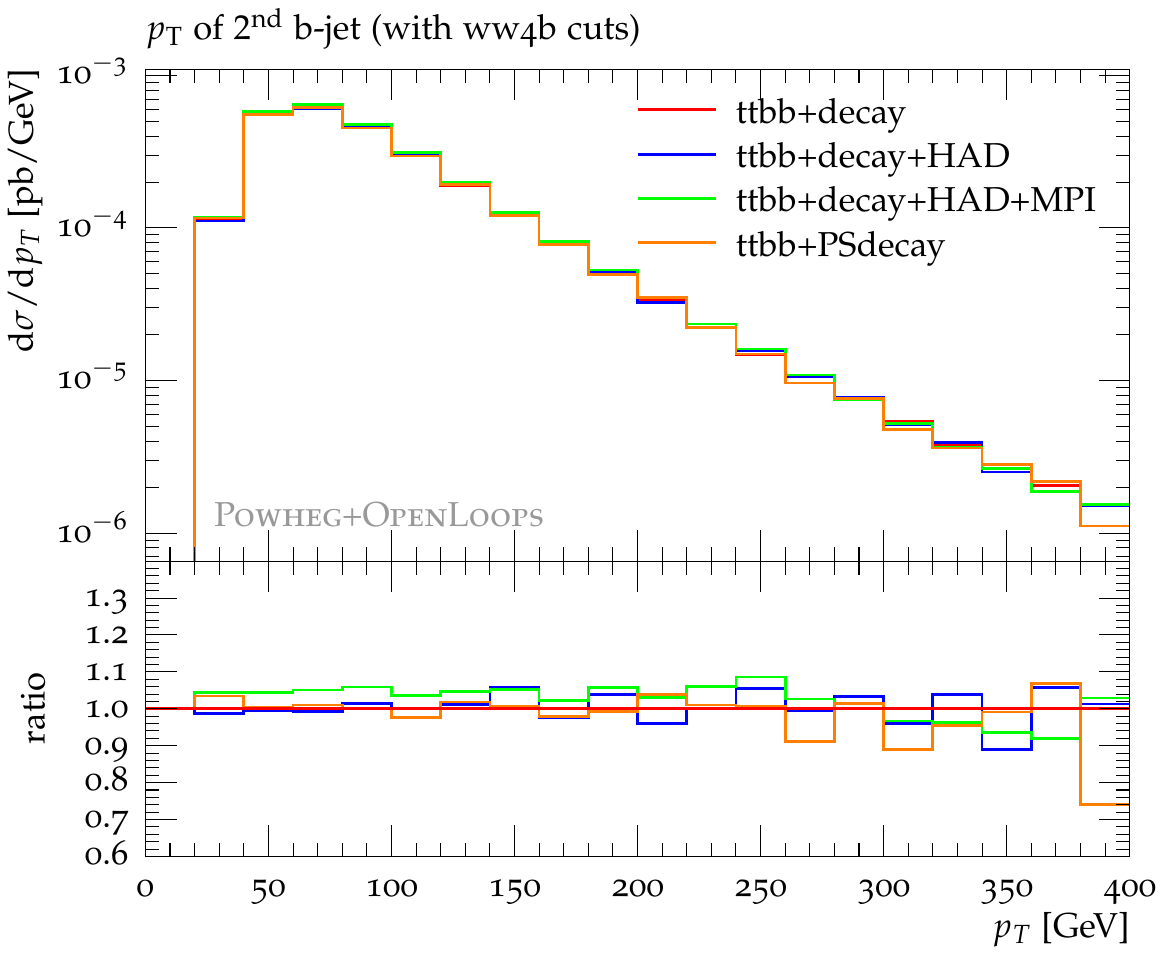}
\label{fig_agnostic:4dec_PT_B2}}
\subfigure[]{
\agnosticplotR{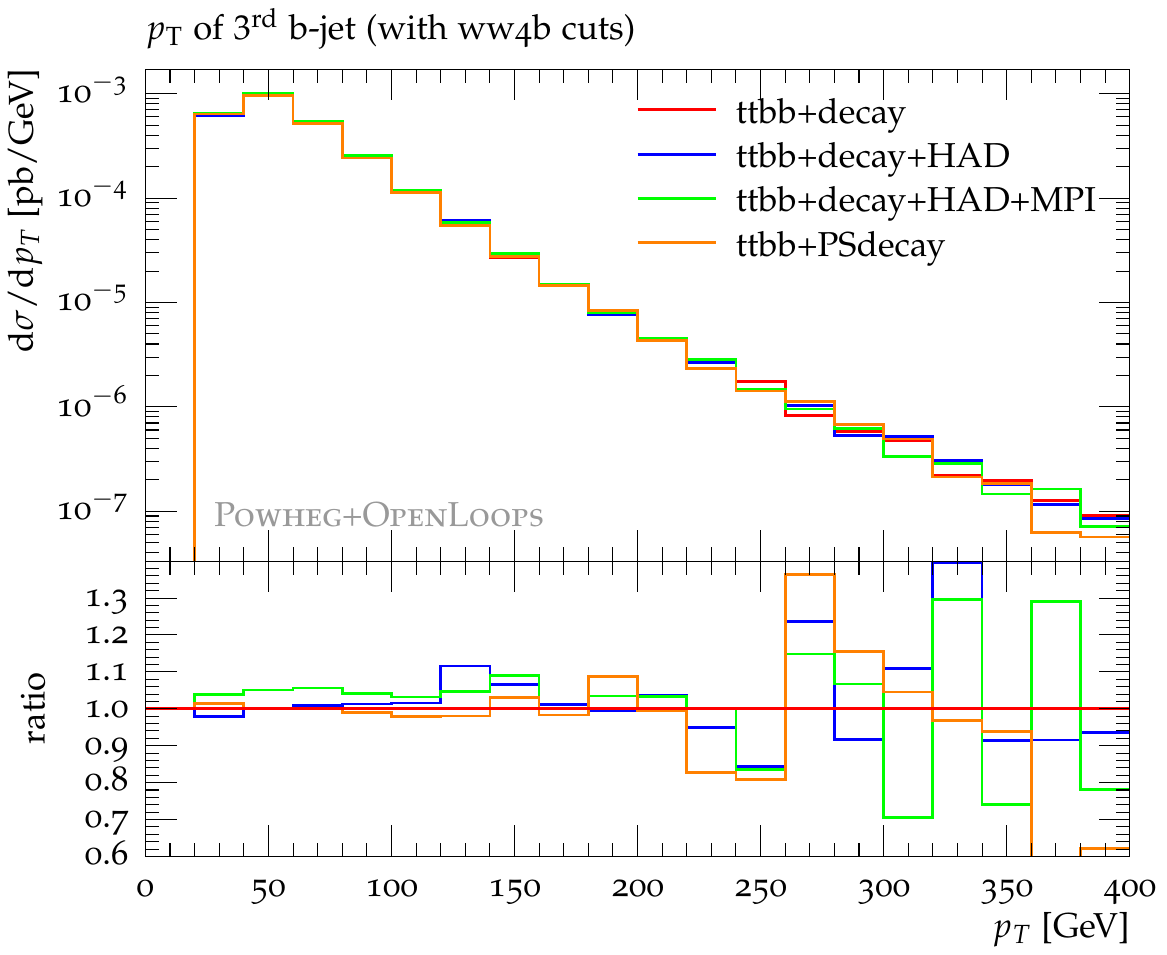}
\label{fig_agnostic:4dec_PT_B3}}
\caption{
Predictions for $pp\to \ttbar+b$-jets at $\sqrt{s}$=13\,TeV after leptonic 
top-quark decays. Four $b$-jets and two leptons within acceptance are required
without any distinction between $b$-jets from $\ttbb$ production and decay.
Distributions in the inclusive number of $b$-jets (a) and in the 
$\pT$ of the first (b), second (c) and third (d) $b$-jet.
All results are based on \PowhegPythia, and in the
lower frame nominal NLOPS predictions with spin-correlated decays without (ttbb+decay) 
and with hadronisation (ttbb+decay+HAD) 
and multi-parton interactions (ttbb+decay+HAD+MPI)
are compared to corresponding ones 
with \Pythia decays (ttbb+PSdecay).
}
\label{fig:agnostic1}
\end{figure}

\begin{figure}[t!]
\centering
\subfigure[]{
\agnosticplotL{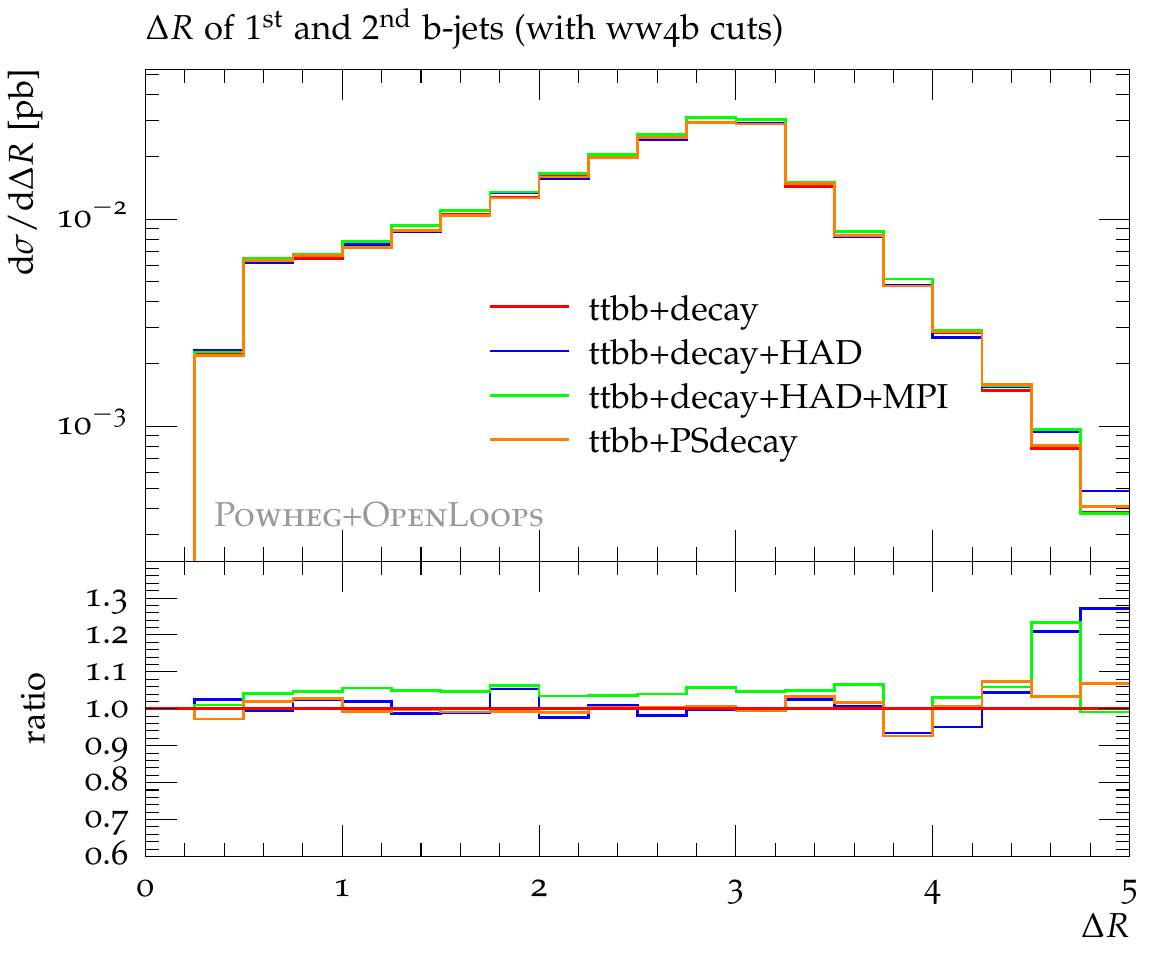}
\label{fig_agnostic:4dec_DR_B1B}}
\subfigure[]{
\agnosticplotR{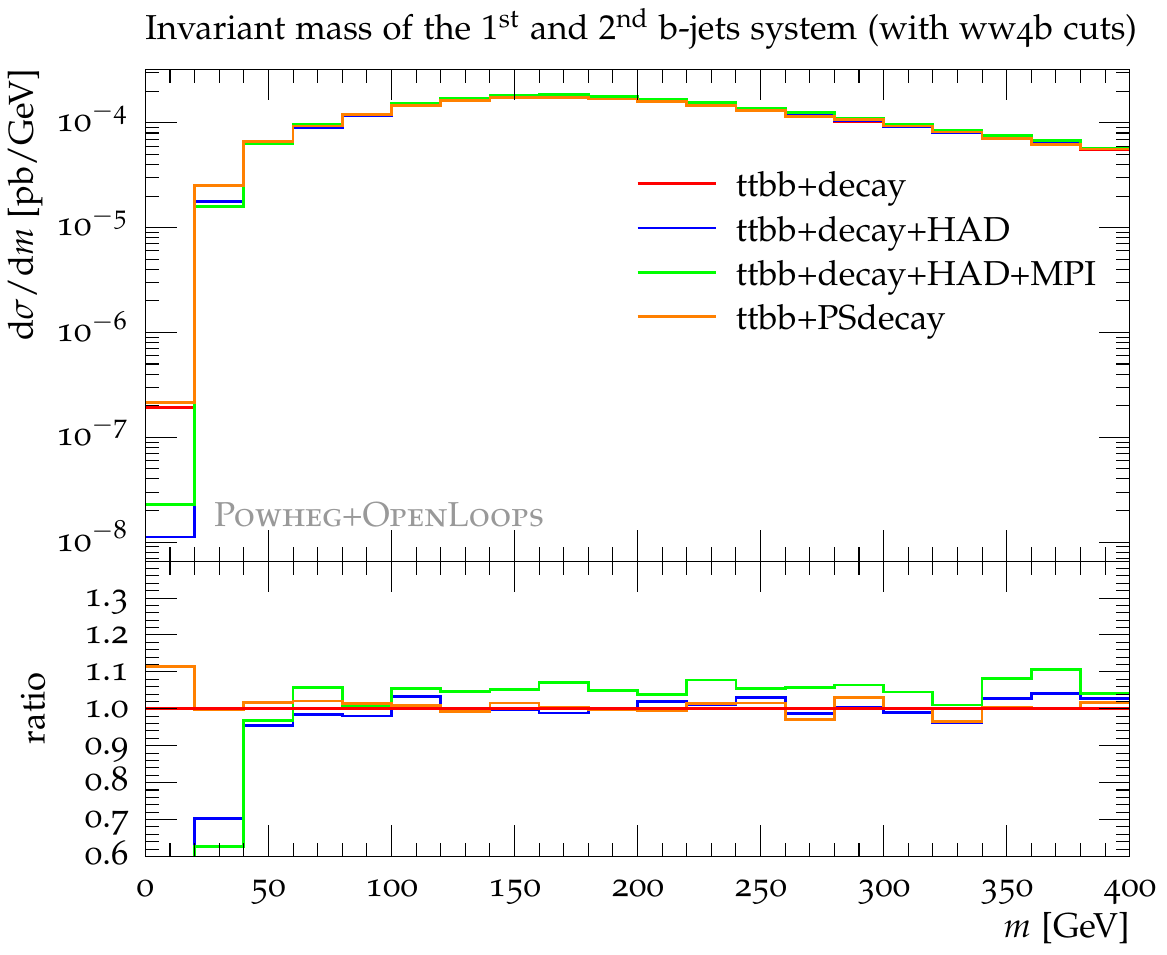}
\label{fig_agnostic:4dec_M_B1B2}}
\subfigure[]{
\agnosticplotL{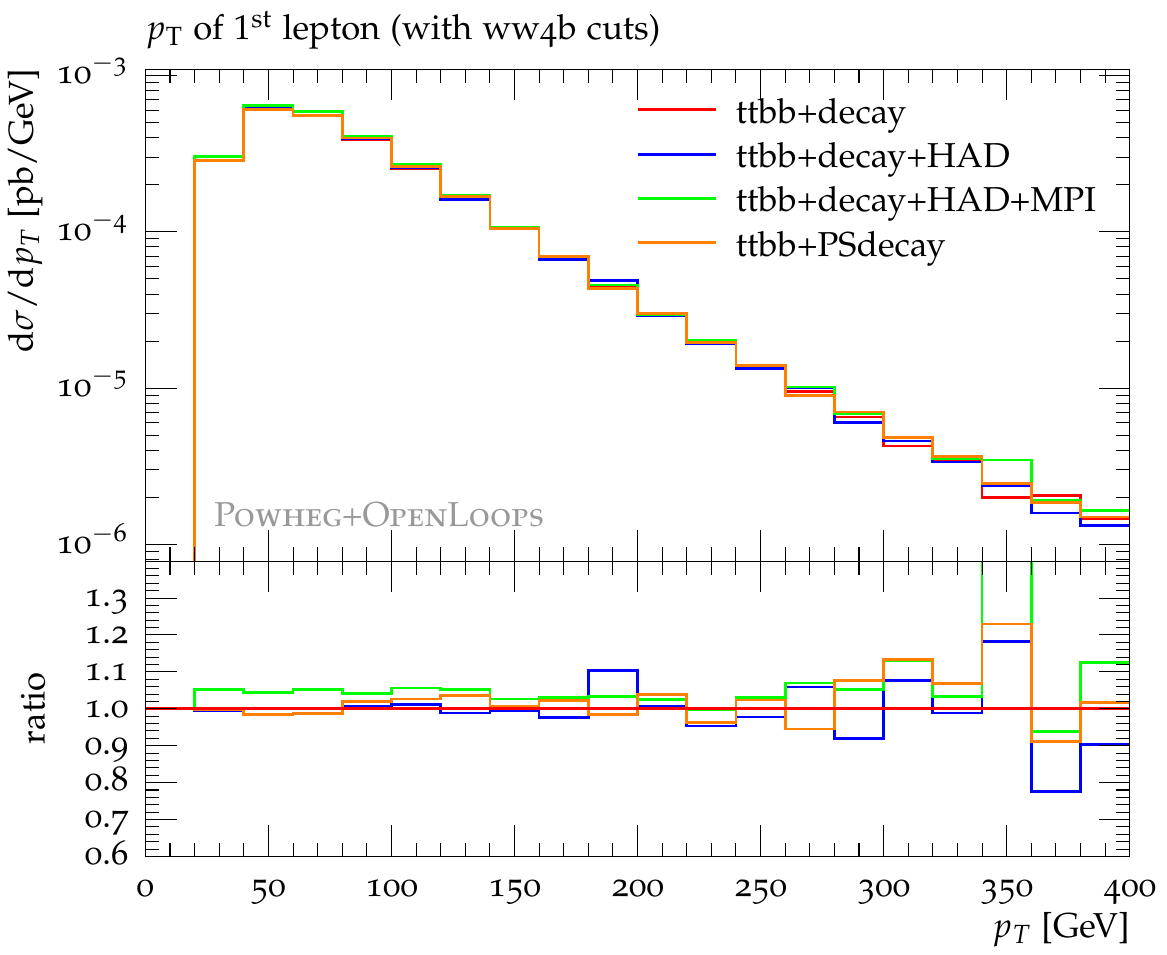}
\label{fig_agnostic:4dec_PT_l1}}
\subfigure[]{
\agnosticplotR{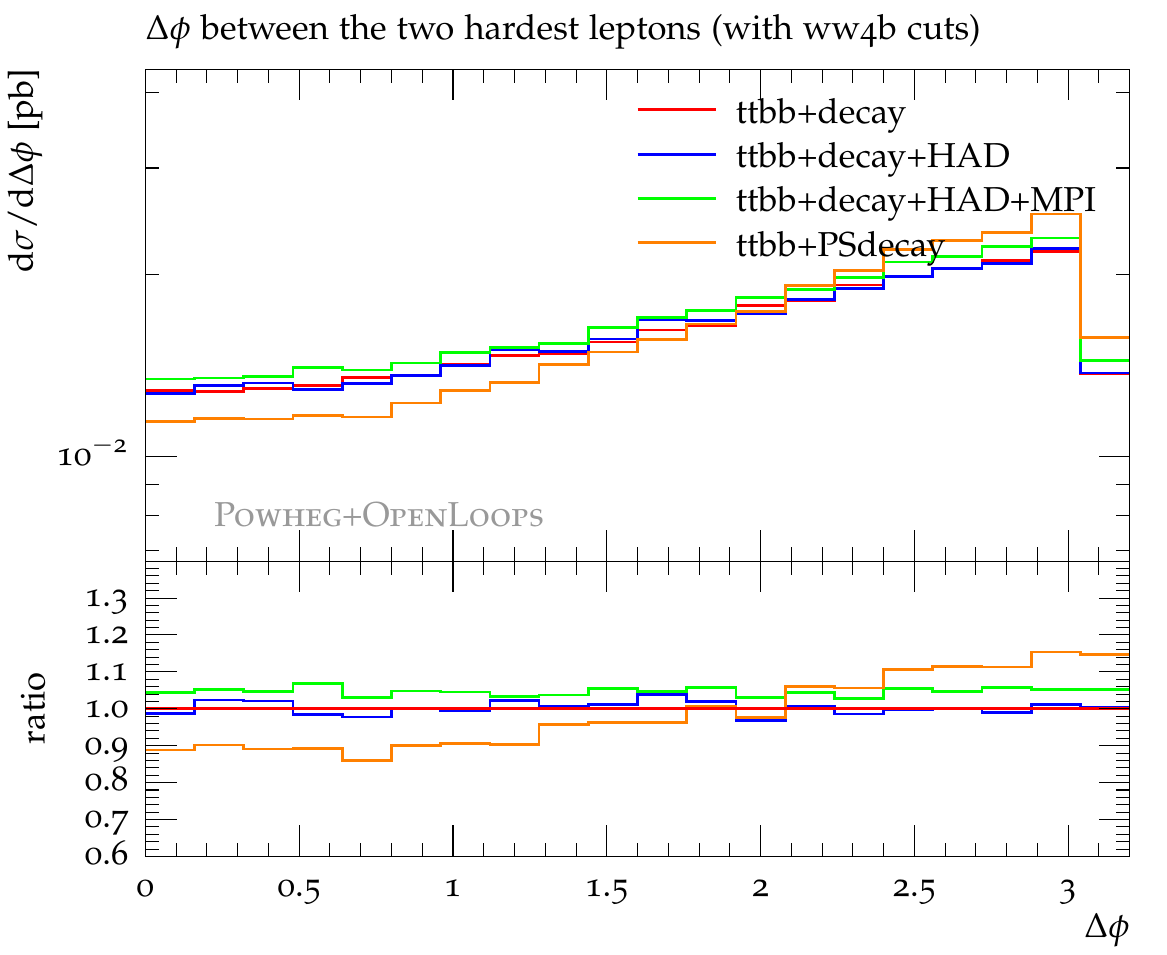}
\label{fig_agnostic:4dec_DPHI_ll}}
\caption{
Distributions in the invariant mass (a) and 
$\Delta R$ (b) of the first and second $b$-jet, in the 
$\pT$ of the leading lepton (c), and in the azimuthal $\Delta \phi$
separation of the two charged leptons (d).
Predictions and ratios as in \reffi{fig:agnostic1}.
}
\label{fig:agnostic2}
\end{figure}

As shown in \reffi{fig:decayed1}, $\ttbar+b$-jet
observables with stable top quarks and reconstructed top decays turn out to
agree quite well:
$b$-jet cross sections and distributions deviate by only
5--10\%, and also in the light-jet $\pt$-distribution 
decay effects hardly exceed 10\%.
These differences can be understood as indirect effect of the acceptance
cuts on top-decay products, which result from the correlation between the kinematics
of the $\ttbar$ system and the additional jets.

Keeping in mind that realistic $b$-jet observables consist of a
combinatorial superposition of $b$-jets from $\ttbb$ production and from top
decays, the fact that  Monte Carlo truth acceptance cuts on top decays have
only a minor effect on the production of $b$-jets suggests that the
essential features observed in $pp\to \ttbb$ production, such as
double-splitting effects, are expected to show up also in the presence of
top decays.

In order to assess the importance of spin correlations, in
\reffi{fig:decayed1} we also compare spin-correlated top decays to isotropic
decays generated by \Pythia.  At the level of reconstructed $\ttbb$
observables this comparison does not reveal any significant effect of spin
correlations.

A more realistic analysis of $\ttbb$ production and decay is presented in
\reffis{fig:agnostic1}{fig:agnostic2}, where $b$-jet and 
leptonic observables are defined at the level of the full final state,
and two charged leptons and
four $b$-jets within the acceptance cuts \refeq{eq:jetcuts}--\refeq{eq:leptoncuts}
are required, without any distinction between $\ttbb$ production and decay.

Comparing spin-correlated and isotropic top decays, in $b$-jet
observables we find no significant deviation, and significant
spin-correlation effects show up only in the azimuthal correlation of the
two charged leptons.

In \reffis{fig:agnostic1}{fig:agnostic2} we also assess the relative impact
of hadronisation and multi-parton interactions (MPI).  It turns out that $b$-jet observables are very
stable with respect to hadronisation, with differences between parton and
hadron level that do not exceed the few percent level. The same holds for MPI 
effects.

The above results indicate that insights on the QCD dynamics of 
$\ttbb$ production gained through studies with stable top quarks at parton level
should hold true also in the presence of top decays and hadronisation.


\section{Summary and conclusions}
\label{se:conclusions}

Searches for $\ttbar H$ production in the $H\to \bbbar$ channel call for a
precise theoretical description of the irreducible $\ttbar+b$-jet
background.
To shed light on the QCD dynamics that governs this nontrivial multi-scale
process, in the first part of this paper we have analysed the relative
importance of the various mechanisms that lead to the radiation of
$b$-quarks off $pp\to \ttbar$ events.
To this end we have compared the role of $pp\to \ttbb$ 
topologies involving
initial-state and final-state $g\to \bbbar$ splittings.
Using a naive diagrammatic splitting, as well as gauge-invariant collinear
approximations, we have demonstrated that the $\ttbar+b$-jet cross section is
strongly dominated by $b$-jet production via final-state $g\to \bbbar$
splittings.  This holds both for phase space regions with two or only one resolved $b$-jets.
These findings support the usage of NLOPS generators based on $pp\to \ttbb$
matrix elements in the four-flavour scheme, while we have pointed out that
$\ttbar+b$-jet predictions based on $\ttbar+$multi-jet merging rely 
very strongly on the parton-shower modelling of $g\to \bbbar$ splittings.

Motivated by these observations we have introduced a new $pp\to \ttbb$
\Powheg generator in the 4F scheme.  This tool is based on the
\PowhegBoxRes framework, and all relevant matrix elements are computed with
\OpenLoops.
When applied to a multi-scale process like $pp\to\ttbb$, the 
\Powheg method can lead to subtle technical issues.  
In particular, we have pointed out that
the FKS mappings which generate the recoil associated with the first \Powheg
emission can enhance the amplitude of the underlying $\ttbb$ Born process 
in a way that leads to anomalously large weights as compared to the 
behaviour expected from the factorisation 
of soft and collinear radiation.
Fortunately, such anomalies arise only from events with finite transverse momenta
and not in the soft and collinear limits.
Moreover, the \PowhegBox framework
disposes of a mechanism that automatically attributes
such events to the so-called finite remnant, where QCD radiation is handled
as in fixed-order NLO calculations.  
This mechanism, which is controlled by the $\hbzd$ parameter in \refeq{eq:bzddef},
plays an important role 
for the efficiency of event generation. 
Moreover, it permits to avoid artefacts that can result from the
application of QCD factorisation and resummation far away from their
validity domain.

We have discussed predictions of the new \Powheg generator and theoretical
uncertainties for various $\ttbar+b$-jet cross sections and distributions
at the 13\,TeV LHC.
At variance with previous studies, in order to provide a better picture of
the perturbative convergence, we have evaluated QCD corrections using the
same $\as$ value and the same PDFs at LO and NLO.
The resulting NLO $K$-factors turn out to be close to two, even if the
renormalisation scale is chosen in a way that is expected to absorb large
logarithms associated with the running of $\as$.  The question 
of the origin of such large higher-order effects and the search for possible remedies, 
such as improved scale choices, deserve to be addressed 
in future studies.

Scale uncertainties at fixed-order NLO amount to 25--30\%
and are dominated by renormalisation-scale variations.
At NLOPS they tend to increase in a similar way as in 
\MadgraphaMC, while in \Sherpa they tend to decrease~\cite{deFlorian:2016spz}.
However this behaviour may be an artefact of the
incomplete implementation of scale variations in NLOPS generators.

Comparing predictions at NLO, LHE and LOPS level reveals significant shower
effects at the level of 10\% in the ttbb cross section and up to 30\% or
more for the invariant-mass and $\Delta R$ distributions of $b$-jet pairs.
These effects can be attributed to double $g\to \bbbar$
splittings~\cite{Cascioli:2013era} and are qualitatively and
quantitatively consistent with the findings of~\citeres{Cascioli:2013era,deFlorian:2016spz}.
For the $p_\rT$-distribution of light-jet radiation, 
the predictions of the new \Powheg generator are
quite close to fixed-order NLO and also
quite stable with respect to variations of the parameters 
$\hdamp$ and $\hbzd$, which 
separate real radiation into singular and finite parts.
This good stability is guaranteed by the 
$\hbzd$-dependent mechanism mentioned above.

To assess pure shower uncertainties we have compared \Powheg samples 
generated with \Pythia8 and \Herwig7.  
In addition, we have considered systematic uncertainties
due to the modelling of $g\to \bbbar$ splittings and the choice of $\as$ in
\Pythia.  At NLOPS, all shower uncertainties
turn out to be rather small and clearly subleading with respect to QCD scale
variations.
As a further independent estimate of matching and shower uncertainties we
have compared NLOPS $\ttbb$ generators based on \PowhegPythia and
\Sherpa finding remarkable agreement both for $\ttbar+b$-jet cross sections
and distributions.
We have also shown that matching and shower uncertainties increase
considerably if NLO corrections are not taken into account.  The same holds
for NLOPS generators of inclusive $\ttbar$ production as compared to 
$\ttbb$ generators.

Finally, we have presented predictions 
for $pp\to\ttbb$ with spin-correlated
top decays. In this context we have show that hadronisation and MPI effects
are almost negligible. Thus, the key features of the QCD dynamics 
of $\ttbb$ production at parton level are expected to 
hold true also at 
particle level after top decays.

The new $\ttbb$ \Powheg generator will be made publicly available in the near
future, and its application to experimental analyses may lead to 
significant steps forward in the understanding of the QCD dynamics of
$\ttbar+b$-jet production and in the control of the theoretical
uncertainties that plague $\ttbar H(\bbbar)$ searches.

\acknowledgments 
TJ would like to thank Emanuele Bagnaschi, Silvia Ferrario Ravasio, Alessandro Vicini and especially 
Paolo Nason for enlightening discussions.
This research was supported in part by the Swiss National Science
Foundation~(SNF) under contracts BSCGI0-157722, PP00P2-153027,
and CRSII2-160814, as well as from 
the Research Executive Agency of the European Union under the Grant Agreement
PITN--GA--2012--316704~({\it HiggsTools}).

\bibliography{paper}

\end{document}